\documentclass[referee,longauth]{aa}

\usepackage{graphicx}
\usepackage{txfonts}
\usepackage{lscape}
\usepackage{longtable}

\begin{document}

\title{
AKARI mid-infrared slitless spectroscopic survey of star-forming galaxies at $z\lesssim 0.5$
}

\titlerunning{AKARI MIR slitless spectroscopic survey of galaxies at $z\lesssim 0.5$}

\subtitle{}

\author{Y.~Ohyama
\inst{\ref{inst1}}
T.~Wada
\inst{\ref{inst2}}
H.~Matsuhara
\inst{\ref{inst2}}
T.~Takagi
\inst{\ref{inst2},\ref{inst3}}
M.~Malkan
\inst{\ref{inst4}}
T.~Goto
\inst{\ref{inst5},\ref{inst6}}
E.~Egami
\inst{\ref{inst7}}
H.-M.~Lee
\inst{\ref{inst8}}
M.~Im
\inst{\ref{inst8}}
J.-H.~Kim
\inst{\ref{inst8},\ref{inst9}}
C.~Pearson
\inst{\ref{inst10},\ref{inst11},\ref{inst12}}
H.~Inami
\inst{\ref{inst13},\ref{inst14}}
S.~Oyabu
\inst{\ref{inst15}}
F.~Usui
\inst{\ref{inst16},\ref{inst17}}
D.~Burgarella
\inst{\ref{inst18}}
F.~Mazyed
\inst{\ref{inst18}}
M.~Imanishi
\inst{\ref{inst19}}
W.-S.~Jeong
\inst{\ref{inst20}}
T.~Miyaji
\inst{\ref{inst21}}
J.~D\'iaz~Tello
\inst{\ref{inst21}}
T.~Nakagawa
\inst{\ref{inst2}}
S.~Serjeant
\inst{\ref{inst11}}
T.~T.~Takeuchi
\inst{\ref{inst15}}
Y.~Toba
\inst{\ref{inst1},\ref{inst23}}
G.~J.~White
\inst{\ref{inst10},\ref{inst11}}
H.~Hanami
\inst{\ref{inst22}}
\and
T.~Ishigaki
\inst{\ref{inst22}}
}

\institute{Academia Sinica, Institute of Astronomy and Astrophysics, 11F of Astronomy-Mathematics Building, AS/NTU, No. 1, Sec. 4, Roosevelt Rd., Taipei 10617, Taiwan, R.O.C. \\
\email{ohyama@asiaa.sinica.edu.tw}\\
\label{inst1}
\and
Institute of Space and Astronautical Science, Japan Aerospace Exploration Agency, 3-1-1 Yoshinodai, Chuo-ku, Sagamihara, Kanagawa 252-5210, Japan\\
\label{inst2}
\and
Japan Space Forum, Shin-Ochanomizu Urban Trinity Bldg. 2F 3-2-1, Kandasurugadai, Chiyoda-ku, Tokyo 101-0062, Japan\\
\label{inst3}
\and
Division of Astronomy and Astrophysics, University of California, Los Angeles, 430 Portola Plaza, Box 951547, Los Angeles, CA 90095-1547, USA\\
\label{inst4}
\and
Dark Cosmology Centre, Niels Bohr Institute, University of Copenhagen, Juliane Maries Vej 30, 2100 Copenhagen \O, Denmark\\
\label{inst5}
\and
National Tsing hua University, No. 101, Section 2, Kuang-Fu Road, Hsinchu 30013, Taiwan, R. O. C.\\
\label{inst6}
\and
Steward Observatory, University of Arizona, 933 N. Cherry Ave, Tucson, AZ 85721, USA\\
\label{inst7}
\and
Astronomy Program, Department of Physics and Astronomy, Seoul National University, Shillim-Dong, Kwanak-Gu, Seoul 151-742, Republic of Korea\\
\label{inst8}
\and
Subaru Telescope, National Astronomical Observatory of Japan, 650 North A\`ohoku Place, Hilo, HI 96720, USA\\
\label{inst9}
\and
RAL Space, STFC Rutherford Appleton Laboratory, Chilton, Didcot, Oxfordshire, OX11 0QX, UK\\
\label{inst10}
\and
Department of Physics and Astronomy, The Open University, Walton Hall, Milton Keynes, MK7 6AA, UK\\
\label{inst11}
\and
Oxford Astrophysics, Denys Wilkinson Building, University of Oxford, Keble Rd, Oxford OX1 3RH, UK\\
\label{inst12}
\and
National Optical Astronomy Observatory, 950 North Cherry Avenue, Tucson, AZ 85719, USA\\
\label{inst13}
\and
CRAL, Observatoire de Lyon, 9, avenue Charles Andr\'e, 69561 Saint Genis Laval, France\\
\label{inst14}
\and
Graduate School of Science, Nagoya University, Furo-cho, Chikusa-ku, Nagoya, Aichi 464-8602, Japan\\
\label{inst15}
\and
Graduate School of Science, The University of Tokyo, 7-3-1 Hongo, Bunkyo-ku, Tokyo 113-0033, Japan\\
\label{inst16}
\and
Center for Planetary Science, Graduate School of Science, Kobe University, 7-1-48 Minatojima-Minamimachi, Chuo-Ku, Kobe 650-0047, Japan\\
\label{inst17}
\and
Laboratoire d'Astrophysique de Marseille, P\^ole de l'\'Etoile Site de Ch\^ateau-Gombert 38, rue Fr\'ed\'eric Joliot-Curie 13388, Marseille cedex 13, France \\
\label{inst18}
\and
National Astronomical Observatory of Japan, 2-21-1 Osawa, Mitaka, Tokyo 181-8588, Japan\\
\label{inst19}
\and
Korea Astronomy and Space Science Institute, 776 Daedeokdae-ro, Yuseong-gu, Daejeon 34055, Republic of Korea\\
\label{inst20}
\and
Instituto de Astronom\'ia Campus Ensenada, Universidad Nacional Aut\'onoma de M\'exico, Km. 103 Carretera Tijuana-Ensenada, Ensenada Baja California, M\'exico. C.P. 22860\\
\label{inst21}
\and
Physics Section, Faculty of Humanities and Social Sciences, Iwate University, Morioka 020-8550, Japan\\
\label{inst22}
\and
Department of Astronomy, Kyoto University, Kitashirakawa-Oiwake-cho, Sakyo-ku, Kyoto 606-8502, Japan
\label{inst23}
}
\authorrunning{Y. Ohyama et al.}


\abstract
{
Deep mid-infrared (MIR) surveys have revealed numerous strongly star-forming galaxies at redshift $z\lesssim2$.
Their MIR fluxes are produced by a combination of continuum and Polycyclic Aromatic Hydrocarbon (PAH) emission features.
The PAH features can dominate the total MIR flux, but are difficult to measure without spectroscopy.
}
{
We aim to study star-forming galaxies by using a blind spectroscopic survey at MIR wavelengths to understand evolution of their star formation rate (SFR) and specific SFR (SFR per stellar mass) up to $z\simeq0.5$, by paying particular attention to their PAH properties.
}
{
We conducted a low-resolution ($R\simeq 50$) slitless spectroscopic survey at 5--$13\ \mu$m of $9\ \mu$m flux-selected sources ($>0.3$~mJy) around the North Ecliptic Pole with the Infrared Camera (IRC) onboard AKARI.
After removing 11 AGN candidates by using the IRC photometry, we identified 48 PAH galaxies with PAH 6.2, 7.7, and $8.6\ \mu$m features at $z<0.5$.
The rest-frame optical--MIR spectral energy distributions (SEDs) based on CFHT and AKARI/IRC imaging covering 0.37--$18\ \mu$m were produced, and analysed in conjunction with the PAH spectroscopy.
We defined the PAH enhancement by using the luminosity ratio of the $7.7\ \mu$m PAH feature over the $3.5\ \mu$m stellar component of the SEDs.
}
{
The rest-frame SEDs of all PAH galaxies have a universal shape with stellar and $7.7\ \mu$m bumps, except that the PAH enhancement significantly varies as a function of the PAH luminosities.
We identified a PAH-enhanced population at $z\gtrsim0.35$, whose SEDs and luminosities are typical of luminous infrared galaxies.
They show particularly larger PAH enhancement at high luminosity, implying that they are vigorous star-forming galaxies with elevated specific SFR.
Our composite starburst model that combines a very young and optically very thick starburst with a very old population can successfully reproduce most of their SED characteristics,
although we could not confirm this optically thick component from our spectral analysis.
}
{}


\maketitle

\section{Introduction}\label{introduction}

Mid-infrared (MIR) extragalactic studies have been providing new insights about galaxies in the distant universe, for three main reasons:
first, about half of the total energy throughout cosmic history is emitted between the MIR and far-infrared (FIR) wavelengths (e.g., \citealt{elbaz02,lefloch05,dole06,caputi07,goto11a}).
Second, the effect of dust extinction is much less prominent at MIR wavelengths when compared to optical (OPT) and near-infrared (NIR) wavelengths, and is a good spectral region for measuring activity from star formation as well as active galaxy nuclei (AGNs), even in the presence of copious amounts of dust.
Third, under limited technology at the time of AKARI \citep{murakami07} and before, in particular about the large cryogenic space telescope for sharper diffraction-limited resolution and the sensitive detector system, the MIR spectral region has been more sensitive to flux from distant astronomical sources than the FIR one.
The importance of deep MIR extragalactic surveys was first recognised by the discovery of strong evolution from $15\ \mu$m source counts by using ISOCAM \citep{isocam} onboard the {\it Infrared Space Observatory} ({\it ISO}; \citealt{iso}).
The rapidly evolving population was found as an excess of $15\ \mu$m sources at a flux of 0.1--0.5~mJy (e.g., \citealt{elbaz99,serjeant00}; see also \citealt{lagache04,wada08,pearson10}).
Later, extensive studies at MIR and other wavelengths helped to define the global spectral energy distribution (SED) shapes across the OPT--NIR--MIR--FIR for various types of luminous galaxies (such as AGNs, starburst galaxies, luminous infrared galaxies (LIRGs), and ultra-luminous infrared galaxies (ULIRGs); e.g., \citealt{spinoglio95, pearson01,lagache04,rowan04,lefloch05}).
Such studies clearly indicated that most of these galaxies, excluding AGNs, show prominent emission features in their MIR spectra, which are believed to originate in Polycyclic Aromatic Hydrocarbons (PAHs; e.g., \citealt{lutz98,xu98}).
The luminosity of the PAH features, as well as that of the underlying hot dust continuum, has been used as a measure of star formation rate (SFR), using conversions from MIR luminosity to the FIR one, where the bulk of the energy from star-forming regions is emitted (e.g., \citealt{genzel98,rigopoulou99,farrah07,shipley16}).

The unprecedented sensitivity of {\it Spitzer} at MIR wavelengths has greatly improved our understanding of cosmic galaxy evolution, with help of various diagnostics of galaxies for their activities up to $z\sim 2$--4 provided by both MIPS imager \citep{rieke04} at $24\ \mu$m and IRS spectrometer (\citealt{hock04}; e.g., \citealt{sajina07,yan07,pope08,wu10}; see also \citealt{spoon07}).
Cosmic star formation history, or star formation rate density (SFRD), in galaxies and/or AGNs has been particularly examined (e.g., \citealt{menendez07,farrah08,pope08,nordon12}. See also \citealt{goto10,goto11a,goto11b}).
In many studies, the analysis has relied on SED templates and/or models for scaling from the MIR wavelengths to the bulk of the dust emission in the FIR wavelengths.
The scaling is based mostly on studies of nearby galaxies and AGNs; there is no guarantee that it is appropriate at higher redshifts.
In some rare cases, extremely deep {\it Spitzer} FIR photometry was used to directly find global SEDs even at higher redshifts ($z\sim 2$--3; e.g., \citealt{lefloch05,bavouzet08,murphy11}).
Recent {\it Hershel} \citep{pilbratt10} FIR photometry has improved the situation (e.g., \citealt{berta11,elbaz11,gruppioni13,magnelli13}).
It turned out that their NIR--MIR--FIR SEDs are systematically different from scaled-up versions of local SED templates of presumably the same activity type (e.g., \citealt{menendez07,farrah08,pope08,elbaz11,murphy11,magdis12,nordon12}).
\cite{nordon12} argued that this introduces significant offsets in measuring SFRD.
They claimed that the MIR spectral features are not simple tracers of SFR, but that their power is modulated by changing physical conditions in the Inter Stellar Medium (ISM) or in Photo-Dissociation Regions (PDRs) (see also \citealt{elbaz11}).
This appears reasonable because such an effect has been indeed observed in spatially resolved SINGS studies of local galaxies \citep{sings}, as well as GOALS studies of luminous infrared galaxies \citep{goals}, where a range of MIR spectral features is seen within individual objects \citep{dale06}.
We note, however, that their integrated properties over the galaxy scale greatly smear out these local variations \citep{bavouzet08}.

Another serious problem in analysing galaxy evolution has been the uncertainty in the assumed $K$-correction used to derive rest-frame quantities (e.g., rest-frame MIR luminosity function and SFRD, \citealt{lefloch05,caputi07,bavouzet08,nordon12}).
The $K$-correction to rest-frame monochromatic MIR luminosity is particularly large for redshifted ($z\gtrsim 1$) galaxies because of the contribution of the PAH features.
{\it Spitzer} observed mainly in its IRAC \citep{irac} $8\ \mu$m and MIPS $24\ \mu$m bands, which miss the most prominent PAH features around $7.7\ \mu$m at $0.3<z<1.8$.
Without completely reliable redshift information, interpreting the observed in-band fluxes is not straightforward at MIR.
This also introduced a complicated selection function for {\it Spitzer} colour selection for higher-$z$ sources.
For example, sensitive MIPS $24\ \mu$m surveys favour both $z\sim 2$ star-forming galaxies with the redshifted prominent PAH~$7.7\ \mu$m feature and AGNs with very red continuum emission.
Distinguishing these possibilities has called for IRS spectroscopy (e.g., \citealt{yan04,yan07,pope08}).
Although the SED templates have been empirically calibrated to reproduce various observed correlations among broad-band photometric data (e.g., \citealt{chary01,lagache04}), the uncertainties in $K$-corrections are still large.
In particular, to make the MIR part of the templates realistic showing complex MIR spectral features, observed MIR spectra of small number of representative galaxies are often implanted on empirical low-resolution SED templates that are based on a simple synthetic model of dust emission \citep{dale01} or a galaxy stellar evolutionary model in combination with radiative transfer calculation through dusty circumstellar regions \citep{polletta07}.

AKARI was a cryogenic space infrared telescope that observed in the NIR, MIR, and FIR spectral regions \citep{murakami07}.
In addition to its primary mission to perform an all-sky survey \citep{murakami07,ishihara10}, some time was spent on the deeper pointing-mode observations of some specified targets during an intermittence of the scanning.
Multi-band deep extragalactic imaging surveys at NIR and MIR (2--$24\ \mu$m) were performed toward the North Ecliptic Pole (NEP) region by using the pointing mode (the AKARI NEP surveys; \citealt{maruma06} for the summary) with the wide-field Infrared Camera (IRC; \citealt{onaka07}).
The AKARI NEP surveys included as many contiguous bands as possible, with the data covering the entire wavelength range continuously with 9 filters at a spectral resolution of $R\simeq 5$.
With this filter set, one can discriminate range of SED types, including AGNs with red continuum-dominated SED, normal and starburst galaxies with hot dust continuum and PAH features, luminous infrared galaxies with strong silicate absorption (peaking at $9.7\ \mu$m), up to $z\sim 2$ (see, e.g., \citealt{takagi05,wada08,takagi10,hanami12}).
As \cite{takagi10} demonstrated (see also \citealt{hanami12,kim12}), the MIR colours of redshifted sources show extreme diversity, due to combination of the MIR features from a range of the SED types and redshift, and such rich multi-colour information can be used to extract various information about the nature of the sources.
\cite{takagi07,takagi10} and \cite{hanami12} have utilised SED fitting techniques to extract the activity type, redshift, extinction, MIR and FIR luminosities, and so on, from the complex SEDs with much less ambiguity than before.
During the course, they also showed that some observed SEDs could not be well reproduced by simple models.
Even if the fit seems successful, we need to be cautious, because it relies on local SEDs or very simple physical models to fit observations of redshifted galaxies in which physical conditions may be different from the local ones.
Because of the rapid evolution of galaxies peaking at $z\sim 2$ (e.g., \citealt{elbaz99,serjeant00,lagache04}), evolution of the SEDs should be examined and taken into account to interpret various observables, such as source counts and monochromatic MIR luminosities.
Spectroscopy of the MIR photometric sample of galaxies in the similar wavelength range would provide a recipe to properly interpret the photometric properties.

The IRC had not only multi-band imaging capability, but also a wide-band low-resolution spectroscopy capability \citep{ohyama07}.
This was possible because the IRC includes transmissive direct-view dispersers (a prism and grisms) in the filter wheel, as well as the broad-band imaging filters.
In the spectroscopy mode, by using the short slits at an edge of the field of view (FOV) of the IRC, spectroscopic studies of active galaxies have been extensively done.
Especially, the $3.3\ \mu$m PAH feature has been utilised to trace the star formation activities and to diagnose the AGN activity \citep{imanishi08,imanishi10,woo12,castro14,ichikawa14,yano16}.
In addition to the regular slit spectroscopy, the IRC could perform slitless spectroscopy:
all sources that can be imaged within its $\simeq 10'\times 10'$ FOV are dispersed by either a prism or grism.
This slitless mode was particularly well-suited to blind spectroscopic surveys of point-like sources:
The survey can be unbiased because there is no pre-selection of sources from, e.g., colour or flux at other wavelengths.
Instead, the sources are simply selected after the spectroscopy observations on the basis of their fluxes at the same wavelengths as for the spectroscopy.
This is particularly important for studying the MIR evolution of galaxies, because, as noted earlier, it is quite difficult to define a fair sample that is independent of types (including AGN) and strengths of the activities, and/or redshift for statistical studies.
Sensitive observations with this mode at MIR wavelengths are available only from space to avoid high atmospheric background, and the IRC was the first instrument that provided us with this unique observing opportunity.

In this paper, we first describe the design and observation of our survey program in Sect.~\ref{survey_section}, and data reduction procedure in Sect.~\ref{data_reduction_section}.
Our spectral PAH fit is described in Sect.~\ref{pahfit_section}, and the results are analysed in Sect.~\ref{spectral_characteristics_results}.
The OPT--NIR--MIR broad-band photometry is also compiled for the spectroscopic sample in Sect.~\ref{photo_data}, and its basic characteristics are analysed in Sect.~\ref{photo_characteristics_results}.
In particular, we photometrically classify activity types, normal/starburst galaxies and AGNs, in Sect.~\ref{photo_classification_results}.
For galaxies with the PAH features detected in their spectra, we analyse their colour-redshift relations in Sect.~\ref{colour_redshift_diagrams}, and construct their rest-frame SEDs in Sect.~\ref{rest_frame_seds}.
We then compare their spectroscopic and photometric properties in Sect.~\ref{spec_photo_comp}.
In particular, we compare PAH luminosities measured in both photometric and spectroscopic ways in Sect.~\ref{pah_luminosity_comp_section}, and characterise the variation of the rest-frame SED shape by the spectroscopic properties in Sect.~\ref{sed_variation_result}.
We here identify the PAH-enhanced population at $z\simeq 0.35$--0.5 as a distinctive subgroup of the PAH galaxies.
Next we compared the observed rest-frame SEDs with various SED templates and models in Sect.~\ref{SED_fit_result}.
We discuss implications of the SED variation of the PAH galaxies, in particular the PAH-enhanced population, for the star-formation properties in Sect.~\ref{SFR_sSFR_discussion}.
We finally summarise advantages and limits of our slitless spectroscopic and photometric data analysis in Sect.~\ref{method_advantage_limit}.
Conclusions are given in Sect.~\ref{conclusion_section}.
We use $H_0=70$ km s$^{-1}$ Mpc$^{-1}$, $\Omega_{\rm m}=0.3$, and $\Omega_{\rm \lambda}=0.7$ throughout this paper.

\section{Observations and data}

\subsection{The {\it SPICY} survey: basic design and observations}\label{survey_section}

We conducted our IRC slitless spectroscopic survey, ``slitless SpectroscoPIC surveY of galaxies'' ({\it SPICY}), to study strongly evolving population discovered with {\it ISO} that shows an excess in source count at a flux range of 0.1--0.5~mJy at $15\ \mu$m (e.g., \citealt{elbaz99,serjeant00,lagache04,wada08,pearson10}; Sect.~\ref{introduction}).
Given the typical observed flux ratio between $15\ \mu$m and $9\ \mu$m (Sect.~\ref{photo_classification_results} below), this flux range corresponds to 0.5~mJy or smaller at $9\ \mu$m, where the IRC provides the best MIR sensitivity. 
We targeted the NEP with IRC spectroscopy, where deep extragalactic photometric studies with AKARI/IRC, NEP-Deep (coverage: 0.57~deg$^2$; $5\ \sigma$ sensitivity at 7--$12\ \mu$m: $33.6\ \mu$Jy; \citealt{wada08,takagi12,murata13}) and NEP-Wide (coverage: 5.4~deg$^2$; $5\ \sigma$ sensitivity at 7--$12\ \mu$m: $67.3\ \mu$Jy; \citealt{lee09,kim12}) surveys, were conducted.
This {\it SPICY} survey thus provides NIR--MIR spectra of sources that have also been detected with the IRC photometric surveys, enabling direct comparisons of the observed characteristics, redshift and PAH luminosity in particular, measured in both photometric and spectroscopic ways.
Another advantage of targeting the NEP is that we can utilise many associated surveys at different wavelengths, ranging from X-ray \citep{krumpe15}, ground-based OPT--NIR \citep{jeon10,ko12,jeon14,oi14}, to sub-millimetre \citep{geach17}.

\begin{figure*}[t] 
\centering
\includegraphics[width=17cm]{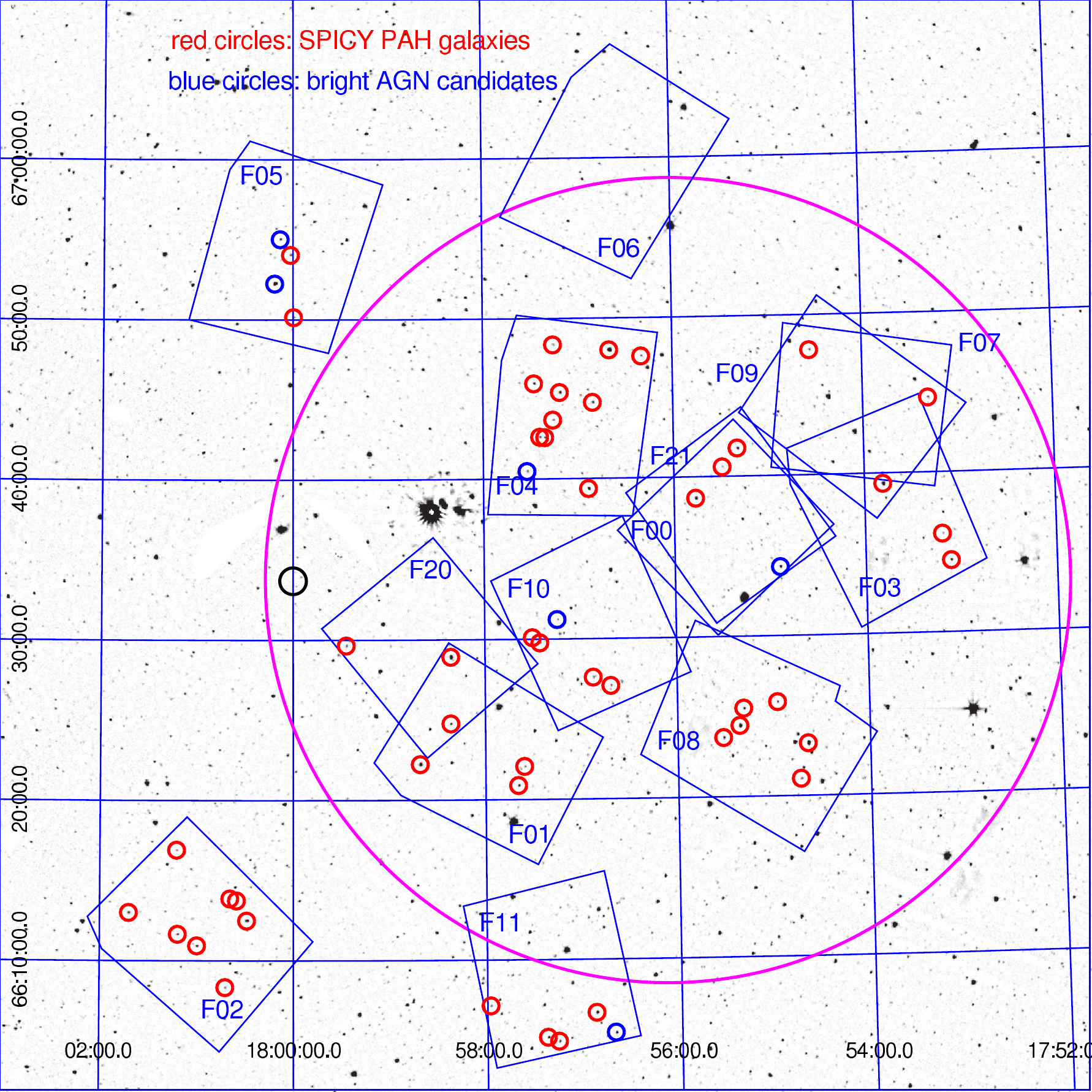}
\caption{{\it SPICY} survey tiles marked on the AKARI/IRC NEP-Wide S9W image \citep{lee09,kim12}.
North is up and east is to the left.
The NEP is marked with a small black circle, and the NEP-Deep survey field is indicated with a big magenta circle.
Blue rectangles show approximate locations of individual tiles.
The galaxies with detectable PAH features (the PAH galaxies; Sect.~\ref{pahfit_section}) and bright AGN candidates (Sect.~\ref{photo_classification_results}) are marked with small red and blue circles, respectively.
}
\label{SPICY_tiles}
\end{figure*}

We designed the {\it SPICY} survey to utilise the IRC spectroscopy capability for optimum survey outputs (see Appendix~\ref{survey_detail} for the details).
In this paper, we focus only on the short MIR camera within the IRC, ``MIR-S'' \citep{onaka07}, to cover 5--$13\ \mu$m, although the survey was designed to utilise all three cameras of the IRC for its full wavelength coverage (2.5--$24\ \mu$m).
This MIR-S camera is suited for studying galaxies at $z\lesssim 0.5$ by using prominent PAH features at 6--$10\ \mu$m in the rest frame.
We visited the same pointing coordinates 9 or 10 times to achieve the sensitivity goal of 0.5~mJy at $9\ \mu$m for a tile of $\simeq 10'\times 10'$ corresponding to one FOV of this camera.
The actual tile shape, after stacking all individual pointing data, was slightly elongated and distorted because of a slight field rotation among the observations, which is unavoidable due to constraints posed by the satellite's orbit and the NEP coordinates.
We aimed to overlap tiles of all three different cameras of the IRC to increase the overall wavelength coverage.
As a result, 14 tiles were distributed in a non-contiguous way around the NEP, in a form of a complicated shape of folded chains (Fig.~\ref{SPICY_tiles}).
We tried to concentrate on the NEP-Deep field where more observations at other wavelengths are available, but some tiles were made within the surrounding Wide survey field, which encompasses the Deep field.

We used a standard AOT (Astronomical Observing Template) for IRC spectroscopy, AOT04a \citep{onaka07}, for the {\it SPICY} observations.
Within one pointing observation lasting about $\simeq 10$ minutes, several spectroscopic images through dispersers, as well as a few direct images through a broad-band filter, were taken in this AOT.
In the MIR-S camera, both lower-resolution ($R\simeq 50$) dispersers (``SG1'' and ``SG2'' grisms covering 4.6--$9.2\ \mu$m and 7.2--$13.4\ \mu$m, respectively) and a broad-bang filter (``S9W'', covering 6.7--$11.6\ \mu$m with an effective wavelength of $9.0\ \mu$m) were used.
The direct image, also called a reference image, was used to provide wavelength reference points for all sources on the spectroscopic images.
No telescope dithering was made within the AOT, to ensure that the reference image provides accurate wavelengths.
Within one AOT operation, as many as 12 SG1 images, 12 or 16 SG2 images, and 3 S9W images, were taken, and the effective exposure time was 6.36 sec for each image independently of the filters/grisms used.
In total, exposure times were 4796 sec (for 12 images per pointing and 9 pointings)--7106 sec (for 16 images per pointing and 10 pointings) for both SG1 and SG2 spectra per tile.

\subsection{Data reduction}\label{data_reduction_section}

We reduced the {\it SPICY} data by using the IRC spectroscopy ``toolkit'' (collection of software and calibration database) modified for our multi-pointing observations.
The original toolkit was developed by the instrument team\footnote{The spectroscopy toolkit is available at http://www.ir.isas.jaxa.jp/AKARI/Observation/support/IRC/.} to reduce single AOT observation dataset \citep{ohyama07}.
This toolkit is composed of the two parts: an image processing/calibration pipeline, and a spectrum plotting tool.
The former is to reduce raw images to generate calibrated and stacked spectral images, and the latter is to extract and plot one-dimensional spectra.
Both parts are based on calibration database provided with the toolkit.
The pipeline performs standard array image processing such as dark subtraction, flat-fielding, and background subtraction, and also extracts two-dimensional spectral images for individual sources detected on a reference image, and stacks them.
To work on the multiple AOT observation datasets, we modified the pipeline while keeping all original calibration routines and database.
Specifically, we split the original pipeline into two pieces, one for image processing/calibration and source extraction, and another for stacking the pre-calibrated/extracted spectral images.
We then added a simple file organising mechanism between them, in order to collect all individually calibrated images taken at different AOT observations for the same tile before stacking all of them at the same time.

\begin{figure}[t] 
\centering
\resizebox{\hsize}{!}{\includegraphics{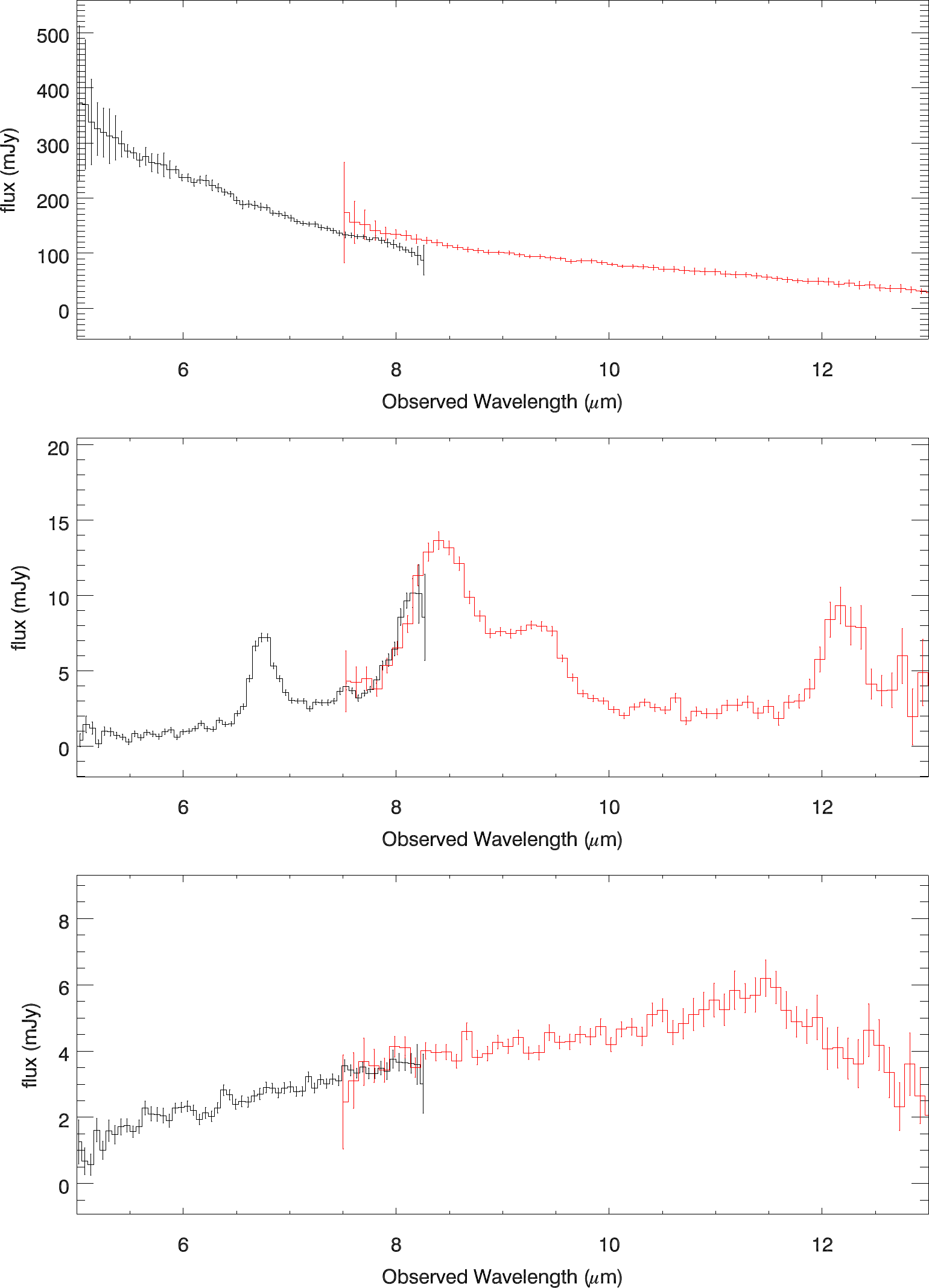}}
\caption{Examples of the {\it SPICY} spectra of three representative types of bright sources.
The SG1 (black) and SG2 (red) spectra are shown separately.
The error bars are for one sigma.
{\it top}: a field star.
{\it middle}: a galaxy with prominent PAH 6.2, 7.7, 8.6, and $11.3\ \mu$m features (F04-0 at $z_{\rm PAH}=0.087\pm 0.001$; see also Fig.~\ref{IRC_spec_PAH_galaxies}).
{\it bottom}: an AGN candidate with a red featureless continuum and possibly a deep silicate $9.7\ \mu$m absorption redshifted to $\simeq 14\ \mu$m (F05-1 at $z_{\rm opt.~spec}=0.4508$; see Sect.~\ref{photo_classification_results}).
Slight flux offset of the star spectrum within the overlapping wavelength range of the SG1 and SG2 spectra (7.5--$8.2\ \mu$m) is due to slight systematic flux calibration error caused by errors in aligning the SG1 and SG2 spectral images.
Increased noise level at $\gtrsim 12.0\ \mu$m in the {\it middle} panel is caused by partial truncation of the SG2 spectrum.
}
\label{spicy_example}
\end{figure}

In order to run the spectroscopy pipeline for sources that are too faint to be clearly detected on a reference image of a single AOT observation, one needs to find their positions by using multiple AOT observations beforehand.
This is because the pipeline cannot apply the wavelength-dependent calibration without knowing the wavelength reference point.
Therefore, we first generated a master reference image by stacking all reference images of multiple AOT observations for a tile.
We used the IRC imaging toolkit \citep{ita08}\footnote{The imaging toolkit is the data reduction software for the IRC imaging datasets developed by the instrument team. It is also available at http://www.ir.isas.jaxa.jp/AKARI/Observation/support/IRC/.} to stack the reference images.
In this imaging toolkit, the flux scale and sky coordinates (RA and DEC) are calibrated after standard array image processing such as dark subtraction, flat-fielding, and background subtraction.
We detected sources in the master reference image, and created a master source catalogue that includes their fluxes and sky coordinates.
To meet our sensitivity goal, sources that are brighter than 0.3~mJy at S9W ($9.0\ \mu$m) were selected for processing with the spectroscopy pipeline.
We also reduced the individual reference images for each AOT observation in the same way as reducing the master reference image.
We then transformed the sky coordinates of the master reference catalogue to the pixel coordinates of individual single-AOT reference images to generate source tables.
We next ran the first part of the spectroscopy pipeline (image processing/calibration and source extraction) for individual AOT observations with their corresponding source tables.
We here used our additional file organising software tool to organise lists of images to be stacked for the same sources taken at different AOT observations, and stacked the pre-calibrated/extracted spectral images for each of the sources by using the second part of the spectroscopy pipeline.
By using the original plotting tool, we finally extracted fully calibrated one-dimensional spectra, and saved the results for further analysis.
Examples of the final processed spectra of bright sources showing very different spectral shapes are shown in Fig.~\ref{spicy_example}.

Due to slitless nature of the observing mode, we obtained spectra of all kinds of compact sources within a FOV.
We removed field stars based on their (very bright and point-like) appearance on the CFHT optical images \citep{hwang07}, and found 171 extragalactic sources (galaxies and AGNs).
We found that all PAH galaxies (sources with detectable PAH features; Sect.~\ref{spectral_characteristics_results}) are extended at the resolution of FWHM$\simeq 1.0''$;
Their stellarity indices, which quantify morphological similarity of objects to point-like sources and are often used for shape-based star--galaxy classification \citep{sextractor}, are $<0.1$ (0 for galaxies and 1 for stars) on the CFHT optical images \citep{hwang07}.
Among all 171 spectra, some fraction of the spectra was not useful in our analysis.
Some of them were heavily contaminated by nearby sources on the slitless spectral images, and some others near the edge of the FOV have only truncated spectra at the edge of the detector array\footnote{
The FOV of the IRC imaging mode occupies almost the entire detector footprint.
By inserting a direct-view disperser (grism) in place of a broad-band filter, the dispersed spectral images of sources near the edge of the imaging FOV extend beyond the detector footprint, resulting in the truncation.
}.
We note that these problems randomly damage the spectra depending on distributions of the sources and their neighbours.
Among the sources with useful spectra, eight sources were observed twice within the overlapping tiles (Sect.~\ref{survey_section}), duplicating the spectra for such sources.
For the brightest four such sources, we adopted ones that are less affected by the contamination.
For the remaining four such sources, we coadded the spectra to improve the signal-to-noise.

\subsection{Spectral PAH fit}\label{pahfit_section}

\subsubsection{Method}\label{pah_fit_method}

\begin{figure*}[t] 
\centering
\includegraphics[width=17cm]{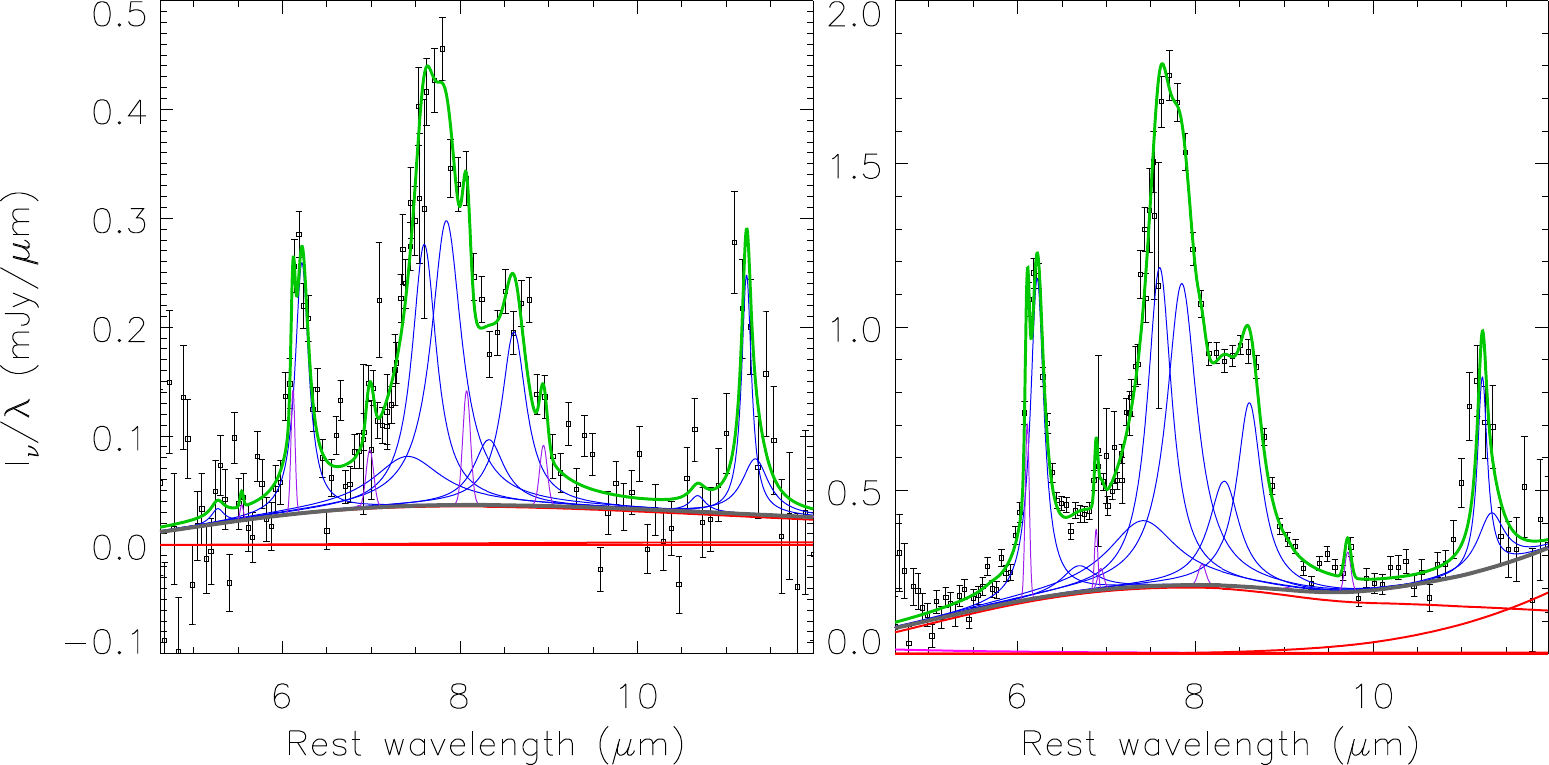}
\caption{Results of the PAHFIT for two nearby {\it SPICY} PAH galaxies.
Individual PAH features (blue), narrow emission lines (magenta), dust continua (multiple components for different temperatures; red), stellar continuum (magenta), sum of the continua (dust and stellar; grey), and sum of all components (continua and spectral features; green) are shown for each galaxy.
}
\label{pahfit_examples}
\end{figure*}

Many {\it SPICY} spectra show 6.2, 7.7, 8.6, and $11.3\ \mu$m PAH features (hereafter, PAH~$6.2\ \mu$m, PAH~$7.7\ \mu$m, PAH~$8.6\ \mu$m, and PAH~$11.3\ \mu$m, respectively), and they were examined by using spectral PAH fitting.
We first examined some bright sources showing prominent PAH features with PAHFIT software \citep{smith07}.
This software relies on a priori information on shapes of individual PAH features (a set of central wavelengths and widths, assuming a Drude profile) and central wavelengths of narrow ionised/atomic/molecular lines.
With external information on redshift, the software finds strengths of these spectral features, as well as the underlying continuum and amount of extinction, by means of a chi-square minimisation.
Figure~\ref{pahfit_examples} shows the results of two nearby bright galaxies as examples.
Here, we assumed a fully mixed extinction geometry and the extinction curve toward the Galactic centre developed by \cite{smith07}.
We adopted redshifts measured with our own PAH fit software (see below).
Although we found reasonably good fits on them, the spectral model of PAHFIT is too detailed for analysing the {\it SPICY} spectra.
This is understandable because PAHFIT was designed for the IRS low-resolution spectra ($R=60$--130; $\simeq 5$--$35\ \mu$m), which provide higher spectral resolution and wider wavelength coverage than those of the {\it SPICY} spectra ($R\simeq 50$; $\simeq 5$--$13\ \mu$m).

\begin{figure*}[t] 
\centering
\includegraphics[width=17cm]{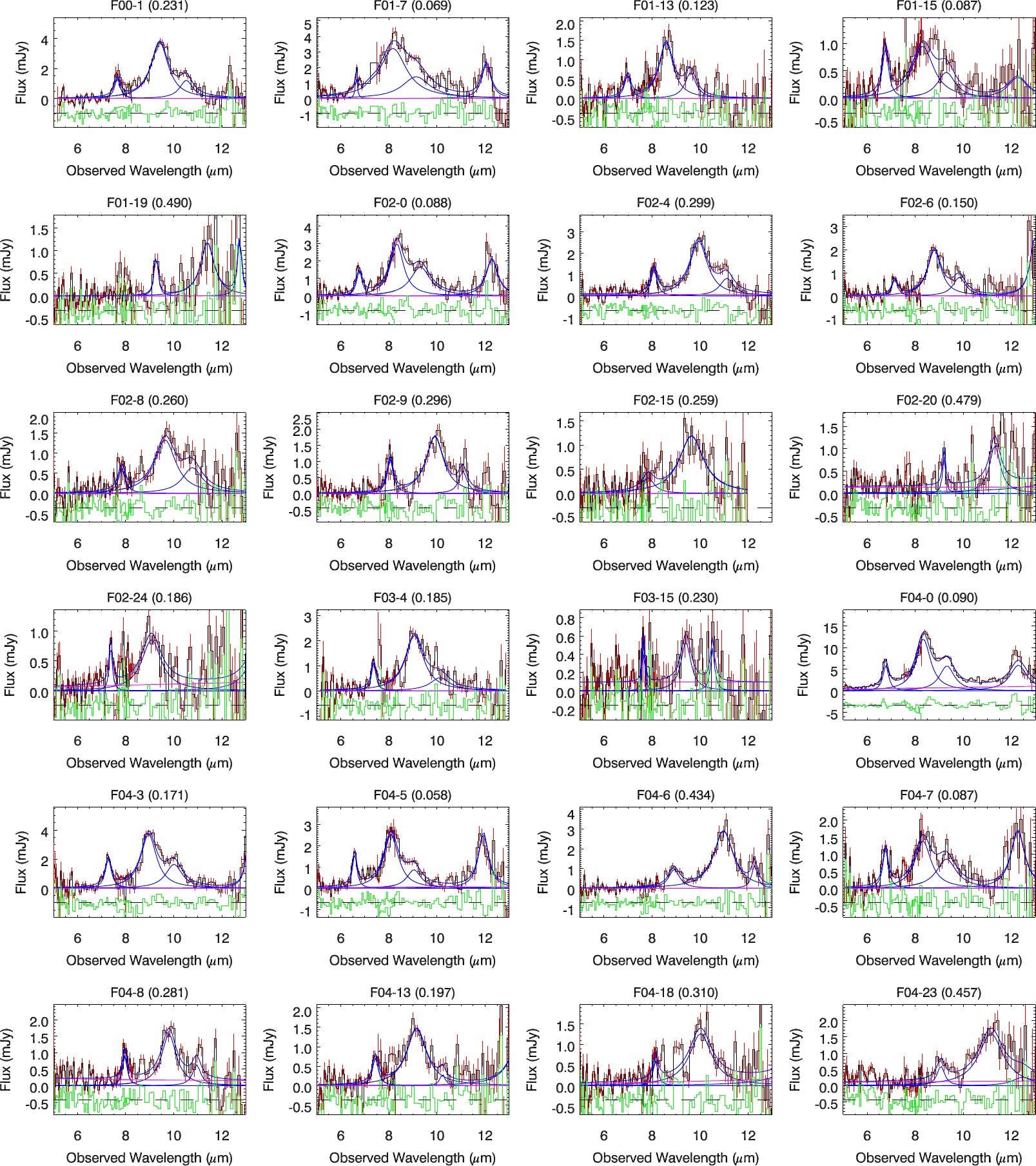}
\caption{The SG1 and SG2 spectra of the {\it SPICY} PAH galaxies with the PAH fit results overlaid.
The observed spectrum is shown in red line with one-sigma error bars, with the fitted individual PAH components (blue), the power-law continuum (magenta), and their sum (blue) in each panel.
The residual of the fit (observed$-$fitted) is shown in green at offset baseline (horizontal broken line) for clarity.
The redshift from the PAH fit is indicated next to the source name in the plot title.
The Y axis is scaled to have the same PAH $7.7\ \mu$m peak hight for all galaxies.
}
\label{IRC_spec_PAH_galaxies}
\end{figure*}

\begin{figure*}[t] 
\centering
\includegraphics[width=17cm]{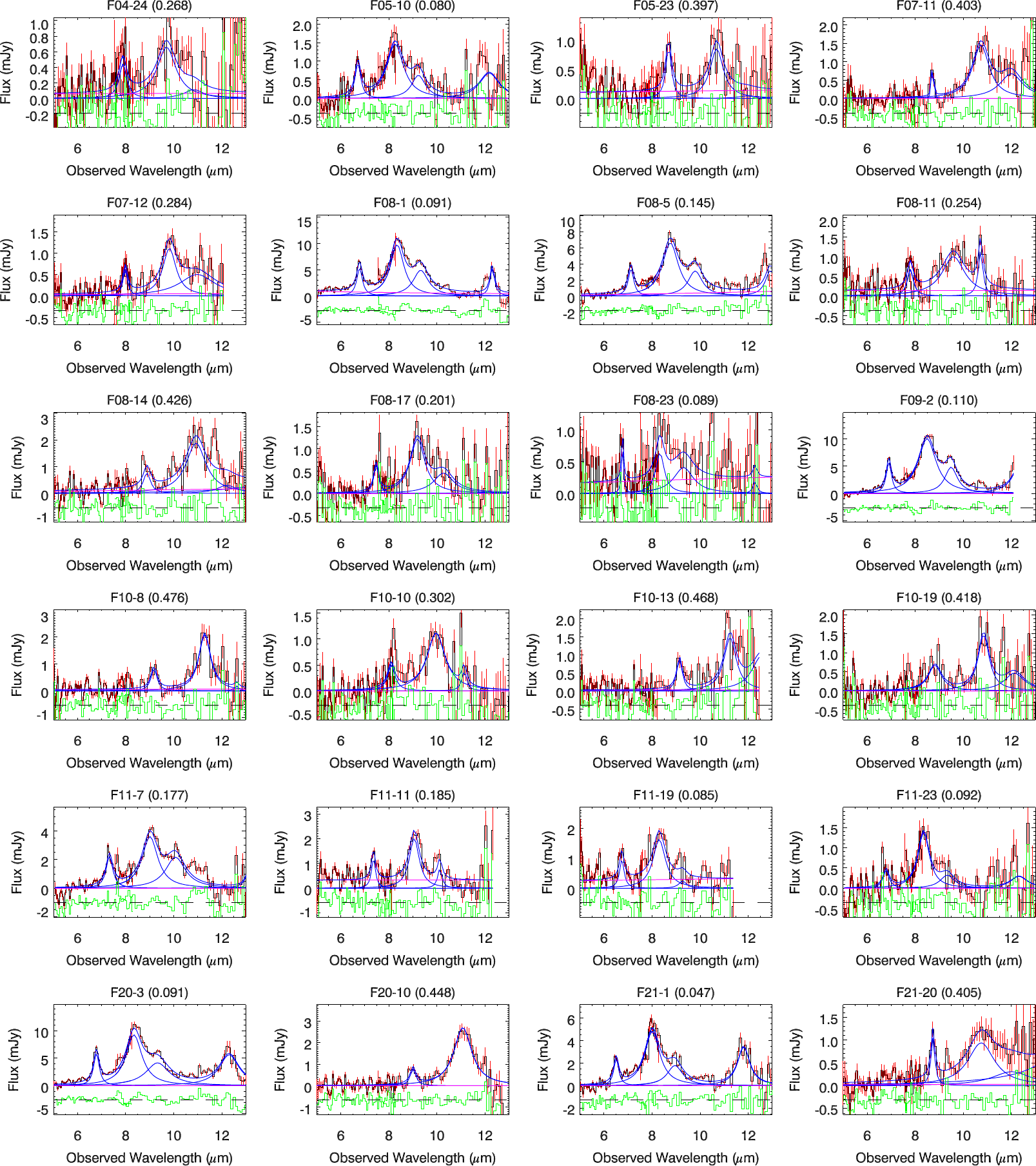}
\caption{Figure~\ref{IRC_spec_PAH_galaxies}. continued.}
\label{IRC_spec_PAH_galaxies_2}
\end{figure*}

We developed our own spectral PAH fit (hereafter, simply ``the PAH fit'') to apply to the {\it SPICY} spectra.
Most of the sources in the slitless surveys are faint
and serendipitously detected.
We aimed to establish a way to robustly extract fundamental properties of star-forming galaxies from the slitless survey data.
To identify sources on the spectral images, it seems more efficient to use (only) the PAH features that are prominent in terms of both line width and flux.
Therefore, we designed this software to detect (even very faint) PAH features on low-resolution IRC spectra, and robustly measure their redshifts and PAH strengths, without any a priori information about the sources.
For this purpose, we adopted a simple spectral model with only four major PAH features (at 6.2, 7.7, 8.6, and $11.3\ \mu$m) at the same redshift on a power-law continuum, to minimise number of the free parameters.
Each PAH feature is known to show an extended wing in its profile that contains much more power than in a Gaussian profile (e.g., \citealt{smith07}).
We compared the Lorentzian and Drude \citep{smith07} profiles by using the PAH fit on the {\it SPICY} spectra, and found only negligible differences, and adopted the Lorentzian profile simply because of its simpler functional form.
Narrow (unresolved) emission lines were not included in the fit, because no such lines were clearly detected in our spectra due to low spectral resolution (Figs.~\ref{IRC_spec_PAH_galaxies},~\ref{IRC_spec_PAH_galaxies_2}).
Free parameters in our fit are the shape parameters (Lorentzian amplitudes and width (half width at half maximum; HWHM)) of the four PAH features, slope and amplitude of the power-law continuum, and redshift.
We required the width of all PAH features to be larger than the instrumental spectral resolution.
The fit was made by means of a chi-square minimisation.

\subsubsection{Tests}\label{pah_fit_test}

\begin{figure*}[t] 
\centering
\includegraphics[width=17cm]{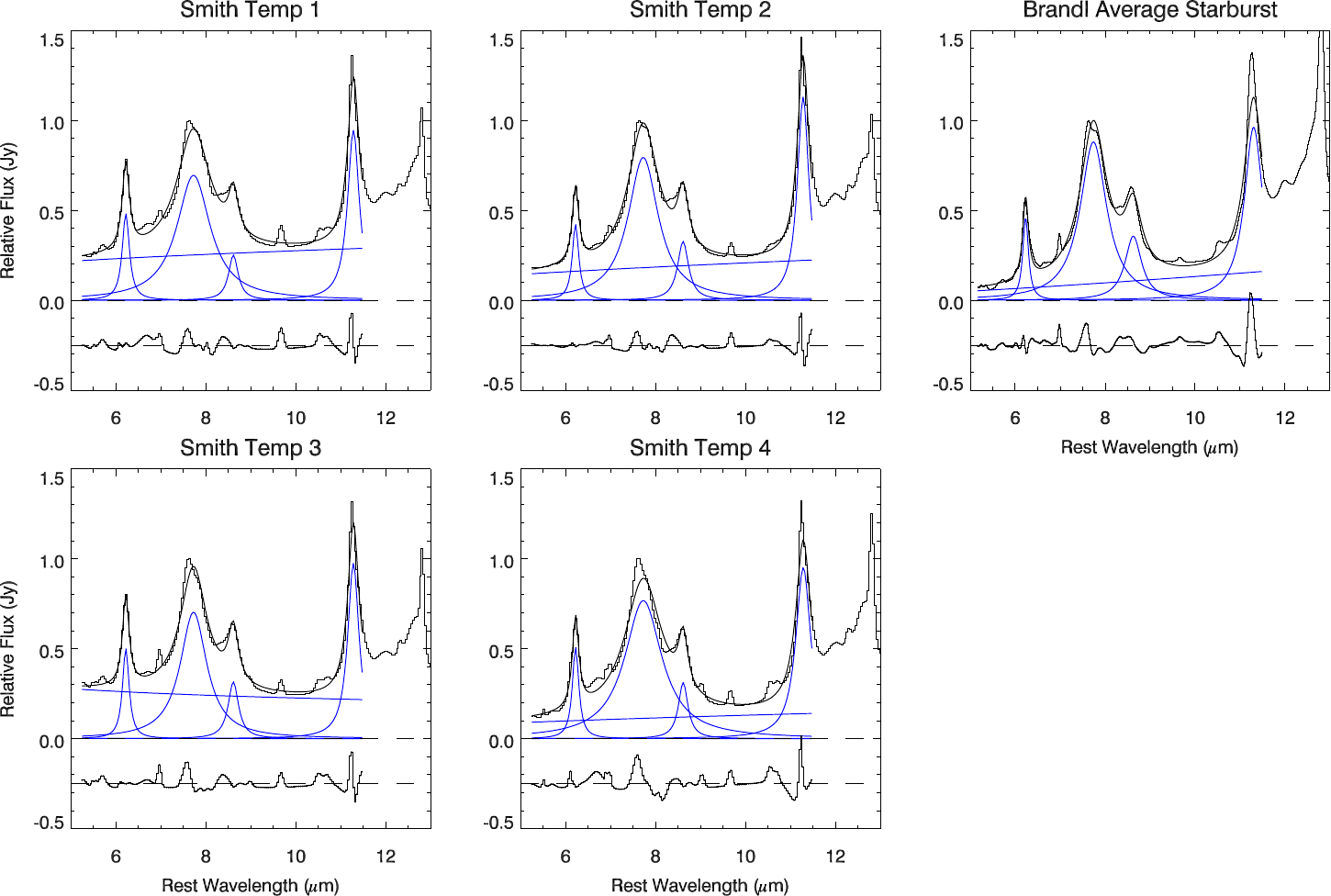}
\caption{Representative IRS spectra of star-forming galaxies with the PAH fit results overlaid.
An averaged spectrum of nearby starburst galaxies of \citet{brandl06} ({\it top right}) and four ``noise-free galaxy spectral templates'' of \cite{smith07} (remaining four panels) are shown in black lines.
The individual PAH features (blue), the power-law continuum (blue), and their sum (black) are overlaid in each panel.
The residual of the fit (observed$-$fitted) is shown at offset baseline (horizontal broken line) for clarity.
}
\label{PAH_fit_over_IRS}
\end{figure*}

We tested the PAH fit on ``noise-free galaxy spectral templates'' of \cite{smith07} and an averaged spectrum of nearby starburst galaxies of \cite{brandl06} (Fig.~\ref{PAH_fit_over_IRS}), both of which were obtained with the IRS low-resolution mode.
We note that \cite{smith07} constructed the four templates to cover ranges of PAH inter-band flux ratios in their sample.
We found the following problems in the fit, and revised the fitting procedure to apply to the {\it SPICY} spectra.
One problem is the systematic residual (observed$-$fitted) features, in particular around the peaks of the PAH~$7.7\ \mu$m and $11.3\ \mu$m, as well as at narrow emission lines of, e.g., [Ar~{\sc ii}] at $7.0\ \mu$m and H$_2$ 0--0 $S$(3) at 9.7~$\mu$m.
Such residuals are naturally expected because the PAH~$7.7\ \mu$m complex is actually composed of multiple PAH components \citep{smith07}, and these narrow lines are not included in the spectral model.
Therefore, we decided to ignore this problem.
Another problem is that a pseudo plateau at red side of the PAH~$11.3\ \mu$m up to $\sim 13\ \mu$m, which is made by a collection of weak PAH features (see, e.g., Fig.~4 of \citealt{smith07}), cannot be well reproduced with our simple spectral model.
To mitigate this problem, we decided to fit the spectra only below $11.5\ \mu$m in the rest frame just to include the peak of the PAH~$11.3\ \mu$m and a bit of its red side.
To work on the actual redshifted spectra, we need to tailor the fitting wavelength according to the source redshift by following the two steps:
First, we fit the entire spectral region to find approximate redshift ($z_{\rm 0}$).
This redshift is almost accurate because its measurement relies mostly on more prominent PAH $6.2\ \mu$m and $7.7\ \mu$m.
Second, we restrict the fitting wavelength up to $11.5\times(1+z_{\rm 0})\ \mu$m to find the final fit results.

We also tested the PAH fit on the {\it SPICY} spectra, and found that the fit sometimes returns unusually wide PAH~$11.3\ \mu$m profile width when compared to the results on the IRS templates described above, particularly when the signal-to-noise is low.
This is most likely because we cannot firmly constrain the PAH~$11.3\ \mu$m profile due to limited wavelength coverage, particularly when this feature comes near the end of our wavelength coverage due to redshift.
This feature goes out of the wavelength coverage at $z\gtrsim 0.2$.
Therefore, we limited the allowed range of this width in the fit just to include the measured widths on the IRS templates, and considered the fit valid even when the best fit is found at the limits.
We found, by changing this allowed width range, that other fit parameters (parameters of other PAH features and redshift) are insensitive to this fit condition.
Because of such problem in the profile fitting, we decided not to discuss the PAH~$11.3\ \mu$m properties in detail in this paper.

We did not include extinction in the PAH fit on the {\it SPICY} spectra, and here discuss potential problems in neglecting the extinction and how to mitigate them in our following analysis.
We tested the extinction effect with an extinction fitting in the PAH fit, and found that the extinction cannot be well constrained for many sources due to limited wavelength coverage at long side of the redshifted silicate absorption feature.
We adopted an extinction curve toward Galactic centre of \cite{chiar06} that shows a prominent broad (8--$12\ \mu$m) silicate absorption feature peaking at $9.7\ \mu$m, and assumed a screen geometry.
At $z<0.15$, we found that the optical depth at $9.7\ \mu$m is typically $\tau(9.7\ \mu$m$)\lesssim 0.01$ (or $A_\mathrm{V}\lesssim 0.1$ mag) for bright sources, and $\tau(9.7\ \mu$m$)\lesssim 0.1$ (or $A_\mathrm{V}\lesssim 1$ mag) with very large uncertainties for fainter sources.
At $z\gtrsim 0.2$, we could not constrain the extinction for most cases, because peak of the PAH~$11.3\ \mu$m goes out of our wavelength coverage, and PAH-free region at red side of the PAH~$8.6\ \mu$m comes at the end of the coverage.
The silicate absorption profile is not fully covered at this redshift range, and quality of the extinction measurement becomes worse particularly when the signal-to-noise is low.
Given the uncertainties of the extinction, measurements of the PAH~$7.7\ \mu$m and PAH~$6.2\ \mu$m are more robust than that for the PAH~$11.3\ \mu$m.
This is because the silicate absorption profile extends to $>11.3~\mu$m, whereas it does not to $<7.7~\mu$m.
In contrast, the redshift measurement is little affected by the extinction, because it relies much more on fits over the PAH~$6.2\ \mu$m and $7.7\ \mu$m.
Therefore, we decided not to include extinction in the fit for all sources, and will focus mostly on the luminosities of both PAH~$6.2\ \mu$m and $7.7\ \mu$m, as well as redshift, in our following analysis.
See also Sect.~\ref{113_77_variation} for more discussion about the extinction based on photometric information.

\begin{figure} 
\centering
\resizebox{\hsize}{!}{\includegraphics{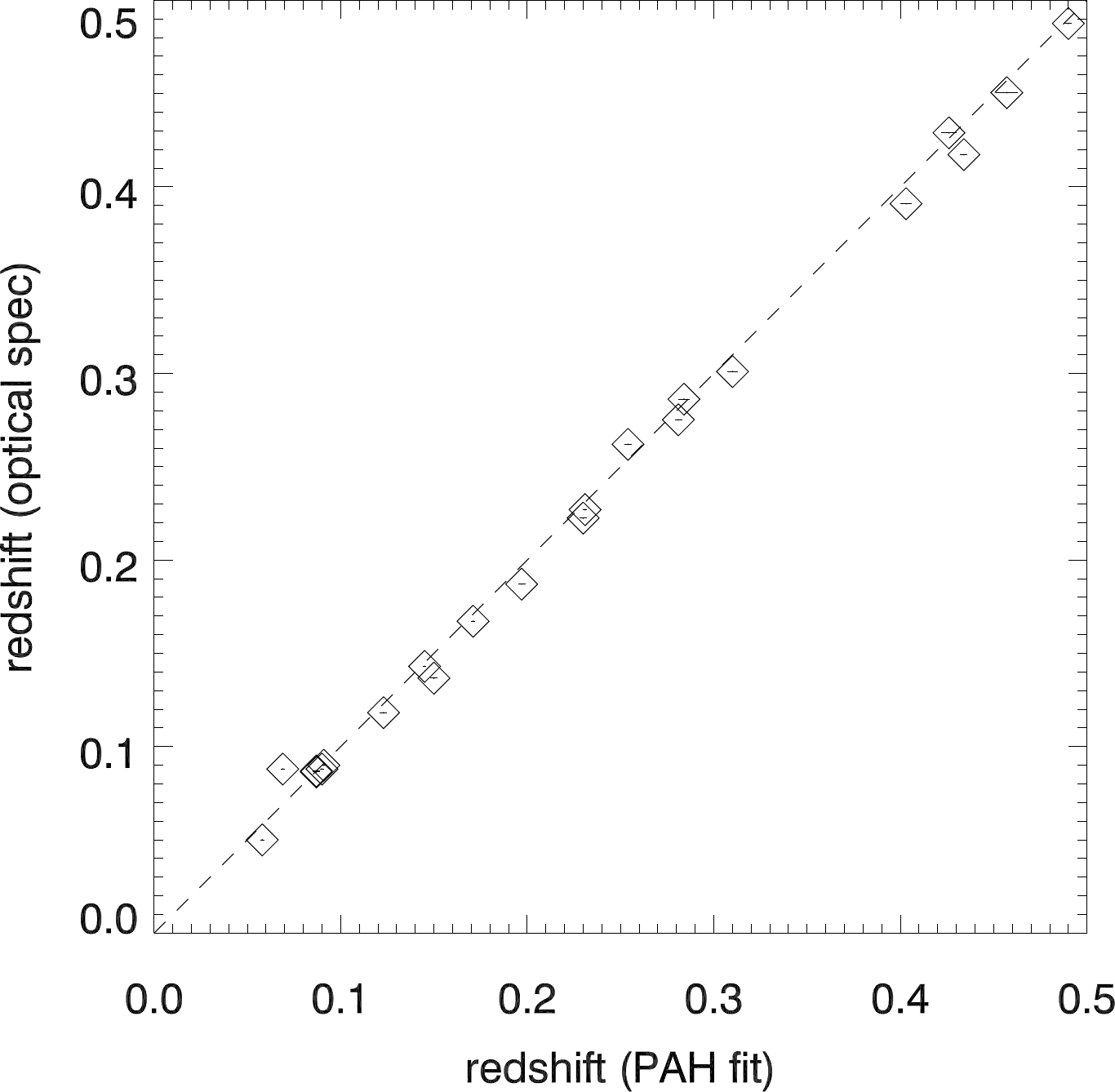}}
\caption{Comparison of the redshifts between the PAH fit and optical spectroscopy for the PAH galaxies.
Error of the optical spectroscopic redshift is not shown, but we plot only the sources with good redshift measurements (``quality flag'' of 3 or 4 in \citealt{shim13}; see text).
The diagonal broken line is for the same redshifts in the two measurements.}
\label{redshift_consistency_check}
\end{figure}

We finally examined accuracy of redshifts from the PAH fit by comparing to those from the optical spectroscopy.
We identified 32 galaxies with detectable PAH features in the {\it SPICY} spectra with secure\footnote{We adopted the optical redshifts if their quality flags are either 4 (secure; identified by more than two features) or 3 (acceptable and almost good; identified by two features) according to \cite{shim13}.} optical redshifts either by \cite{shim13}, \cite{oi14}, or our own observations (Table~\ref{table3}).
In our own observations, we performed optical spectroscopy with the OSIRIS at the Gran Telescope Canarias (GTC) in 10A and 14A semesters (PI: T.~Miyaji).
We found that redshifts from the PAH fit are accurate at a level of 1\% or less in $\mathrm{d}z/(1+z)$ (Fig.~\ref{redshift_consistency_check}).

\subsubsection{Results}\label{pah_fit_results}

We fitted all extragalactic {\it SPICY} spectra with the PAH fit, and identified 48 ``PAH galaxies'' with successful PAH fits (Figs.~\ref{IRC_spec_PAH_galaxies},~\ref{IRC_spec_PAH_galaxies_2}).
We refer to those without PAH detections (i.e., sources with unsuccessful PAH fits) as ``non-PAH galaxies''.
Some of the non-PAH galaxies are bright and photometrically similar to the PAH galaxies (Sect.~\ref{photo_classification_results}), but we could not detect their PAH features due to problems in the spectral data (contamination and truncation; Sect.~\ref{data_reduction_section}).
Other non-PAH galaxies show either lower signal-to-noise spectra or intrinsically weak PAH features (e.g., elliptical galaxies without prominent PAH features or AGNs), and the PAH fit failed to detect the features.
Table~\ref{table1} summarises numbers of sources in each category.
Table~\ref{table2} shows source identification, cross-identification with the AKARI/IRC NEP-Wide catalogue \citep{lee09,kim12}, and the {\it SPICY} S9W ($9.0\ \mu$m) photometry of the PAH galaxies, as well as cross-identification with the CFHT optical NEP photometric catalogue \citep{hwang07}.

\begin{figure*}[t] 
\centering
\includegraphics[width=17cm]{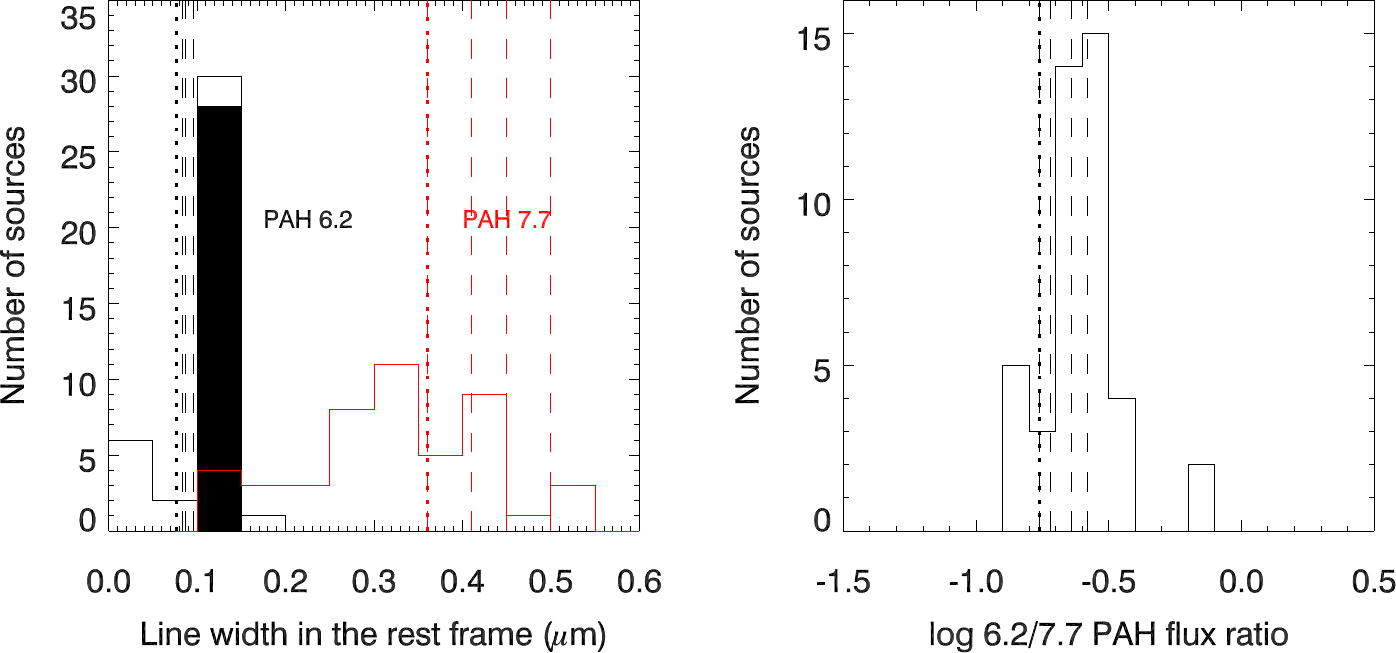}
\caption{Distributions of the PAH shape parameters of the PAH galaxies.
{\it Left}: PAH widths (Lorentzian half width at half maximum (HWHM) in the rest frame) of the PAH~$6.2\ \mu$m (black) and the PAH~$7.7\ \mu$m (red).
The wavelength resolving widths of our instrument were subtracted from the measured widths in quadrature.
When the PAH~$6.2\ \mu$m is unresolved, the instrumental width is used to indicate the upper limit of the intrinsic width (filled histogram).
{\it Right}: inter-band flux ratio of PAH~$6.2\ \mu$m$/$PAH~$7.7\ \mu$m.
In both panels, corresponding measured parameters of the IRS starburst templates of \cite{brandl06} (one template; dotted line) and \cite{smith07} (four templates; dashed lines) are indicated by vertical lines.
}
\label{PAH_fit_param_histogram}
\end{figure*}

We compared the {\it SPICY} PAH galaxies and the IRS spectral templates of nearby star-forming galaxies \citep{brandl06,smith07} by using the PAH fit.
Figure~\ref{PAH_fit_param_histogram} shows distributions of widths of both PAH~$6.2\ \mu$m and PAH~$7.7\ \mu$m, and an inter-band flux ratio of the PAH~$6.2\ \mu$m to the PAH~$7.7\ \mu$m of the {\it SPICY} spectra\footnote{
We do not show parameters involving the PAH~$8.6\ \mu$m (inter-band flux ratios of PAH~$8.6\ \mu$m$/$PAH~$7.7\ \mu$m and PAH~$8.6\ \mu$m$/$PAH~$6.2\ \mu$m, and the PAH~$8.6\ \mu$m widths), and will not discuss them in the following.
This is because they typically show much larger uncertainties than for other PAH features, and this is likely caused by our low spectral resolution ($R\simeq 50$) to clearly isolate this feature on the profile of nearby stronger PAH~$7.7\ \mu$m.
}.
We found that all these parameters are similar to those of the IRS templates.
The PAH~$6.2\ \mu$m is not resolved in most {\it SPICY} spectra, because the wavelength resolving width of our instrument is slightly larger than the widths of the PAH~$6.2\ \mu$m of the IRS templates.
We also measured equivalent widths of the PAH~$6.2\ \mu$m for the {\it SPICY} PAH galaxies to be mostly 0.8--$2.0\ \mu$m (with a median of $1.4\ \mu$m).
They are consistent with those of the IRS spectral templates showing $0.6\ \mu$m (template 3 of \citealt{smith07})--$2.3\ \mu$m (average starburst of \citealt{brandl06}).
They are also consistent with those of typical starburst galaxies reported in literature (e.g., \citealt{brandl06,spoon07,weedman08}).

We measured integrated (under the PAH profile) spectroscopic PAH luminosities of the PAH~$6.2\ \mu$m, $L_\mathrm{PAH}$ ($6.2\ \mu$m), and the PAH~$7.7\ \mu$m, $L_\mathrm{PAH}$ ($7.7\ \mu$m) of the PAH galaxies by using the PAH fit.
We also measured the spectroscopic monochromatic luminosity at the PAH~$7.7\ \mu$m peak, $\nu L_{\nu\ \mathrm{spec}}$ ($7.7\ \mu$m) (hereafter, spectroscopic monochromatic luminosity at $7.7\ \mu$m; also called as peak luminosity of the PAH~$7.7\ \mu$m; \citealt{weedman08}) by using the PAH fit.
This spectroscopic monochromatic luminosity includes contributions of the underlying continuum.
Table~\ref{table3} lists the PAH fit results of the PAH galaxies as well as the optical spectroscopic redshifts.


\subsection{Multi-band photometry}\label{photo_data}

\begin{figure} 
\centering
\resizebox{\hsize}{!}{\includegraphics{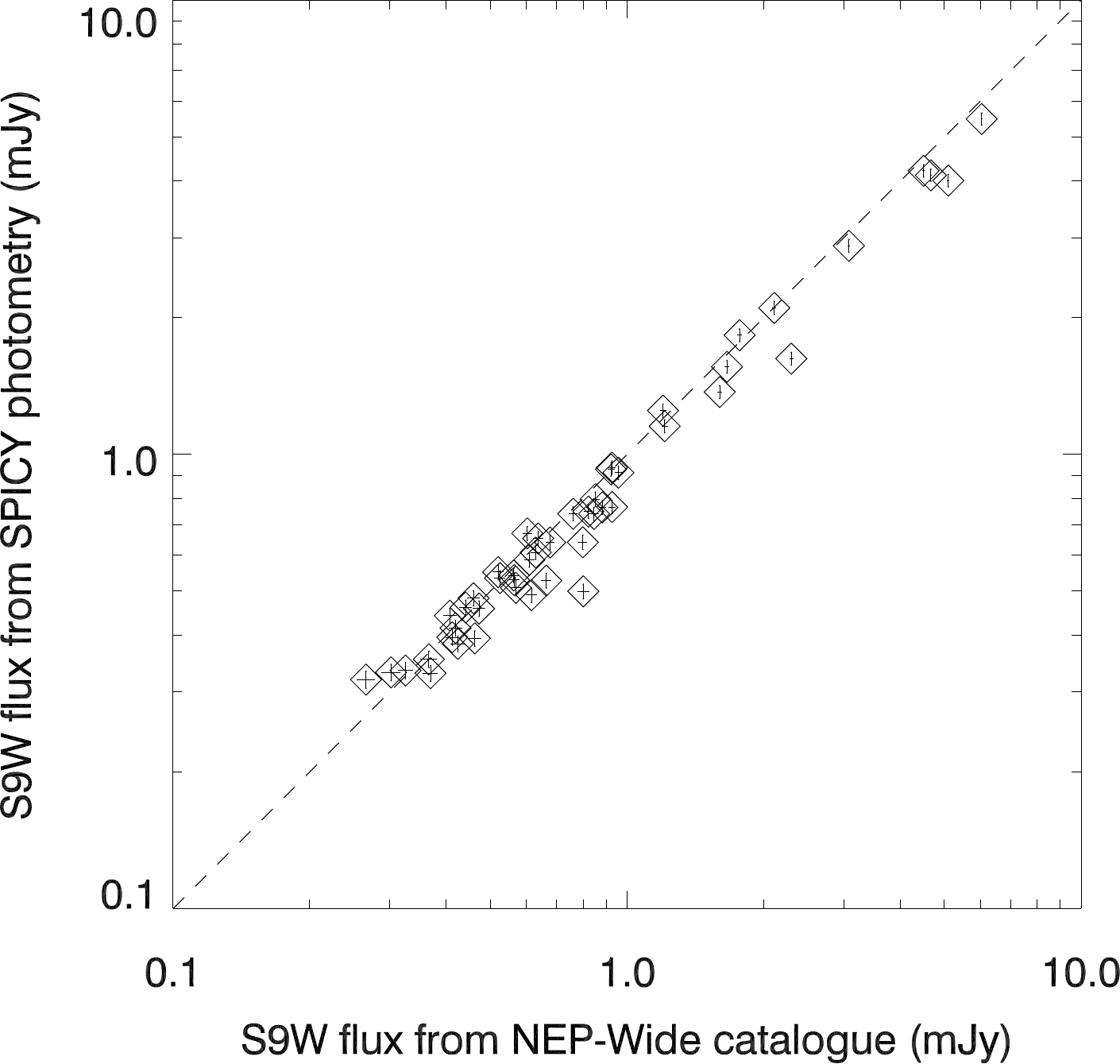}}
\caption{
Comparison between the S9W ($9.0\ \mu$m) {\it SPICY} photometry of the PAH galaxies and that in the NEP-Wide catalogue.
The diagonal broken line is for the same fluxes in the two measurements.
}
\label{s9w_consistency_check}
\end{figure}

We compiled photometric information for all {\it SPICY} extragalactic sources at OPT--NIR--MIR wavelengths.
The AKARI/IRC NEP-Wide photometric catalogue \citep{lee09,kim12} provides eight IRC filter fluxes ($N2$, $N3$, $N4$, $S7$, $S9W$, $S11$, $L15$, and $L18W$\footnote{
The IRC has three optical channels, ``NIR'', ``MIR-S'' (S for short), and ``MIR-L'' (L for long), and each channel has filters whose names start with ``N'', ``S'', and ``L'', respectively.
These initial characters are followed by approximate central wavelengths of the filters in $\mu$m.
Both S9W and L18W filters cover wider wavelength ranges than other filters, and their names include ``W'' (for wide) after the wavelengths.
}) centred approximately at 2.4, 3.2, 4.1, 7.0, 9.0, 11.0, 15.0, and $18.0\ \mu$m, respectively (see \citealt{onaka07} for more filter specifications).
We did not use the L24 ($22.9\ \mu$m) flux in our analysis, because it is much shallower than the others, although it is included in the catalogue.
We compared the {\it SPICY} S9W photometry of the PAH galaxies (Sect.~\ref{pah_fit_results}) with that in the NEP-Wide catalogue, and found that they are consistent with each other (Fig.~\ref{s9w_consistency_check}).
The CFHT optical NEP photometric catalogue \citep{hwang07} provides five optical filter fluxes ($u^{\rm *}$, $g'$, $r'$, $i'$, and $z'$).
In total, as many as 13-band broad-band photometric data, covering 0.37--$18\ \mu$m, are available.
Tables~\ref{table4} and \ref{table5} show the NIR--MIR and OPT photometries of the PAH galaxies, respectively.
Because the photometric catalogues are deeper than the {\it SPICY} spectroscopy ($>0.3$~mJy at S9W), almost all photometric data across OPT--NIR--MIR wavelengths are available, although the $u^{\rm *}$ band is often unavailable due to faintness of the sources and limited sensitivity in this band.


\section{Analysis and results}

\subsection{Spectroscopic characteristics}\label{spectral_characteristics_results}

\begin{figure*}[t] 
\centering
\includegraphics[width=17cm]{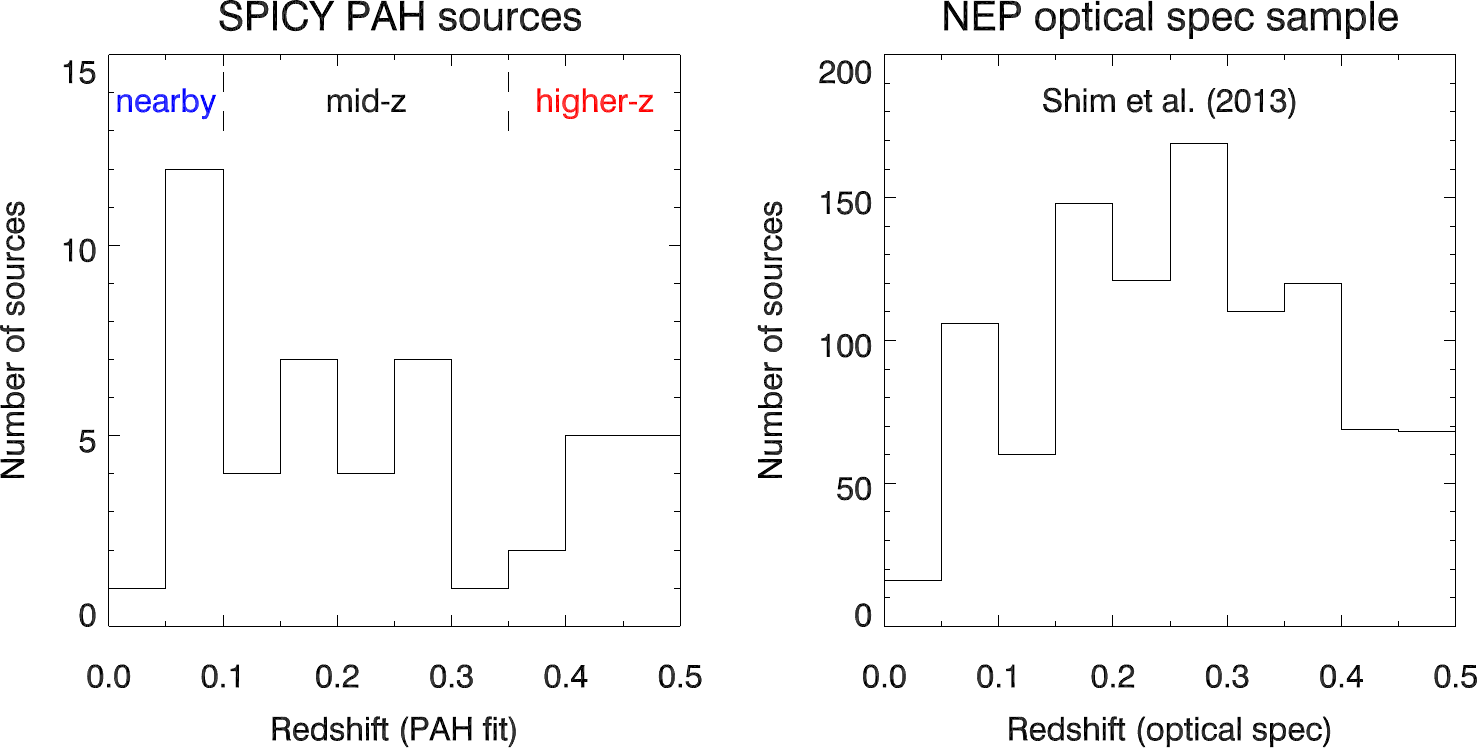}
\caption{Redshift distributions of the {\it SPICY} PAH galaxies ({\it left}) and the optical spectroscopic sample of \cite{shim13} toward the NEP ({\it right}).
In the right panel, we only show the sources with good redshift measurements (``quality flag'' of 3 or 4 in \citealt{shim13}; see text).
We also indicate ranges of the three redshift bins; nearby (including 13 PAH galaxies), mid-$z$ (23), and higher-$z$ (12) bins, in the left panel.
}
\label{redshift_histogram}
\end{figure*}

The redshift distribution of the {\it SPICY} PAH galaxies extends up to $z\simeq0.5$, with a few notable narrow peaks (left panel of Fig.~\ref{redshift_histogram}).
Such peaks are likely due to real galaxy distribution toward the NEP, rather than due to our source selection bias.
The S9W filter (6.7--$11.6\ \mu$m) used for our source selection (Sect.~\ref{survey_section}) is broad and covers most of the redshifted PAH~$7.7\ \mu$m in our redshift coverage ($z=0.0$--0.5).
The similar redshift distribution with three notable peaks ($z\lesssim0.1$, $z\lesssim0.2$, and $z\lesssim0.3$; right panel of Fig.~\ref{redshift_histogram}) is also found in the optical spectroscopic survey toward the NEP by \cite{shim13}, in spite of their slightly different survey coverage and target selection functions from us.
In particular, the first redshift peak is due to a supercluster at $z\simeq0.087$ toward the NEP \citep{ko12}.
At $z\gtrsim0.5$, the red side of the most prominent PAH~$7.7\ \mu$m goes out of our wavelength coverage ($\gtrsim13\ \mu$m) (see Sect.~\ref{sed_variation_result} for more discussion), and no PAH galaxies were detected in this redshift range.
In the following, we group our sample by redshifts into three bins.
Redshift ranges of the bins were defined as $z<0.1$, $0.1<z<0.35$, and $0.35<z<0.5$, and numbers of the sources are 13, 23, and 12 in the respective bins (Table~\ref{table1}).
We respectively call them nearby, mid-$z$, and higher-$z$ bins in the following.
The boundary between the mid-$z$ and higher-$z$ bins was set so that the peak at $z\lesssim 0.3$ is included in the mid-$z$ bin.
The boundary between the nearby and mid-$z$ bins was set so that the peak at $z\lesssim 0.1$ is included in the nearby bin.

\subsection{Photometric characteristics}\label{photo_characteristics_results}

\subsubsection{Broad-band colours and photometric type classification}\label{photo_classification_results}

\begin{figure*}[t] 
\centering
\includegraphics[width=17cm]{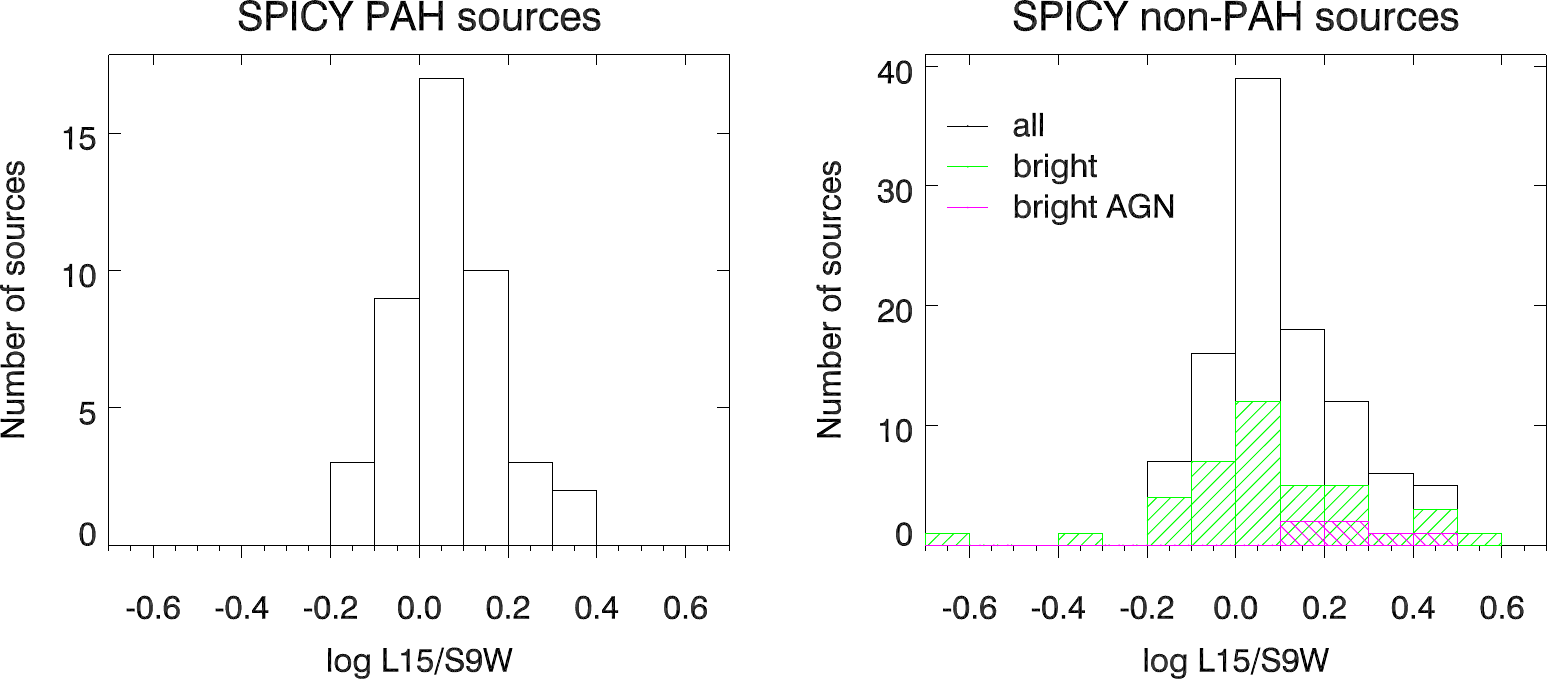}
\caption{Number distributions of the PAH ({\it left} panel) and non-PAH ({\it right} panel) sources as a function of $\log L15\ (15.0\ \mu\mathrm{m})/S9W\ (9.0\ \mu\mathrm{m})$.
In the {\it right} panel, number distributions of bright non-PAH sources and bright AGN candidates are also shown in green and magenta, respectively.
}
\label{s9w_l15_rat_hist}
\end{figure*}

We first compared the S9W ($9.0\ \mu$m) and L15 ($15.0\ \mu$m) fluxes of the {\it SPICY} sources.
The S9W band is the source detection band in our survey, and the L15 band is similar to the ``LW3'' band of the ISOCAM, which was extensively used for the ISOCAM deep extragalactic surveys (Sect.~\ref{introduction}).
Figure~\ref{s9w_l15_rat_hist} shows number distributions of the sources as a function of $\log L15/S9W$.
The non-PAH sources are shown separately for bright and faint sources, because detection of the PAH features depends on the MIR fluxes.
We refer to sources that are brighter than 0.3~mJy in all S7 ($7.0\ \mu$m), S9W ($9.0\ \mu$m), and S11 ($11.0\ \mu$m) bands as ``bright'' sources, because the PAH~$7.7\ \mu$m, if any, is expected to be detected at least in one of the three bands.
All other sources are referred to as faint sources.
Numbers of the bright sources are summarised in Table~\ref{table1}.
We found that $\log L15/S9W$ shows a tight distribution around $\simeq 0$, with a typical scatter of $\pm 0.2$, for both PAH and non-PAH sources (see also Sect.~\ref{colour_redshift_diagrams}), excluding the AGN candidates (see below for the definition and identification of the AGN candidates) showing larger $\log L15/S9W \simeq 0.3$.

\begin{figure*}[t] 
\centering
\resizebox{\hsize}{!}{\includegraphics{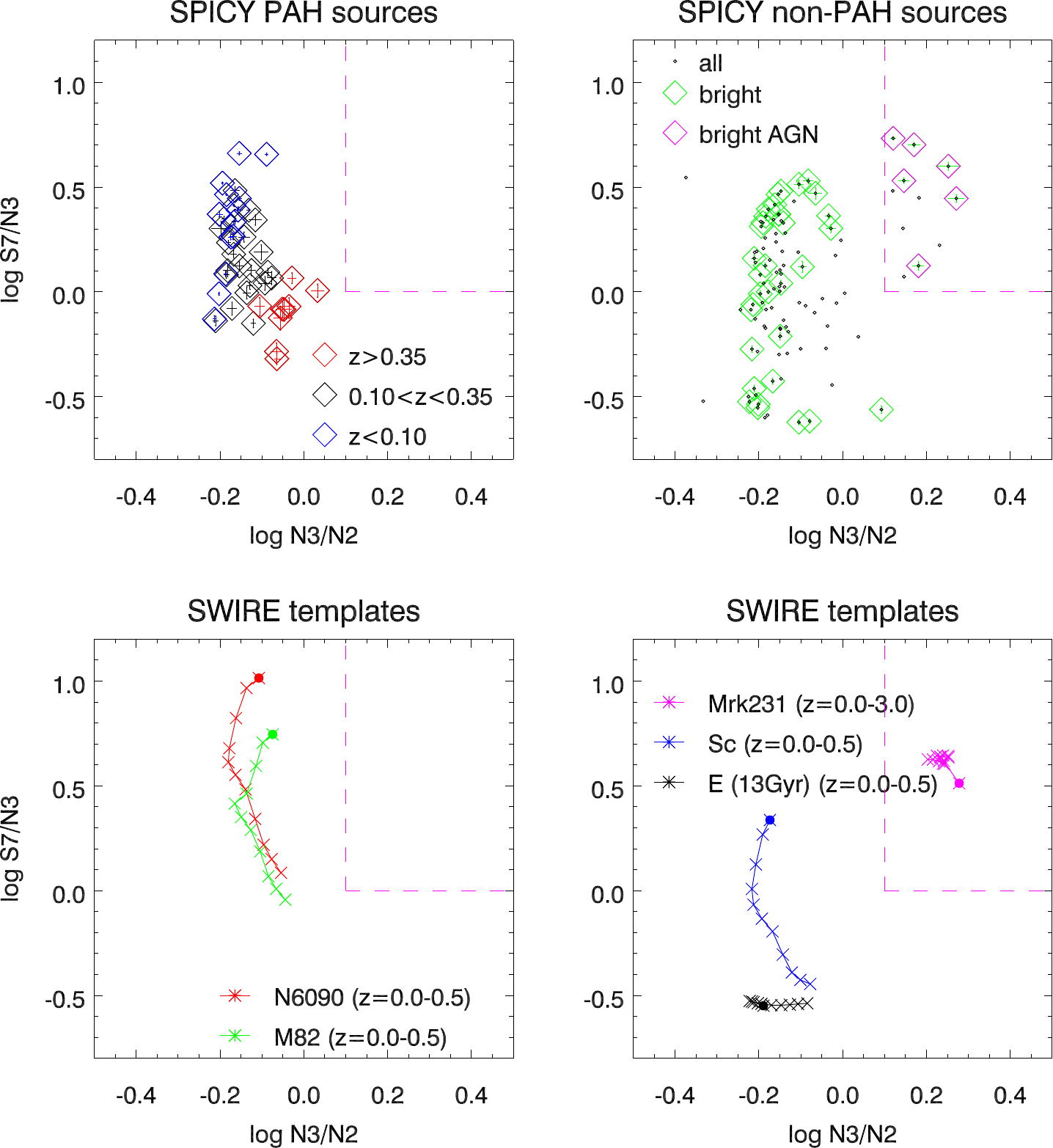}}
\caption{
Flux ratio diagrams of $\log N3$~($3.2\ \mu$m)$/N2$~($2.4\ \mu$m) and $\log S7$~($7.0\ \mu$m)$/N3$~($3.2\ \mu$m) of the {\it SPICY} sources.
{\it Top left}: the flux ratios of the PAH sources are shown with one-sigma error bars in blue, black, and red diamonds for the nearby, mid-$z$, and higher-$z$ sources, respectively.
{\it Top right}: the same diagram as the {\it top left} panel, but for the non-PAH sources (black).
Bright non-PAH sources are shown with one-sigma error bars and diamonds (green).
For the faint non-PAH sources, error bars are omitted for clarify of the figure.
{\it Bottom}: the same diagrams as the {\it top} panels, but for redshifted SWIRE NGC~6090, M82 ({\it left} panel), Sc, E (13~Gyr old), and AGN (Mrk~231) ({\it right} panel) templates with red, green, blue, black, and magenta connected crosses, respectively.
The flux ratios of the AGN template are plotted at $z=0.0$--3.0 in steps of 0.5, whereas those of other templates are plotted at $z=0.0$--0.5 in steps of 0.05.
The flux ratios of the templates at $z=0.0$ are indicated by filled circles in both bottom panels.
A space of the flux ratios to identify photometric AGN candidates (see text for the definition) is outlined by magenta dashed lines in all panels.
Bright photometric AGN candidates are shown with magenta diamonds in the {\it top right} panel.
}
\label{NIR_MIR_colour_colour}
\end{figure*}

\begin{figure}
\centering
\resizebox{\hsize}{!}{\includegraphics{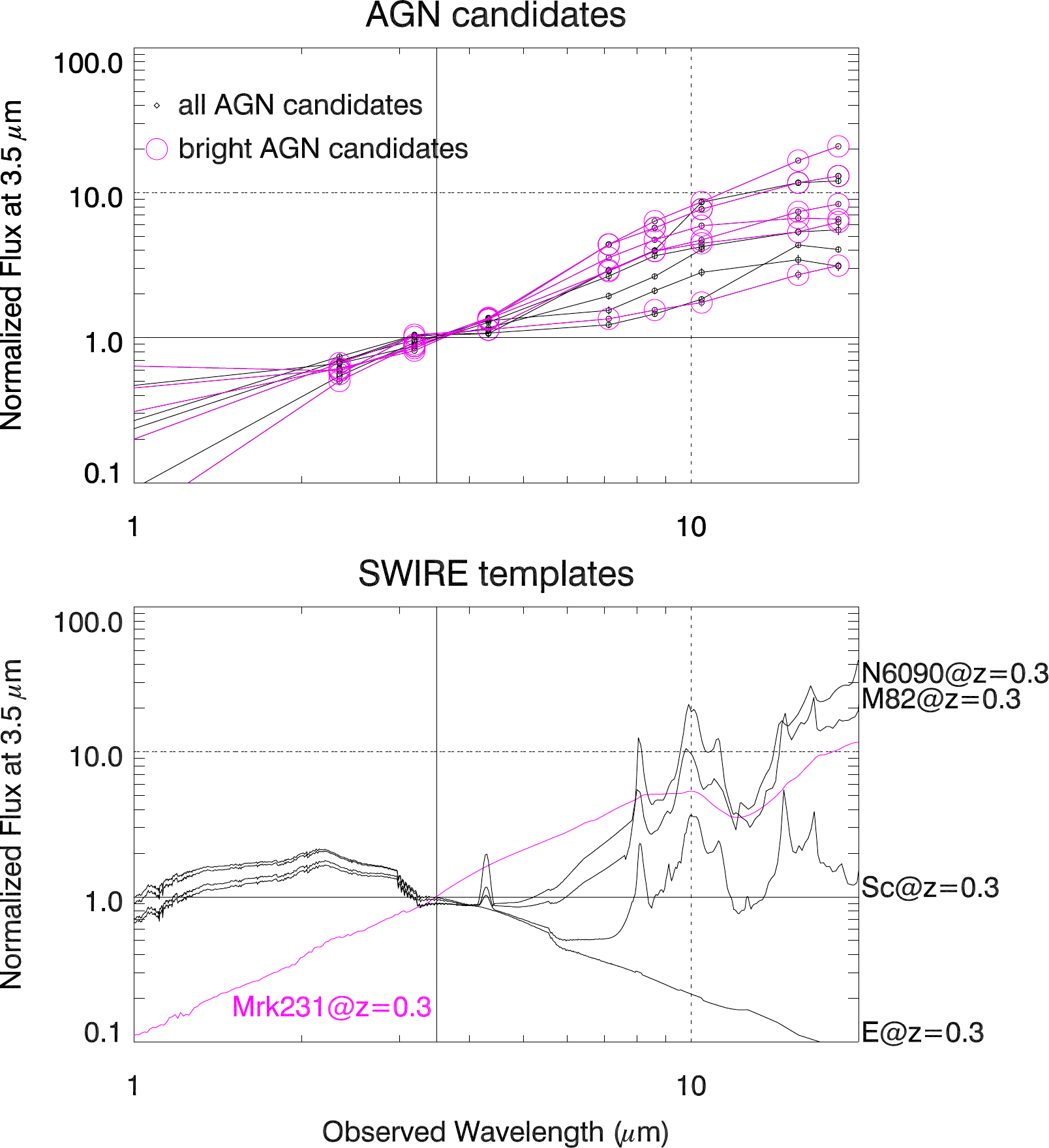}}
\caption{Observed SEDs of the AGN candidates and comparison with various SED templates.
{\it Top}: SEDs of the AGN candidates normalised at $3.5\ \mu$m.
The bright AGN candidates are marked with magenta circles.
{\it Bottom}: similarly normalised SWIRE SED templates of AGN (Mrk~231; magenta) and galaxies of E (13~Gyr old), Sc, M82, and NGC~6090 (black) redshifted to $z=0.3$.
The normalisation wavelength is indicated by a black solid vertical line in each panel.
}
\label{observed_SEDs_AGNcan}
\end{figure}

We then analysed the relationship among broad-band colours, or flux ratios, of the {\it SPICY} sources to broadly characterise their SED types.
We compared $N3$~($3.2\ \mu$m)$/N2$~($2.4\ \mu$m) and $S7$~($7.0\ \mu$m)$/N3$~($3.2\ \mu$m) between the observations and the SED templates (Fig.~\ref{NIR_MIR_colour_colour}).
\cite{takagi10} adopted this colour combination by following the colour diagram of [3.6]$-$[5.8] vs. [4.5]$-$[8.0] with {\it Spitzer}/IRAC to identify AGN candidates (\citealt{lacy04}; see also \citealt{takagi07}).
We adopted SED templates of elliptical (E), late type spiral (Sc), starburst (M82), LIRG (NGC~6090\footnote{NGC~6090 is a strongly interacting LIRG (e.g., \citealt{scoville00}) with moderately strong starburst activity ($\log L_{\rm IR}=11.56\ (L_{\sun})$; \citealt{sanders03}; after being converted to our assumed cosmology).
The SED template of this galaxy was constructed by a physical starburst model ``GRAZIL'' \citep{silva98}, and therefore a possible AGN contribution to this galaxy is not considered.}),
and AGN (Mrk~231) from the SWIRE library \citep{polletta07}, and calculated their colours for a range of redshift.
Most PAH galaxies are blue at NIR ($-0.3\lesssim \log N3/N2\lesssim 0.0$) and blue--moderately red at NIR--MIR ($-0.2\lesssim \log S7/N3\lesssim 0.5$).
Their $S7/N3$ colour gets bluer with redshift, as predicted by the templates of Sc, starburst, and LIRG.
Some of the non-PAH sources show very blue $S7/N3$ ($\log S7/N2<-0.2$) or red $N3/N2$ ($\log N3/N2>0.1$) colours.
The former can be reproduced by the templates of Sc, E, or their intermediate types (e.g., Sa).
The latter can be reproduced only by the AGN template in a wide redshift range (at least up to $z\sim 3$; Fig.~\ref{NIR_MIR_colour_colour}).
It is known that red continuum-dominated power-law-like SED at OPT--MIR ($\alpha\sim -1$ where $f_{\nu}\propto\nu^{\alpha}$; e.g., \citealt{emr,elvis94}) makes an AGN red in both NIR$/$NIR and MIR$/$NIR colours in the very wide redshift range (e.g., \citealt{lacy04,donley12}).
To confirm this, we examined observed SEDs of the latter sources (Fig.~\ref{observed_SEDs_AGNcan}), and found that they indeed show red continuum-dominated, AGN-like SEDs at NIR--MIR wavelengths.
In contrast, all kinds of galaxy templates (E, Sc, starburst, and LIRG) predict blue NIR colour ($\log N3/N2<0.0$) at $z=0.0$--0.5, because the NIR bands cover the stellar bump peaking at $1.6\ \mu$m (see also Fig.~\ref{redshift_colour_swire}).
\cite{takagi10} confirmed that most of such $N3/N2$-red sources in their AKARI/IRC NEP-Deep photometric sample are indeed AGNs with optical spectroscopy.
Therefore, red NIR colour is a clear AGN signature in our redshift coverage ($z=0.0$--0.5).

We defined the AGN candidates in the $N3$~($3.2\ \mu$m)$/N2$~($2.4\ \mu$m) and $S7$~($7.0\ \mu$m)$/N3$~($3.2\ \mu$m) diagram (Fig.~\ref{NIR_MIR_colour_colour}).
We required that the AGN SED template is clearly separated from a range of the galaxy templates in this colour-colour diagram, and that all PAH galaxies are classified as non-AGNs.
We set two colour conditions for the AGNs: $\log N3/N2>0.1$ and $\log S7/N3>0.0$.
The secondary colour condition is useful to separate galaxies and AGNs at $0.35<z<0.5$, when both types are closer in $N3/N2$ while they are more separated in $S7/N3$.
We identified 11 AGN candidates (including six bright ones), and Table~\ref{table1} includes their numbers.
In the following, we pay more attention to the bright AGN candidates, because we are sure about intrinsic faintness of their PAH features.
The number fraction of the bright AGN candidates is 9\% ($=6$ bright AGN candidates$/$(45 bright non-PAH galaxies including the bright AGN candidates$+$24 bright PAH galaxies)) among all bright sources.
Table~\ref{table6} shows source identification, cross-identification with the AKARI/IRC NEP-Wide catalogue \citep{lee09,kim12}, and the {\it SPICY} S9W ($9.0\ \mu$m) photometry of the bright AGN candidates, as well as cross-identification with the CFHT optical NEP photometric catalogue \citep{hwang07} and the optical type classification by \cite{shim13}.
Tables~\ref{table7} and \ref{table8} show their NIR--MIR and OPT photometries, respectively.


We finally cross-checked our AGN classification with information from other wavelengths.
The X-ray luminosities for the sources found in the X-ray catalogue of \cite{krumpe15} are consistent with our classification.
We found only one X-ray source (F09-2) among the PAH galaxies.
Its 0.5--7~keV luminosity is $\approx 8\times 10^{40}$ erg s$^{-1}$ and can be accounted for by a pure star-forming activity.
Among 6 bright AGN candidates, we found three (F00-3=ANEPDCXO004, F10-20=ANEPDCXO221, and F04-20=ANEPDCXO138) X-ray sources.
For the two such sources with measured redshifts, F00-3 and F10-20, their 0.5-7~keV luminosities ($\approx 2\times 10^{43}$ and $\approx 5\times 10^{42}$ erg s$^{-1}$, respectively) are in a range of typical Seyfert galaxies.
The remaining three bright AGN candidates are outside of the X-ray survey area.
Our AGN classification seems also valid when compared to the optical emission-line ratio diagnostics results of \cite{shim13}.
Four (out of all six) bright AGN candidates with the optical spectroscopy are classified as ``TYPE1'' AGNs.
Figure~\ref{spicy_example} shows the {\it SPICY} spectrum of the brightest spectroscopically confirmed AGN, F05-1, at $z=0.4508$ (quality flag$=4$).
This object indeed shows a featureless red continuum at 5--$11.5\ \mu$m as the NIR--MIR SED indicates (Fig.~\ref{observed_SEDs_AGNcan}), followed by a pronounced decline in the spectrum around 11.5--$13\ \mu$m, which is likely due to redshifted silicate $9.7\ \mu$m absorption.
We caution that this identification is tentative because the absorption peak is not seen within the wavelength coverage.
If this is indeed due to silicate absorption, such characteristic MIR spectrum is typically found in type-2 AGNs (e.g., \citealt{hao07}).
None of the 32 (out of all 48) PAH galaxies with optical spectroscopy (Sect.~\ref{pah_fit_test}) are classified as AGNs.

\subsubsection{Colour-redshift diagrams of the PAH galaxies}\label{colour_redshift_diagrams}

\begin{figure*}[t] 
\centering
\includegraphics[width=17cm]{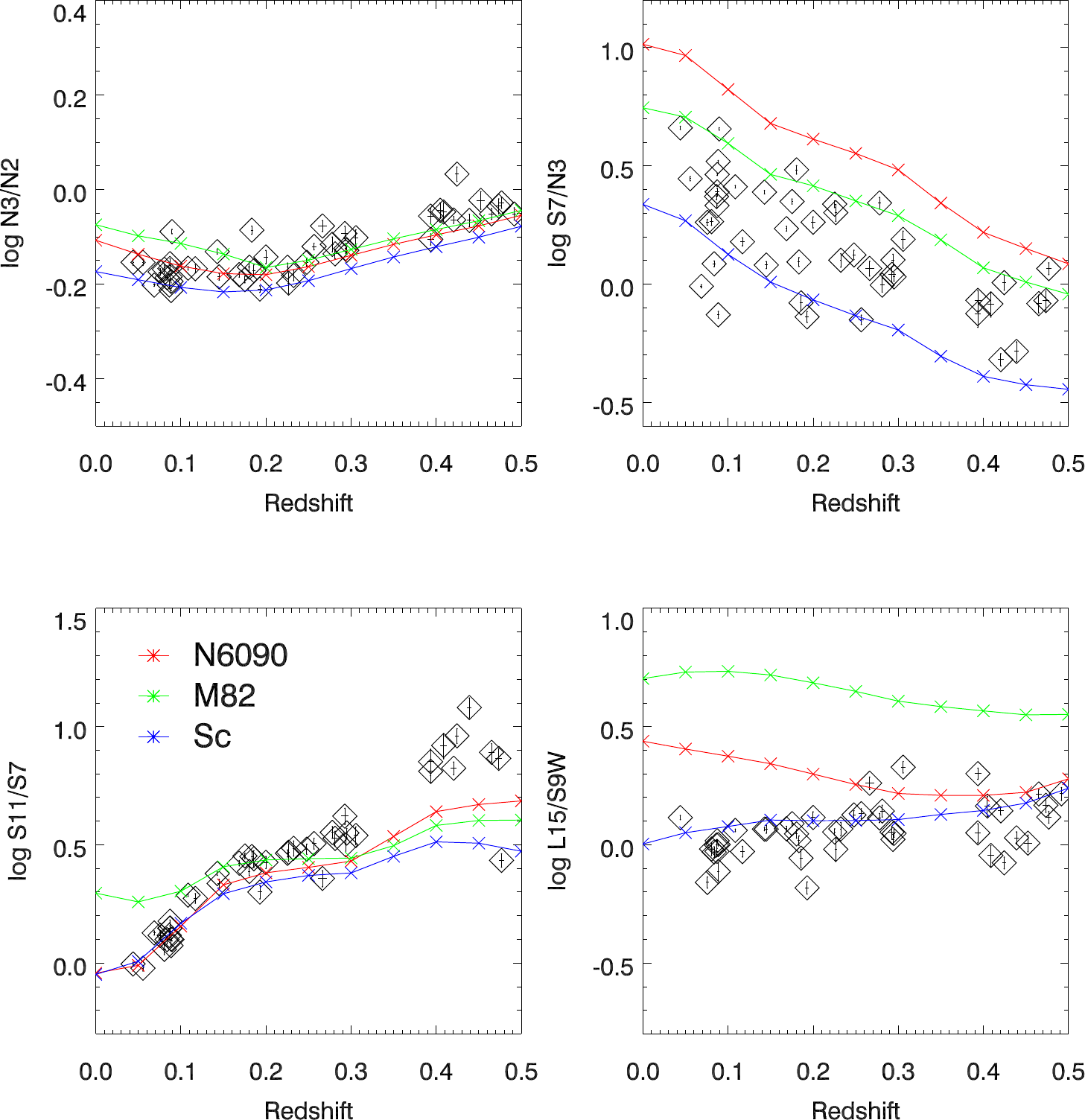}
\caption{NIR and MIR flux ratios of $N3$~($3.2\ \mu$m)$/N2$~($2.4\ \mu$m), $S7$~($7.0\ \mu$m)$/N3$~($3.2\ \mu$m), $S11$~($11.0\ \mu$m)$/S7$~($7.0\ \mu$m), and $L15$~($15.0\ \mu$m)$/S9W$~($9.0\ \mu$m) of the PAH galaxies as a function of redshift.
Error bars are for one sigma.
The flux ratios of the redshifted SWIRE SED templates of Sc (blue), M82 (green), and NGC~6090 (red) are overlaid with connected crosses.
To enable easier visual comparison among the panels, the Y axis covers the same 1.8 dex-wide range in the flux ratios, excluding the panel of $N3/N2$ (covering half the range).
}
\label{redshift_colour_swire}
\end{figure*}

The colours of the {\it SPICY} PAH galaxies depend strongly on redshift in different ways for different colours.
Figure~\ref{redshift_colour_swire} compares the flux ratios of $N3$~($3.2\ \mu$m)$/N2$~($2.4\ \mu$m), $S7$~($7.0\ \mu$m)$/N3$~($3.2\ \mu$m), $S11$~($11.0\ \mu$m)$/S7$~($7.0\ \mu$m), and $L15$~($15.0\ \mu$m)$/S9W$~($9.0\ \mu$m) as a function of redshift between the PAH galaxies and the SED templates.
We again adopted the SWIRE SED templates of normal/star-forming galaxies; Sc, starburst (M82), and LIRG (NGC~6090).
We found very similar results as \cite{takagi10} and \cite{hanami12}, who analysed colour-redshift diagrams for their AKARI/IRC NEP-Deep photometric sample up to larger redshift ($z\sim 1$--2).
In the NIR, both a stellar bump peaking at $1.6\ \mu$m and stellar CO $v=2$--0 absorptions with their bandhead at $2.3\ \mu$m affect the colours.
The templates predict that the $N3/N2$ becomes smaller (bluer) with increasing redshift at $z\lesssim 0.2$, because the $2.3\ \mu$m absorption feature moves from the N2 to N3 bands.
The ratio then becomes larger (redder) with increasing redshift at $z\gtrsim 0.2$, because the $1.6\ \mu$m feature moves from the N2 to N3 bands.
The observations seem to follow such a colour trend at $z=0.0$--0.5, although the overall change in this NIR colour is relatively small (by $\lesssim 0.2$ dex) in our redshift coverage ($z=0.0$--0.5).
In the MIR, because the prominent PAH features around $7.7\ \mu$m move across the bands, the flux ratios show extreme diversity in the colour-redshift diagrams.
The $S7/N3$ and $S11/S7$ become smaller (bluer) and larger (redder), respectively, with increasing redshift.
In particular, the $S11/S7$ changes by as much as a factor of 10 in our redshift coverage.
This is primarily because the PAH~$7.7\ \mu$m moves from the S7 to S11 bands with increasing redshift.
In contrast, the $L15/S9W$ shows little systematic change (by $\lesssim 0.2$ dex in our redshift coverage) with redshift.
This is because the most prominent PAH~$7.7\ \mu$m is included within the S9W band in most of our redshift coverage.
The M82 SED template seems to show too much hot dust continuum at long side of the PAH~$7.7\ \mu$m to overestimate the $L15/S9W$ in most of our redshift coverage (see also Sect.~\ref{sed_fit_swire}; Fig.~\ref{rest_SEDs_model_temp}).

\subsubsection{Rest-frame SEDs of the PAH galaxies}\label{rest_frame_seds}

\begin{figure*}[t] 
\centering
\includegraphics[width=17cm]{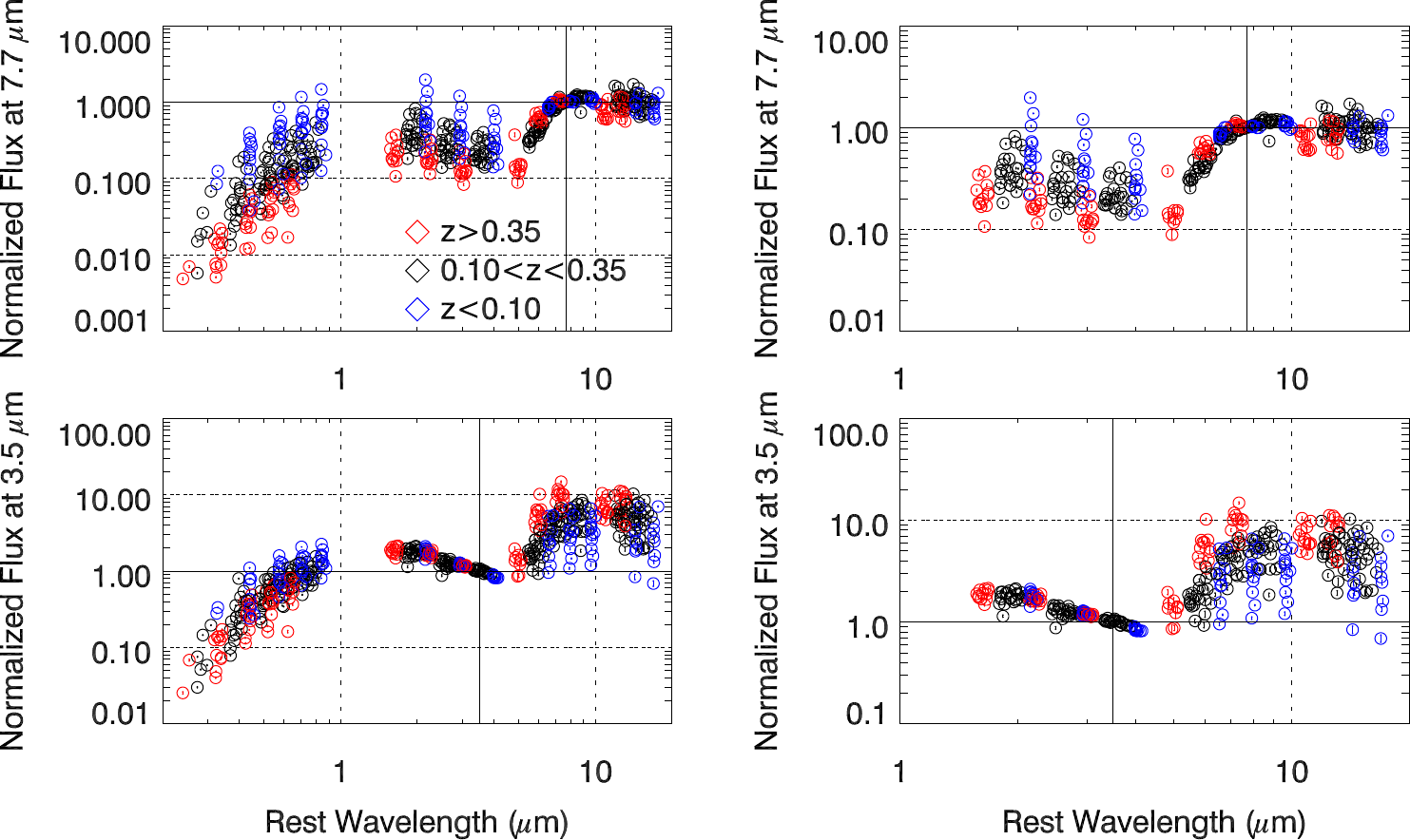}
\caption{Rest-frame SEDs of the PAH galaxies normalised at $7.7\ \mu$m (MIR-normalised; {\it top}) and $3.5\ \mu$m (NIR-normalised; {\it bottom}) with one-sigma error bars.
The normalisation wavelengths are indicated by solid vertical lines in all panels.
The {\it left} and {\it right} panels show the same SEDs but for the entire OPT--NIR--MIR and only the NIR--MIR wavelengths, respectively.
The SEDs are shown in different colours for different redshift bins: blue, black, and red for the nearby, mid-$z$, and higher-$z$ galaxies, respectively.
}
\label{normalized_rest_SEDs}
\end{figure*}

\begin{figure*}[t] 
\centering
\includegraphics[width=17cm]{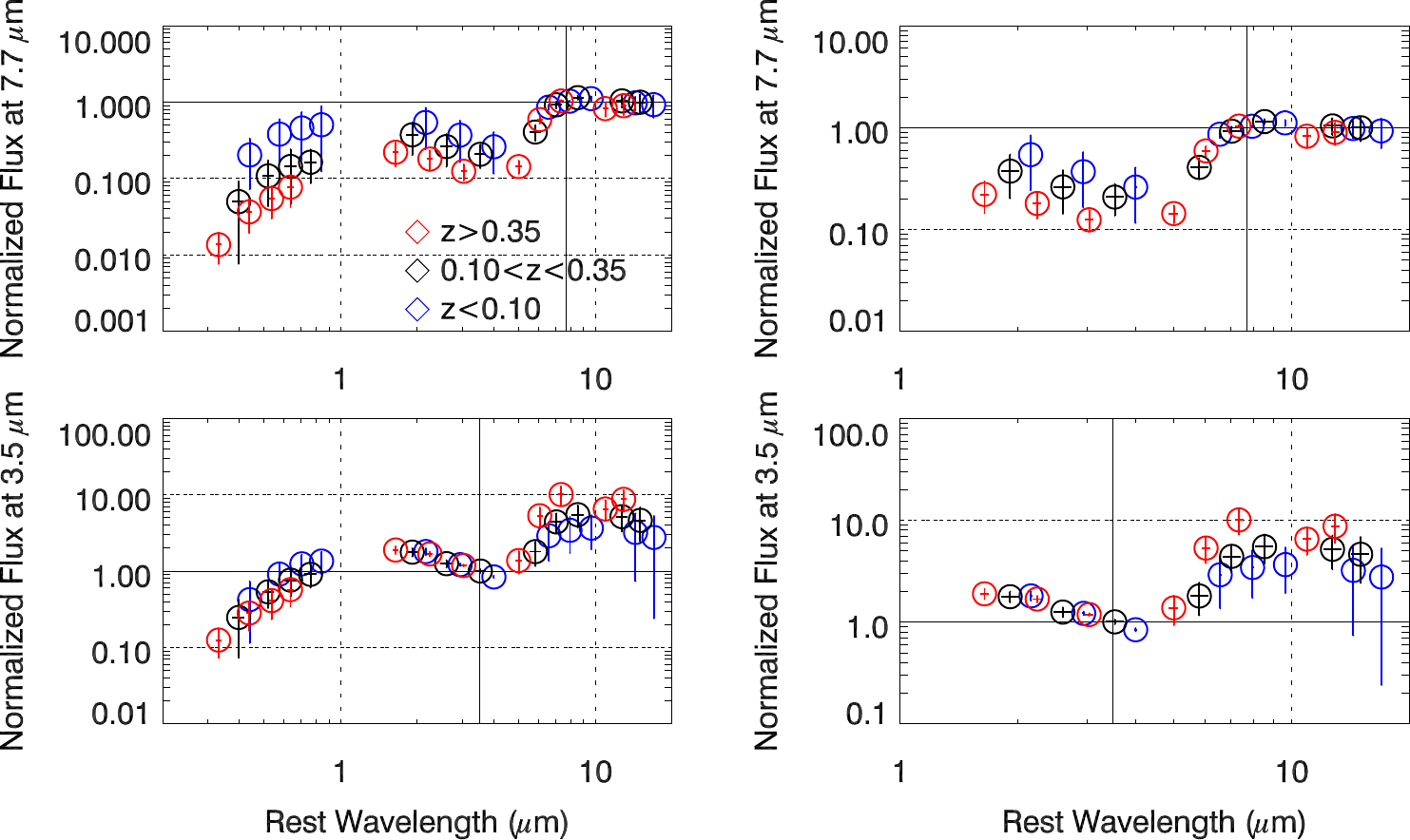}
\caption{The same SED plots as Fig.~\ref{normalized_rest_SEDs}, but for median-averaged ones (see text).
Error bars represent one-sigma scatter of the normalised photometric data along both the flux and wavelength directions around the medians.
}
\label{normalized_rest_SEDs_bin}
\end{figure*}

We examined rest-frame SEDs of the {\it SPICY} PAH galaxies constructed by utilising 13 OPT--NIR--MIR broad-band photometric data and redshift from the PAH fit.
As we show in the following, the rest-frame SEDs have almost identical shapes for a range of redshift, making our analysis easier and probably more fundamental, in contrast to the observed-frame information that shows diversity.
The observed photometric data were simply de-redshifted and normalised to the rest-frame $7.7\ \mu$m and $3.5\ \mu$m fluxes for the MIR- and NIR-normalised rest-frame SEDs, respectively (Fig.~\ref{normalized_rest_SEDs}).
Similar $3.5\ \mu$m-normalised rest-frame SEDs were presented by \cite{hanami12} for their AKARI/IRC NEP-Deep photometric sample at $0.4<z<2$.
The MIR normalisation wavelength was chosen at a peak of the PAH~$7.7\ \mu$m, and flux there was estimated by simply interpolating fluxes of two filters that encompass the wavelength.
The NIR normalisation wavelength was chosen as the longest wavelength within the stellar SED bump that the N4 band can cover even at $z>0.35$.
The $3.5\ \mu$m flux was estimated by using linear interpolation/extrapolation among the three NIR bands in log wavelength--log flux space for sources at $z<0.35$, because the stellar spectrum there can be approximated by a simple Rayleigh-Jeans power-law.
The resultant SEDs indeed show such a power-law form at NIR in the rest frame.
For sources at $z>0.35$, we used only N3 and N4 bands for the interpolation, because the N2 band covers the peak of the stellar SED at $1.6\ \mu$m, and the Rayleigh-Jeans approximation is inappropriate there.
We also constructed the median-averaged rest-frame SEDs by taking median values of normalised fluxes for each filter and redshift bin (Fig.~\ref{normalized_rest_SEDs_bin}).
Missing flux data, if any, were ignored in taking the median.

The normalised rest-frame SEDs clearly illustrate their characteristic two-component shape.
The MIR-normalised rest-frame SEDs are composed of the MIR bump (5--$10\ \mu$m, PAH peak) and the OPT--NIR one (0.3--$4\ \mu$m, red giant peak), with a clear dip around $\simeq 4\ \mu$m between the two bumps.
We found that most sources have very similar MIR bump shape for a range of redshift, indicating that the PAH features look almost identical among the sources at this spectral resolution.
In the meantime, the OPT--NIR bump becomes weaker with respect to the MIR one at higher redshifts.
The NIR-normalised rest-frame SEDs clearly show that the OPT--NIR bump shape, particularly at NIR, is also almost identical among the sources for a range of redshift, showing a broad bump peaking at $\sim 1$--$2\ \mu$m as expected for the stellar SEDs.
Therefore, each of the SED bumps has its own shape with little systematic change as a function of redshift, and sources at higher redshifts show systematically more enhanced MIR bump with respect to the stellar SED.

\subsection{Combined spectroscopic and photometric characteristics of the PAH galaxies}\label{spec_photo_comp}

\subsubsection{PAH luminosities}\label{pah_luminosity_comp_section}

\begin{figure*}[t] 
\centering
\includegraphics[width=17cm]{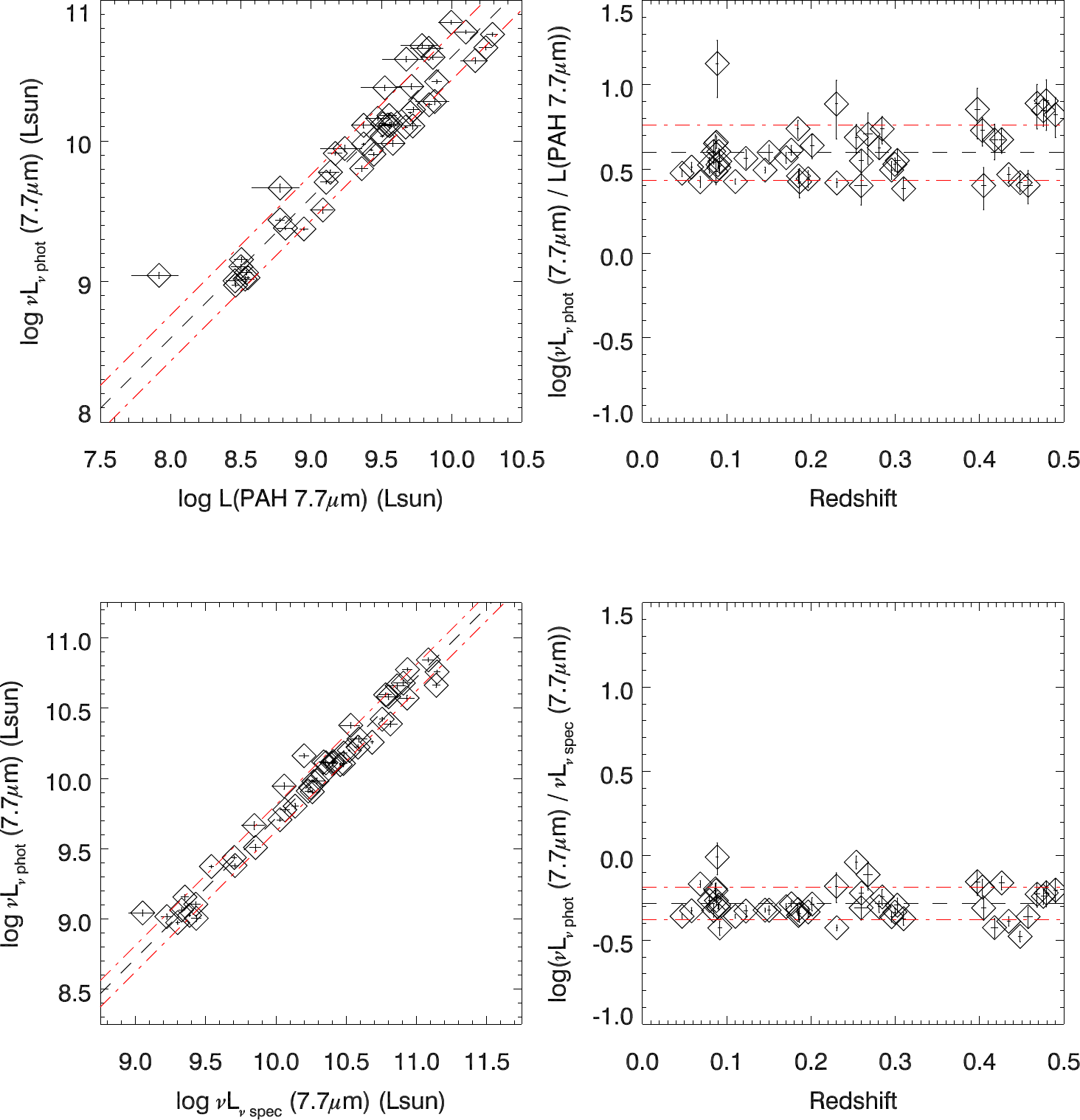}
\caption{Comparisons between the spectroscopic and photometric PAH~$7.7\ \mu$m luminosities of the PAH galaxies.
The photometric monochromatic luminosity at $7.7\ \mu$m, $\nu L_{\nu\ \mathrm{photo}}\ (7.7\ \mu\mathrm{m})$, is compared with the spectroscopic PAH~$7.7\ \mu$m luminosity, $L_\mathrm{PAH}\ (7.7\ \mu\mathrm{m})$ ({\it top}), and with the spectroscopic monochromatic luminosity at $7.7\ \mu$m, $\nu L_{\nu\ \mathrm{spec}}\ (7.7\ \mu\mathrm{m})$ ({\it bottom}).
Their luminosity ratios are plotted as a function of redshift in the {\it right} panels.
Error bars are for one sigma.
In all panels, black broken lines represent the best-fit linear functions with fixed unity slopes, and red dot-broken lines indicate the one-sigma scatters.
}
\label{pah_luminosity_comp_figure}
\end{figure*}

We compared the photometric and spectroscopic PAH luminosities of the {\it SPICY} PAH galaxies to test accuracy and robustness of our photometric measurement of the PAH luminosity, and to bridge the results from the PAH fit and the analysis based on the rest-frame SEDs.
We already measured the spectroscopic PAH luminosities, $L_\mathrm{PAH}$ ($6.2\ \mu$m) and $L_\mathrm{PAH}$ ($7.7\ \mu$m), and spectroscopic monochromatic luminosity at $7.7\ \mu$m, $\nu L_{\nu\ \mathrm{spec}}$ ($7.7\ \mu$m), in Sect.~\ref{pahfit_section}.
In addition, we measured the ``photometric'' monochromatic luminosity at $7.7\ \mu$m, $\nu L_{\nu\ \mathrm{photo}}$ ($7.7\ \mu$m) (hereafter, photometric monochromatic luminosity at $7.7\ \mu$m) while constructing the MIR-normalised rest-frame SEDs (Table~\ref{table9}).
This photometric monochromatic luminosity includes contributions of the underlying continuum.
Both the spectroscopic and photometric monochromatic luminosities at $7.7\ \mu$m are intended to measure luminosities at the PAH~$7.7\ \mu$m peak by using the {\it SPICY} spectra and broad-band SEDs, respectively.
Figure~\ref{pah_luminosity_comp_figure} compares the three kinds of the PAH~$7.7\ \mu$m luminosities, and shows good linear correlations among them, with a slope of unity.
Their ratios show no dependence on redshift, indicating little systematics in the photometric measurement of the PAH peak fluxes, although the PAH~$7.7\ \mu$m is covered with different filters depending on redshift.
The resultant scaling relations are:
$\log \nu L_{\nu\ \mathrm{photo}}\ (7.7\ \mu\mathrm{m})=(0.60\pm 0.16)+\log L_\mathrm{PAH}\ (7.7\ \mu\mathrm{m})$\footnote{
This rather large offset ($0.60\pm 0.16$ in log) is mostly due to different definitions of the two luminosities.
For $\nu L_{\nu}$ ($7.7\ \mu$m), we multiplied the peak flux of the PAH~$7.7\ \mu$m profile (including the underlying continuum) by the peak frequency (corresponding to $7.7\ \mu$m in wavelength).
For $L_\mathrm{PAH}$ ($7.7\ \mu$m), we integrated under the PAH~$7.7\ \mu$m profile by using the best-fit Lorentzian function.
This integration can be analytically made as $\pi\times$ (amplitude of the PAH~$7.7\ \mu$m profile)$\times$(HWHM of the profile; corresponding to $\simeq 0.3\ \mu$m in wavelength; Fig.~\ref{PAH_fit_param_histogram}).
The former luminosity is, by definition, larger than the latter by a factor of $\simeq\pi\times0.3\ \mu$m$/7.7\ \mu$m (plus a contribution of the continuum).
}
and $\log \nu L_{\nu\ \mathrm{photo}}\ (7.7\ \mu\mathrm{m})=(-0.28\pm 0.09)+\log \nu L_{\nu\ \mathrm{spec}}\ (7.7\ \mu\mathrm{m})$.
Here, quoted uncertainties are for one-sigma scatter of the ratio between the luminosities.
Such good correlations imply that our photometric measurement of the PAH feature provides a good estimate of the PAH luminosities.


\begin{figure} 
\centering
\resizebox{\hsize}{!}{\includegraphics{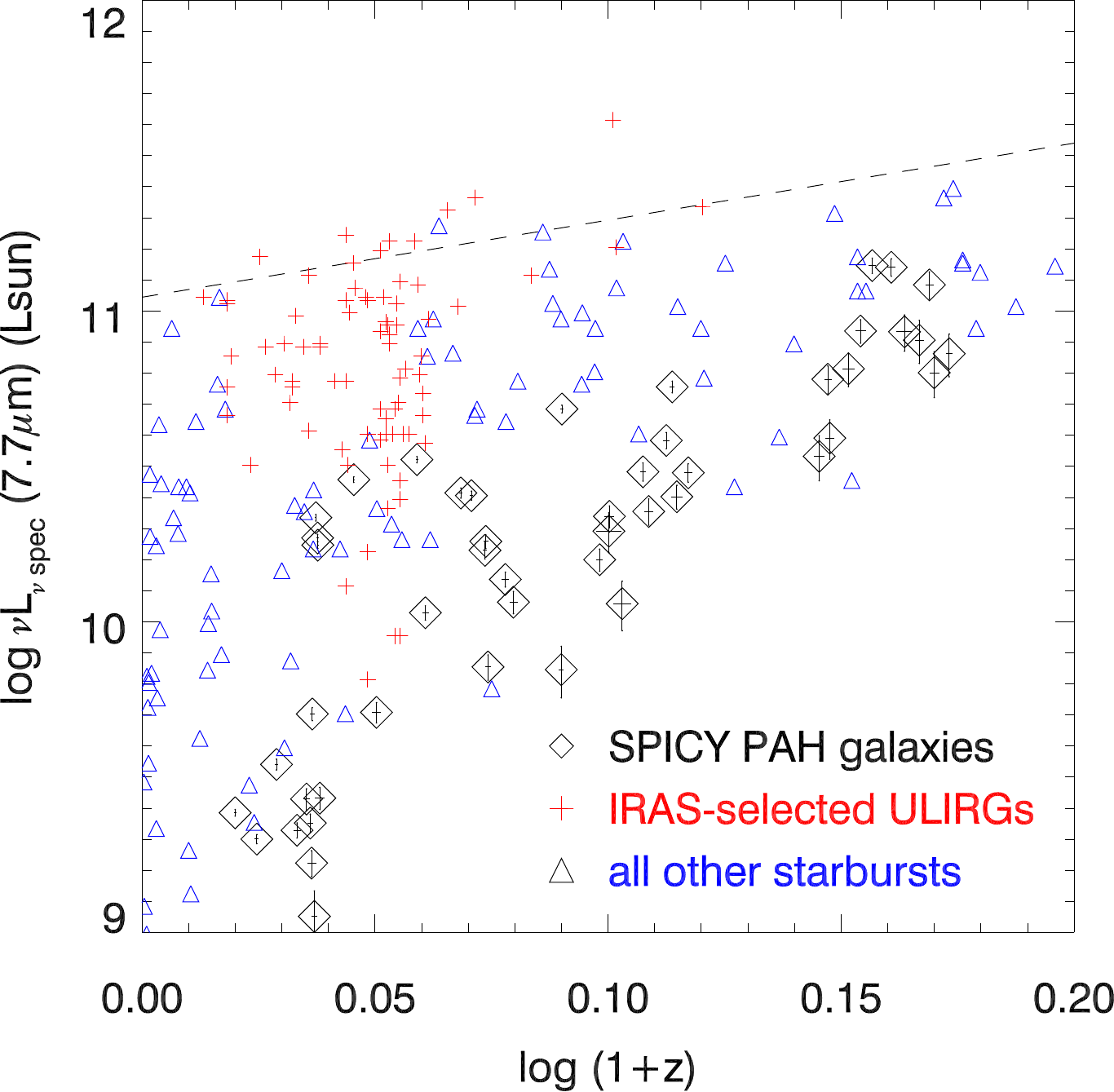}}
\caption{Spectroscopic monochromatic luminosities at $7.7\ \mu$m, $\nu L_{\nu\ \mathrm{spec}}$ ($7.7\ \mu$m), as a function of redshift.
The {\it SPICY} PAH galaxies are shown with black diamonds with one-sigma error bars, and the IRS galaxies in a sample of \cite{weedman08} are shown with red pluses and blue triangles for starburst-component of the {\it IRAS}-selected ULIRGs and all other starburst galaxies, respectively.
A black broken line represents a limit of ``maximum starburst'' of \cite{weedman08} fitted at $z=0.0$--2.5.
}
\label{PAH_luminosity_redshift}
\end{figure}

We characterised the PAH luminosity of the {\it SPICY} PAH galaxies by comparing to a representative IRS sample in the similar redshift range.
Figure~\ref{PAH_luminosity_redshift} compares the spectroscopic monochromatic luminosities at $7.7\ \mu$m of the {\it SPICY} PAH galaxies and those of the star-forming galaxies observed with the IRS low-resolution mode in a sample of \cite{weedman08}.
We neglected the effect of different spectroscopic resolutions between the IRS and IRC, because the PAH~$7.7\ \mu$m is intrinsically broad and resolved by both instruments.
This IRS sample is a collection of wide variety of star-forming galaxies showing strong PAH features, including nearby bright ULIRGs, nearby starburst galaxies, MIPS $24\ \mu$m-selected sources in, e.g., the NOAO Bo\"otes survey and the First Look Survey (see \cite{weedman08} for the details of their sample).
The {\it SPICY} PAH galaxies occupy near lower end of the $\nu L_{\nu\ \mathrm{spec}}$ ($7.7\ \mu$m) distribution of the IRS sample, i.e., the {\it SPICY} PAH galaxies are systematically less luminous for a given redshift.

We estimated the IR luminosities (8--$1000\ \mu$m; $L_{\rm IR}$) of the {\it SPICY} PAH galaxies from the photometric monochromatic luminosities at $7.7\ \mu$m by using a calibration from \cite{takagi10} (see also \citealt{hock07}).
We found that the PAH galaxies show $10\lesssim \log L_{\rm IR}\ (L_{\sun})\lesssim 11.6$ (Fig.~\ref{nuLnu77_vs_L35}), and therefore have the luminosity of LIRGs ($11<\log L_{\rm IR}\ (L_{\sun})<12$) or less.
We note that the AKARI/IRC NEP-Deep photometric samples at $z\gtrsim 1$--2 are typically more IR luminous with $\log L_{\rm IR}\ (L_{\sun})\gtrsim 12$ (\citealt{takagi10,hanami12}; see also \citealt{goto10,goto11a}).
The inferred SFR by using the conversion between $L_{\rm IR}$ and SFR \citep{kennicutt98} is $\gtrsim 30\ M_{\rm \sun}$~yr$^{-1}$ for luminous PAH galaxies at $z\gtrsim 0.35$.

\subsubsection{SED shape variation and its interpretation with spectroscopy information}\label{sed_variation_result}

The normalised rest-frame SEDs of the {\it SPICY} PAH galaxies show a hint of systematic variation of the SED shape as a function of redshift (Sect.~\ref{rest_frame_seds}).
We characterised the shape variation by comparing the photometric monochromatic luminosities and flux ratios measured at four wavelengths in the rest frame.
We already measured the photometric monochromatic luminosities at $7.7\ \mu$m, $\nu L_{\nu\ \mathrm{photo}}$ ($7.7\ \mu$m), and $3.5\ \mu$m, $\nu L_{\nu\ \mathrm{photo}}$ ($3.5\ \mu$m), in constructing the MIR- and NIR-normalised rest-frame SEDs, respectively.
In addition, we measured photometric monochromatic luminosities at $2.0\ \mu$m, $\nu L_{\nu\ \mathrm{photo}}$ ($2.0\ \mu$m), and $11.3\ \mu$m, $\nu L_{\nu\ \mathrm{photo}}$ ($11.3\ \mu$m).
The wavelength of $2.0\ \mu$m is the shortest wavelength that the NIR filters can cover even for the lowest redshift galaxies.
The rest-frame $2.0\ \mu$m flux was measured in the same way as the $3.5\ \mu$m flux, but by using only the N2 ($2.4\ \mu$m) and N3 ($3.2\ \mu$m) bands for all sources.
This is because the $2.0\ \mu$m is close to the $1.6\ \mu$m stellar SED peak, and we cannot assume a simple Rayleigh-Jeans approximation there.
The wavelength of $11.3\ \mu$m is to cover the PAH~$11.3\ \mu$m and dust continuum underneath.
The rest-frame $11.3\ \mu$m flux was measured in the same way as the $7.7\ \mu$m flux, but with the L15 ($15.0\ \mu$m) and L18W ($18.0\ \mu$m) bands at $z>0.25$, and the S11 ($11.0\ \mu$m) and L15 ($15.0\ \mu$m) bands at $z<0.25$.
The rest-frame flux ratios of the $7.7\ \mu$m flux to the $3.5\ \mu$m flux, the $3.5\ \mu$m flux to the $2.0\ \mu$m flux, and the $11.3\ \mu$m flux to the $7.7\ \mu$m flux (hereafter, $F_{\rm rest\ 7.7\ \mu \rm m}/F_{\rm rest\ 3.5\ \mu \rm m}$, $F_{\rm rest\ 3.5\ \mu \rm m}/F_{\rm rest\ 2.0\ \mu \rm m}$, and $F_{\rm rest\ 11.3\ \mu \rm m}/F_{\rm rest\ 7.7\ \mu \rm m}$\footnote{
Although $F_{\rm rest\ 11.3\ \mu \rm m}/F_{\rm rest\ 7.7\ \mu \rm m}$ and $S11$~($11.0\ \mu$m)$/S7$~($7.0\ \mu$m) (Sect.~\ref{colour_redshift_diagrams}) apparently look similar, they are quite different at $z>0$.
For example, $S11/S7$ is closer to $F_{\rm rest\ 7.7\ \mu \rm m}/F_{\rm rest\ 3.5\ \mu \rm m}$ at $0.35<z<0.5$.
}, respectively) were then calculated.
Table~\ref{table9} lists the luminosities and flux ratios of the PAH galaxies.

\begin{figure} 
\centering
\resizebox{\hsize}{!}{\includegraphics{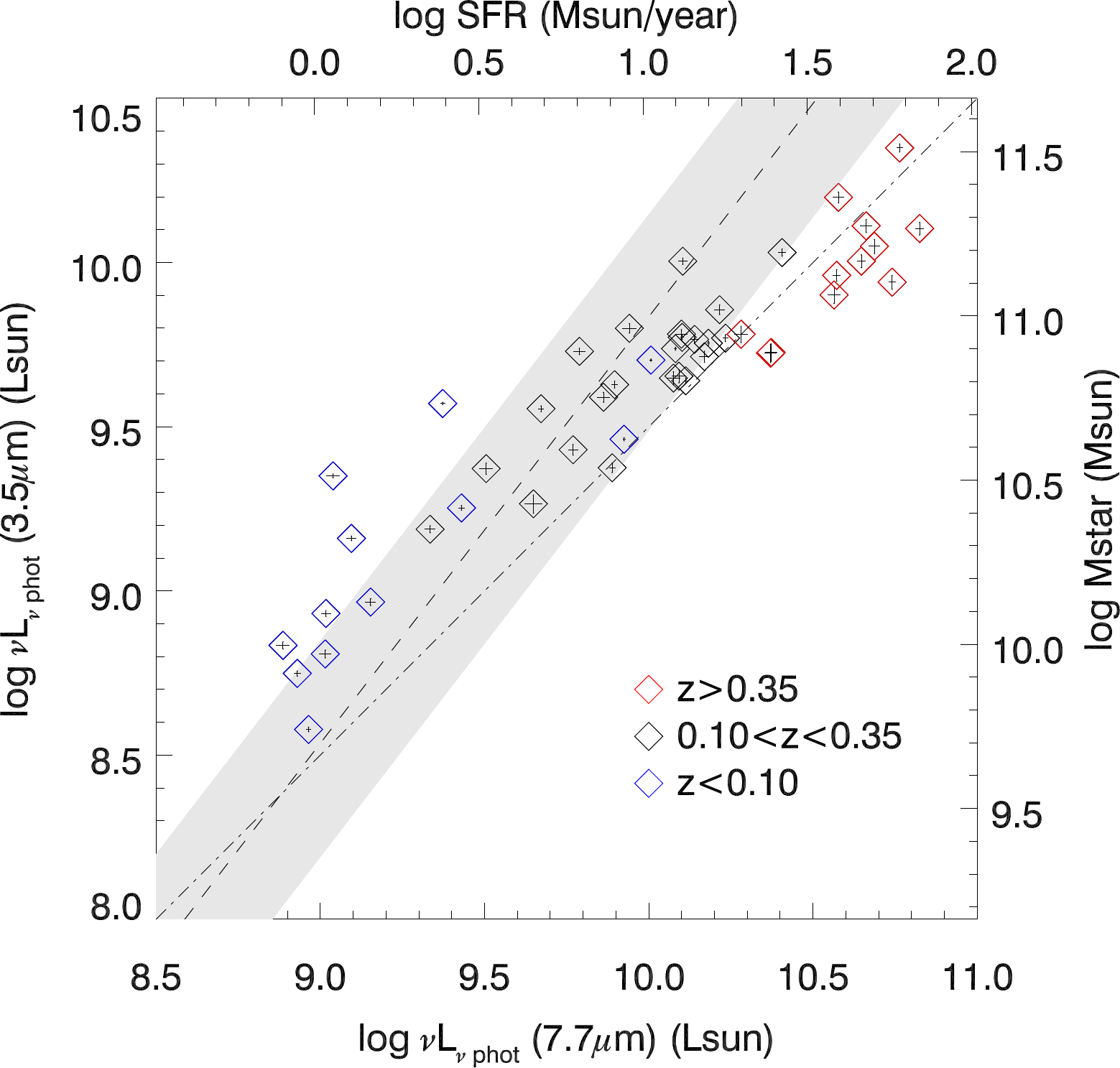}}
\caption{Comparison between the photometric monochromatic luminosities at $7.7\ \mu$m, $\nu L_{\nu\ \mathrm{photo}}$ ($7.7\ \mu$m), and at $3.5\ \mu$m, $\nu L_{\nu\ \mathrm{photo}}$ ($3.5\ \mu$m).
The PAH galaxies are shown with one-sigma error bars in different colours for different redshift bins: blue, black, and red for the nearby, mid-$z$, and higher-$z$ galaxies, respectively.
SFR and stellar mass are estimated from $\nu L_{\nu\ \mathrm{photo}}\ (7.7\ \mu\mathrm{m})$ and $\nu L_{\nu\ \mathrm{photo}}\ (3.5\ \mu\mathrm{m})$, respectively, as explained in the text.
A correlation between stellar mass and SFR for main sequence galaxies for $z\sim 0$ SDSS sample \citep{elbaz07} is indicated by a grey hatched area, showing one-sigma scatter around the best fit line (broken line).
A diagonal dot-broken line for a fixed PAH-to-stellar luminosity ratio ($\log \nu L_{\nu\ \mathrm{photo}}\ (7.7\ \mu\mathrm{m})=0.50+\log \nu L_{\nu\ \mathrm{photo}}\ (3.5\ \mu\mathrm{m})$) is used to identify the PAH-enhanced population (on the {\it lower right} side of this line).
}
\label{nuLnu77_vs_L35}
\end{figure}

\begin{figure*}[t] 
\centering
\includegraphics[width=17cm]{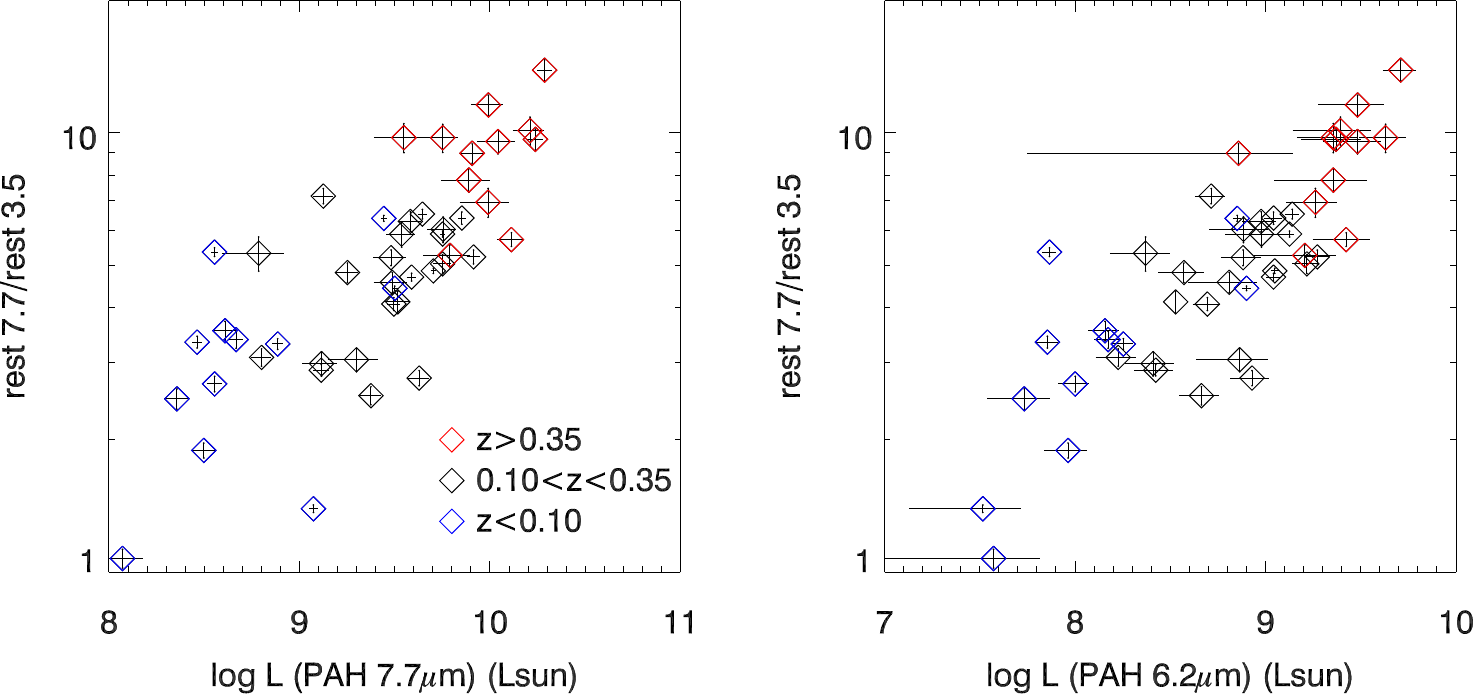}
\caption{Correlations between $F_{\rm rest\ 7.7\ \mu \rm m}/F_{\rm rest\ 3.5\ \mu \rm m}$ and the spectroscopic luminosities of the PAH~$7.7\ \mu$m, $L_\mathrm{PAH}$ ($7.7\ \mu$m) ({\it left}), and PAH~$6.2\ \mu$m, $L_\mathrm{PAH}$ ($6.2\ \mu$m) ({\it right}).
The PAH galaxies are shown with one-sigma error bars in different colours for different redshift bins: blue, black, and red for the nearby, mid-$z$, and higher-$z$ galaxies, respectively.
}
\label{rest_mir_nir_colour_pah_luminosity}
\end{figure*}

We found a significant systematic variation of relative strength of the MIR bump with respect to the OPT--NIR one at $z\gtrsim 0.35$.
Figure~\ref{nuLnu77_vs_L35} compares $\nu L_{\nu\ \mathrm{photo}}$ ($7.7\ \mu$m) and $\nu L_{\nu\ \mathrm{photo}}$ ($3.5\ \mu$m), indicating that the slope of the correlation is not unity.
In particular, we found a population showing stronger $\nu L_{\nu\ \mathrm{photo}}$ ($7.7\ \mu$m) with respect to $\nu L_{\nu\ \mathrm{photo}}$ ($3.5\ \mu$m) at $z>0.35$ when compared to galaxies at $0.1<z<0.35$.
We defined a PAH-enhanced population as a group of galaxies following $\log \nu L_{\nu\ \mathrm{photo}}\ (7.7\ \mu\mathrm{m})>0.50+\log \nu L_{\nu\ \mathrm{photo}}\ (3.5\ \mu\mathrm{m})$ (or, equivalently, $F_{\rm rest\ 7.7\ \mu \rm m}/F_{\rm rest\ 3.5\ \mu \rm m}>7.0$).
We found 10 PAH-enhanced galaxies in total, nine of which are at $z>0.35$ (out of all 12 galaxies in this redshift range; Table~\ref{table1}).
All three remaining PAH galaxies at $z\gtrsim 0.35$ are not PAH-enhanced according to our definition, but they show modest PAH enhancement when compared to $z\sim 0$ SDSS main-sequence galaxies (\citealt{elbaz07}; Fig.~\ref{nuLnu77_vs_L35}).
This figure also shows SFR inferred from $\nu L_{\nu\ \mathrm{photo}}$ ($7.7\ \mu$m) (Sect.~\ref{pah_luminosity_comp_section}) and stellar mass measured by using $\nu L_{\nu\ \mathrm{photo}}$ ($3.5\ \mu$m).
Here, the stellar mass was measured by adopting the NGC~6090 SWIRE SED template, but this mass measurement is insensitive to types of galaxy SEDs adopted; 
the difference from the case with the M82 template is 0.12 dex.
Thus, the trend of the relative enhancement of $\nu L_{\nu\ \mathrm{photo}}$ ($7.7\ \mu$m) over $\nu L_{\nu\ \mathrm{photo}}$ ($3.5\ \mu$m) can be interpreted that SFR becomes larger with respect to the stellar mass at higher redshifts, particularly for the PAH-enhanced population at $z\gtrsim 0.35$ (see Sect.~\ref{SFR_sSFR_discussion} for more discussions).
Similarly, Fig.~\ref{rest_mir_nir_colour_pah_luminosity} shows that $F_{\rm rest\ 7.7\ \mu \rm m}/F_{\rm rest\ 3.5\ \mu \rm m}$ increases as both PAH~$7.7\ \mu$m and $6.2\ \mu$m luminosities increase.
This suggests that the relative strength of the MIR bump is controlled by the PAH luminosity, and that SFR per unit stellar mass (specific SFR or sSFR) becomes larger with SFR.

\begin{figure*}[t] 
\centering
\includegraphics[width=17cm]{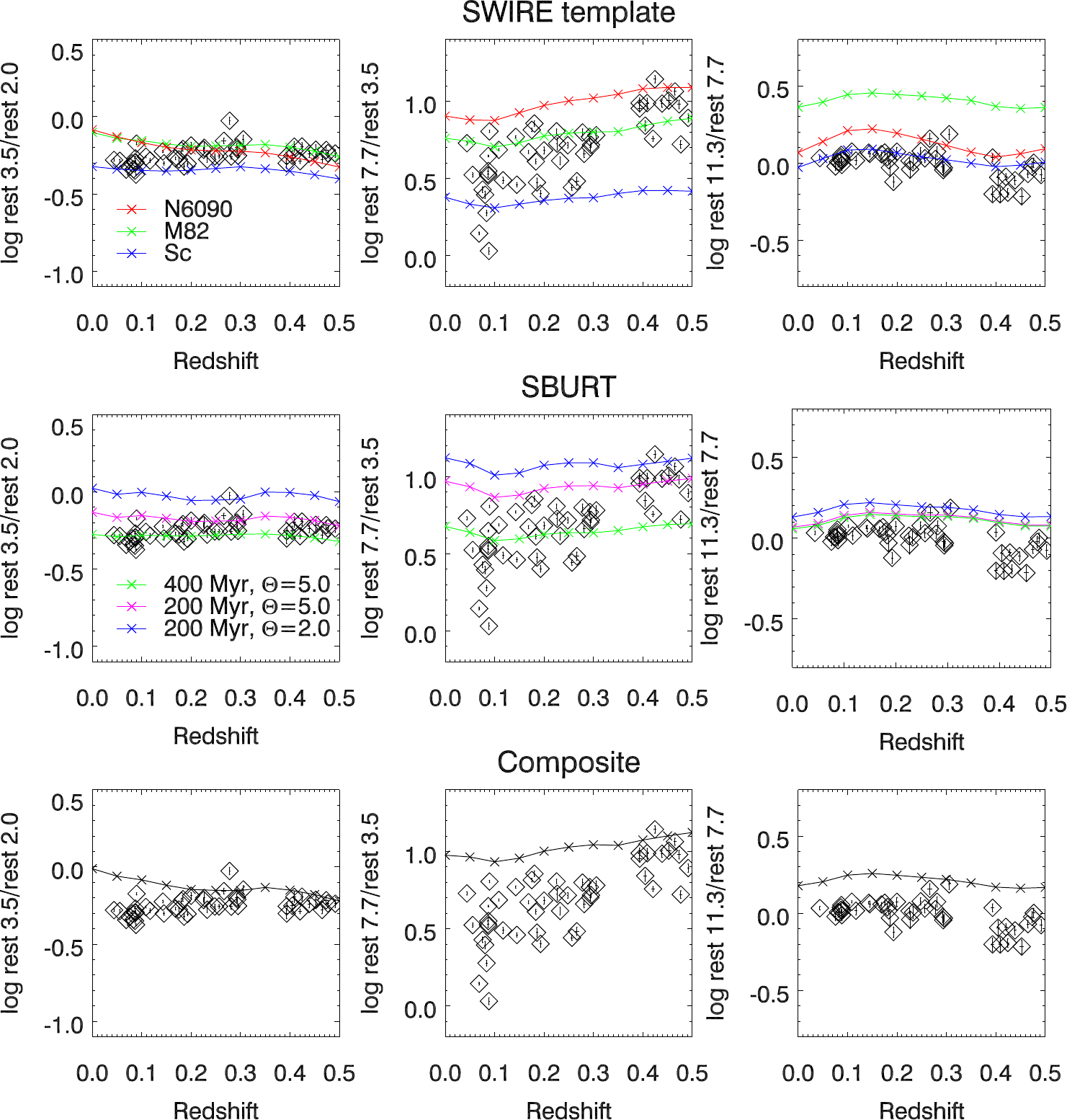}
\caption{
Rest-frame flux ratios of $F_{\rm rest\ 3.5\ \mu \rm m}/F_{\rm rest\ 2.0\ \mu \rm m}$ ({\it left}), $F_{\rm rest\ 7.7\ \mu \rm m}/F_{\rm rest\ 3.5\ \mu \rm m}$ ({\it middle}), and $F_{\rm rest\ 11.3\ \mu \rm m}/F_{\rm rest\ 7.7\ \mu \rm m}$ ({\it right}) as a function of redshift.
The flux ratios of the PAH galaxies and those of the templates/models are shown with diamonds with one-sigma error bars and connected crosses, respectively.
To enable easier visual comparison among the panels, all three kinds of the panels cover the same 1.6 dex-wide range in the flux ratios.
{\it Top}: the SWIRE SED templates of Sc (blue), M82 (green), and NGC~6090 (red).
{\it Middle}: the SBURT models of ($t_{\rm burst}$, $\Theta$)=(400~Myr, 5.0), (200~Myr, 5.0), and (200~Myr, 2.0) in green, magenta, and blue, respectively.
{\it Bottom}: the composite model.
We caution that small changes of the template/model flux ratios are not to predict real changes of the ratios as a function of redshift, but to demonstrate an accuracy of our analysis method (Sect.~\ref{SED_fit_result}).
}
\label{rest_mir_nir_colour_redshift_model_temp}
\end{figure*}

\begin{figure*}[t] 
\centering
\includegraphics[width=17cm]{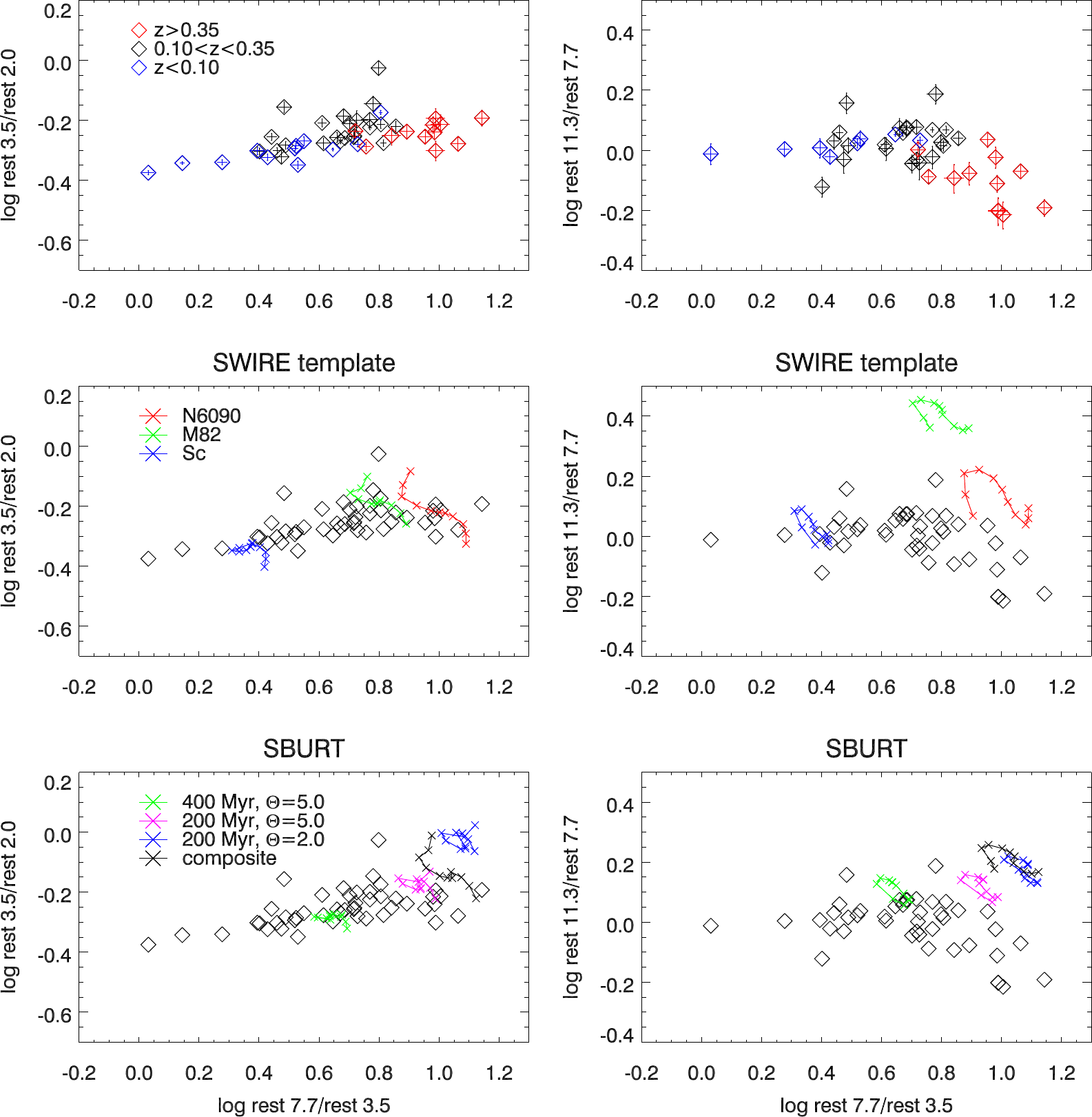}
\caption{
Rest-frame flux ratio diagrams between $F_{\rm rest\ 7.7\ \mu \rm m}/F_{\rm rest\ 3.5\ \mu \rm m}$ and $F_{\rm rest\ 3.5\ \mu \rm m}/F_{\rm rest\ 2.0\ \mu \rm m}$ ({\it left}) and between $F_{\rm rest\ 7.7\ \mu \rm m}/F_{\rm rest\ 3.5\ \mu \rm m}$ and $F_{\rm rest\ 11.3\ \mu \rm m}/F_{\rm rest\ 7.7\ \mu \rm m}$ ({\it right}).
The flux ratios of the observations and templates/models are shown with diamonds and connected crosses, respectively.
{\it Top}: the PAH galaxies in the nearby, mid-$z$, and higher-$z$ redshift bins in blue, black, and red, respectively, with one-sigma error bars.
{\it Middle}: the SWIRE SED templates of Sc (blue), M82 (green), and NGC~6090 (magenta) at $z=0.0$--0.5 in steps of 0.05.
{\it Bottom}: the SBURT models of ($t_{\rm burst}$, $\Theta$)=(400~Myr, 5.0), (200~Myr, 5.0), and (200~Myr, 2.0), and the composite model in green, magenta, blue, and black, respectively.
We caution that small changes of the template/model flux ratios are not to predict real changes of the ratios as a function of redshift, but to demonstrate an accuracy of our analysis method (Sect.~\ref{SED_fit_result}).
}
\label{rest_mir_nir_colours_colours_model_temp}
\end{figure*}

We further explored the SED shape variations of the PAH galaxies by using other rest-frame flux ratios.
The $F_{\rm rest\ 3.5\ \mu \rm m}/F_{\rm rest\ 2.0\ \mu \rm m}$ ratio corresponds to slope of the stellar SED bump at its longer side of its peak ($1.6\ \mu$m), if the stellar SED dominates at NIR.
The $F_{\rm rest\ 11.3\ \mu \rm m}/F_{\rm rest\ 7.7\ \mu \rm m}$ ratio is a flux ratio at two peaks of the PAH $11.3\ \mu$m and $7.7\ \mu$m features, although the ratio can be modified by either hot dust continuum underneath the PAH features, silicate absorption around $9.7\ \mu$m, or red AGN continuum (Sect.~\ref{113_77_variation} for more discussion).
Figure~\ref{rest_mir_nir_colour_redshift_model_temp} shows the redshift dependence of these two rest-frame line ratios as well as $F_{\rm rest\ 7.7\ \mu \rm m}/F_{\rm rest\ 3.5\ \mu \rm m}$.
Figure~\ref{rest_mir_nir_colours_colours_model_temp} shows correlations among these three flux ratios.
As we saw earlier, the $F_{\rm rest\ 7.7\ \mu \rm m}/F_{\rm rest\ 3.5\ \mu \rm m}$ ratio shows enhancement at $z\gtrsim 0.35$ constituting the PAH-enhanced population, whereas it shows significant scatter for a given redshift at $z\lesssim 0.35$.
In the meantime, the $F_{\rm rest\ 11.3\ \mu \rm m}/F_{\rm rest\ 7.7\ \mu \rm m}$ ratio is almost constant at $z\lesssim 0.35$, and it decreases by a factor of $\lesssim 2$ at $z\gtrsim 0.35$ while $F_{\rm rest\ 7.7\ \mu \rm m}/F_{\rm rest\ 3.5\ \mu \rm m}$ increases by a factor of $\sim 2$.
In contrast, the $F_{\rm rest\ 3.5\ \mu \rm m}/F_{\rm rest\ 2.0\ \mu \rm m}$ ratio shows only a small variation with redshift.

We examined possible selection bias on the observed systematic SED variation caused by using the S9W ($9.0\ \mu$m) filter for the source detection.
The most prominent PAH~$7.7\ \mu$m feature is fully covered with the S9W filter up to $z\simeq 0.35$, and its red part goes out of the filter bandpass at $0.35\lesssim z\lesssim 0.5$ although its peak is still included within the filter.
This leads to the selection bias of preferentially detecting the PAH-bright sources at $0.35\lesssim z\lesssim 0.5$, i.e., PAH-enhanced but PAH-faint population would be missed in our sample.
In contrast, PAH-less-enhanced but PAH-bright population (e.g., massive Sa--Sc galaxies) could be detected with the S9W filter, and lack of such population (Figs.~\ref{normalized_rest_SEDs}, \ref{nuLnu77_vs_L35}) indicates that they are indeed rare.
At $z\lesssim 0.35$, both galaxies with and without the PAH enhancement could be detected, because they are brighter in the S9W filter due to their proximity and better coverage of the PAH~$7.7\ \mu$m profile within its bandpass.
This indicates that lack of the PAH-enhanced population at $z\lesssim 0.35$ is not due to the selection bias.
Therefore, it is safe to conclude that the PAH-enhanced population emerges at $z>0.35$ and dominates among the vigorous star-forming populations showing high PAH luminosity.
Another trend of systematically depressed $F_{\rm rest\ 11.3\ \mu \rm m}/F_{\rm rest\ 7.7\ \mu \rm m}$ for galaxies with enhanced $F_{\rm rest\ 7.7\ \mu \rm m}/F_{\rm rest\ 3.5\ \mu \rm m}$ at $z\gtrsim 0.35$ seems not due to any selection biases.
This is because the L15 ($15.0\ \mu$m) band that covers the rest-frame $11.3\ \mu$m at $z\gtrsim 0.35$ was not considered for source selection.

There remains a question if the apparent emergence of the PAH-enhanced population at $z\gtrsim 0.35$ is statistically significant or by chance due to small number statistic.
We tested if the observed relative fractions of the PAH-enhanced population in the higher-$z$ and mid-$z$ redshift bins are statistically significantly different from each other.
We found one and nine PAH-enhanced galaxies out of all 23 and 12 galaxies in the mid-$z$ and higher-$z$ bins, respectively (Table~\ref{table1}).
We performed Fisher's exact test\footnote{This is an exact test of independence of two nominal variables, and works even on small samples.
We used a ``fisher.exact'' function in the ``stats'' package of the ``R'' language.} to calculate the total probability of the observed numbers of the PAH-enhanced population, given the sample size, being as extreme or more extreme if the null hypothesis (expectations of the fractions of the PAH-enhanced population in both redshift bins are the same) is true.
We found $p=2.8\times 10^{-5}$ ($p$: probability that one expects the result being the same as, or more extreme than, the actual observed result when the null hypothesis is true), i.e., this null hypothesis is rejected at $>99$\% significance level.
Therefore, we concluded that the PAH-enhanced population indeed emerges at $z\gtrsim 0.35$, although analysis with larger sample size and more fair sampling at $z\gtrsim 0.35$ is preferred to confirm this conclusion.

\subsection{Template/model SED fit to the rest-frame SEDs of the PAH galaxies}\label{SED_fit_result}

In an attempt to understand the typical SED shape of the PAH galaxies and its systematic variation as demonstrated so far, we compared the observed rest-frame SEDs with the template/model SEDs of known characteristics.
We again adopted the SED templates of Sc, starburst (M82), and LIRG (NGC~6090) from the SWIRE SED library.
In addition, we adopted SEDs of a physical starburst model SBURT \citep{takagi03} to gain insight into the physical parameters of star formation.
The SBURT models have been tested for ultraviolet-selected starbursts and ULIRGs \citep{takagi03,takagi03b}, and applied to, e.g., the sub-millimetre galaxies (e.g., \citealt{takagi04}) and AKARI/IRC NEP-Deep photometric samples (e.g., \citealt{takagi07,takagi10}).
We fitted the averaged SEDs and their systematic variation, rather than fitting SEDs of individual galaxies.
In order to make fair and direct comparisons between the template/model SEDs and the observations, we considered smoothing of the spectral features due to our filter bandpasses.
We simulated observations of redshifted template/model galaxies, by redshifting their SEDs and calculating in-band fluxes through the IRC filters, to make a synthetic photometric catalogue.
Then the rest-frame SEDs were reconstructed from this synthetic catalogue, and the flux ratios of $F_{\rm rest\ 3.5\ \mu \rm m}/F_{\rm rest\ 2.0\ \mu \rm m}$, $F_{\rm rest\ 7.7\ \mu \rm m}/F_{\rm rest\ 3.5\ \mu \rm m}$, and $F_{\rm rest\ 11.3\ \mu \rm m}/F_{\rm rest\ 7.7\ \mu \rm m}$ were calculated by following exactly the same procedure as for the real observations.

Before comparing with the observations, we verified our method of photometrically measuring the SED shape by using the synthetic catalogue of the template/model SEDs.
The reconstructed rest-frame flux ratios from the synthetic catalogue should not change as a function of redshift, although this is not the case in reality.
This is because the spectral resolution of the IRC photometry ($R\simeq 5$) is not high enough to properly sample complicated rich PAH features, and different filters with different spectral resolutions are used to sample such features depending on the redshift.
We found that the synthetic flux ratios change by $\sim 0.2$ dex in our redshift coverage ($z=0.0$--0.5; e.g., Fig.~\ref{rest_mir_nir_colour_redshift_model_temp}), indicating that our analysis methods are reliable within $\sim 0.2$ dex.

\subsubsection{Comparison with the SWIRE SED templates}\label{sed_fit_swire}

\begin{figure*}[t] 
\centering
\includegraphics[width=17cm]{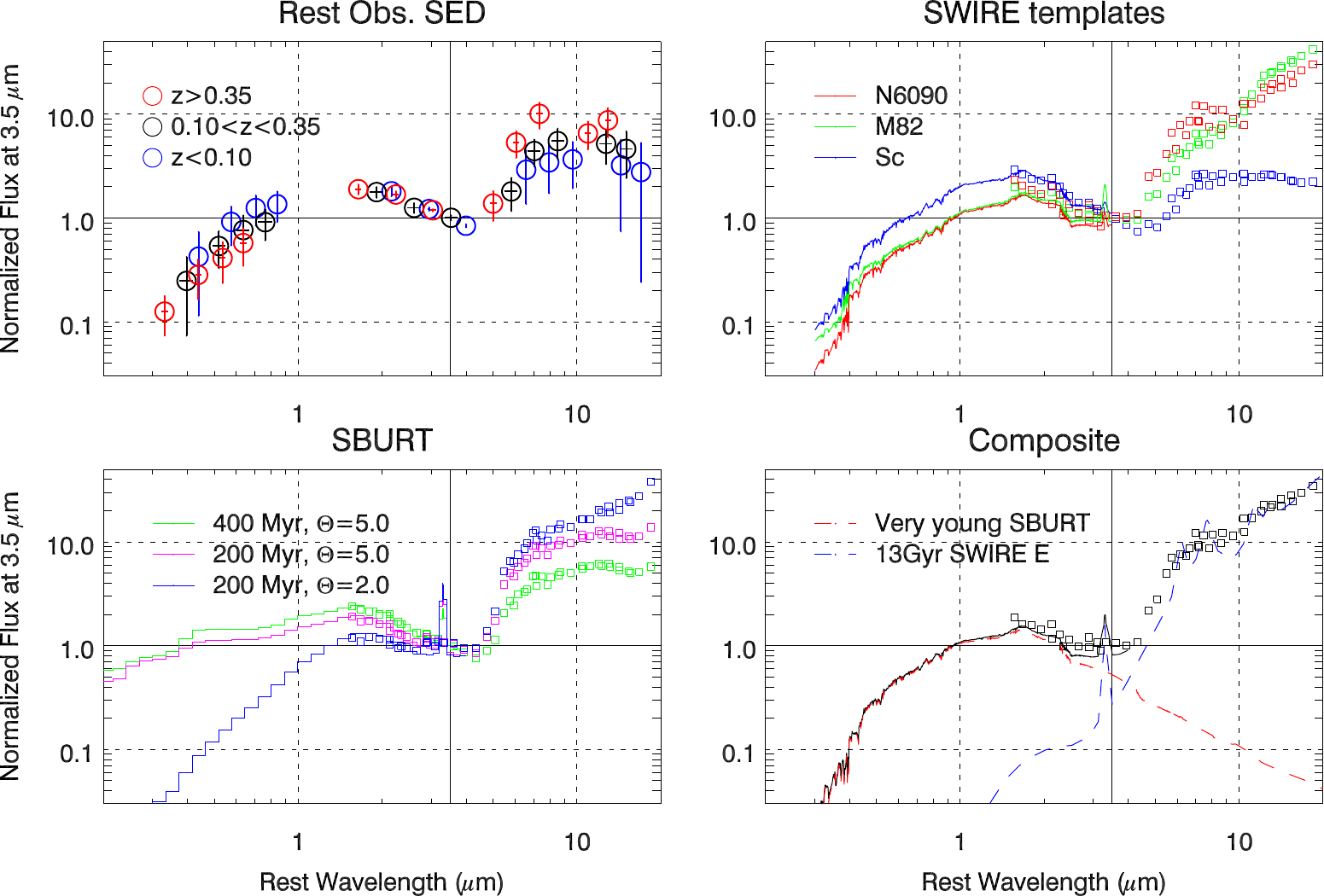}
\caption{Comparisons of the NIR-normalised rest-frame SEDs between the PAH galaxies and the templates/models.
The normalisation wavelength is indicated by a black solid vertical line in each panel.
{\it Top left}: the averaged observed SEDs in each redshift bin from Fig.~\ref{normalized_rest_SEDs_bin}.
{\it Top right}: the SWIRE SED templates of Sc (blue), M82 (green), and NGC~6090 (red).
{\it Bottom left}: the SBURT SED models with ($t_{\rm burst}$, $\Theta$)=(400~Myr, 5.0), (200~Myr, 5.0), and (200~Myr, 2.0) in green, magenta, and blue, respectively.
{\it Bottom right}: two-component composite SED (black) that combines the SWIRE E (13~Gyr old) SED template (red broken line) and a very young and compact SBURT SED model (70~Myr, 1.0; blue broken line).
For the template/model SEDs, the reconstructed SEDs from the synthetic catalogue are shown with small boxes at NIR--MIR wavelengths, whereas the full-resolution original SEDs are shown at OPT--NIR wavelengths.
}
\label{rest_SEDs_model_temp}
\end{figure*}

We compared the observations of the PAH galaxies and the SWIRE SED templates by using the three rest-frame flux ratios (Figs.~\ref{rest_mir_nir_colour_redshift_model_temp},~\ref{rest_mir_nir_colours_colours_model_temp}) and in a form of SEDs (Fig.~\ref{rest_SEDs_model_temp}).
To reproduce the observed range of $F_{\rm rest\ 7.7\ \mu \rm m}/F_{\rm rest\ 3.5\ \mu \rm m}$ at $z\lesssim 0.35$ (2--7), a range of template types between Sc and starburst is needed.
The E and Sa templates show too small $F_{\rm rest\ 7.7\ \mu \rm m}/F_{\rm rest\ 3.5\ \mu \rm m}$ ($\sim 1$).
The larger observed $F_{\rm rest\ 7.7\ \mu \rm m}/F_{\rm rest\ 3.5\ \mu \rm m}$ at $z\gtrsim 0.35$ (7--14) can be reproduced by the LIRG template, although the depressed $F_{\rm rest\ 11.3\ \mu \rm m}/F_{\rm rest\ 7.7\ \mu \rm m}$ cannot be reproduced at the same time.
The starburst template predicts too large $F_{\rm rest\ 11.3\ \mu \rm m}/F_{\rm rest\ 7.7\ \mu \rm m}$ ratio in our redshift coverage due to too steep increase of the red continuum at $>10\ \mu$m (Fig.~\ref{rest_SEDs_model_temp}).
The NIR flux ratio of $F_{\rm rest\ 3.5\ \mu \rm m}/F_{\rm rest\ 2.0\ \mu \rm m}$ can be reproduced by all Sc, starburst, and LIRG templates.

\subsubsection{Comparison with the SBURT models}\label{sed_fit_sburt}

The SBURT model is a simple physical starburst model, in which a star cluster evolves and radiative transfer of the cluster light through dusty medium is solved to predict the observed SED (see \citealt{takagi03} for full details).
Input parameters of the model are dust composition, age of the starburst ($t_{\rm burst}$), and the compactness parameter of the star cluster radial distribution ($\Theta$: the mean stellar density becomes higher for smaller $\Theta$).
For simplicity, we fixed dust composition the Milky-Way type, which produces the most enhanced PAH emission due to larger PAH fraction among three possible choices available within the SBURT, and is suited to fit the PAH-enhanced population.
A starburst happens at the central star cluster following the gas supply from a galactic infall, and the SFR increases until $t_{\rm burst}=100$~Myr, which equals to the time scale of the infall\footnote{An absolute age of 100~Myr here and in the following is a model parameter that can be selected from outside \citep{takagi03}.}.
Both PAH features and hot dust continuum are enhanced as SFR increases due to increased number of UV photons from young stars.
In the meantime, the stellar mass accumulates as the star formation continues, and the stellar continuum emission at NIR increases with $t_{\rm burst}$.
Therefore, the $F_{\rm rest\ 7.7\ \mu \rm m}/F_{\rm rest\ 3.5\ \mu \rm m}$ ratio is larger at $t_{\rm burst}\simeq 100$~Myr, and it becomes smaller at $t_{\rm burst}>100$~Myr.
The optical depth, which is calculated based on amount of the gas for star formation and the cluster compactness, is larger in more compact (smaller $\Theta$) star cluster.
The OPT--NIR colours are redder and $F_{\rm rest\ 7.7\ \mu \rm m}/F_{\rm rest\ 3.5\ \mu \rm m}$ is larger with larger optical depth.
The younger starburst ($t_{\rm burst}<100$~Myr) shows larger $F_{\rm rest\ 3.5\ \mu \rm m}/F_{\rm rest\ 2.0\ \mu \rm m}$ due to enhanced hot dust continuum that contributes to even the $3.5\ \mu$m flux.
In contrast, the $F_{\rm rest\ 11.3\ \mu \rm m}/F_{\rm rest\ 7.7\ \mu \rm m}$ ratio changes only a little over a wide range of the input parameters.
This is because no mechanism of changing PAH inter-band flux ratios is implemented in the model, and change of the dust continuum underneath the PAH features does not affect the flux ratio very much due to the small wavelength separation.

In order to reproduce galaxies with moderately enhanced PAH emission ($F_{\rm rest\ 7.7\ \mu \rm m}/F_{\rm rest\ 3.5\ \mu \rm m}\simeq 7$--10) while keeping the constraint of the NIR colour ($F_{\rm rest\ 3.5\ \mu \rm m}/F_{\rm rest\ 2.0\ \mu \rm m}$), one needs middle-aged ($t_{\rm burst}=200$~Myr) starburst with larger compactness parameter ($\Theta\simeq 5.0$; Figs.~\ref{rest_mir_nir_colour_redshift_model_temp},~\ref{rest_mir_nir_colours_colours_model_temp},~\ref{rest_SEDs_model_temp}).
In order to reproduce galaxies with less-enhanced PAH emission ($F_{\rm rest\ 7.7\ \mu \rm m}/F_{\rm rest\ 3.5\ \mu \rm m}\lesssim 7$), one needs older starbursts with larger compactness parameter ($t_{\rm burst}>200$~Myr, $\Theta\simeq 5.0$).
Such starburst can explain the wide range of $F_{\rm rest\ 7.7\ \mu \rm m}/F_{\rm rest\ 3.5\ \mu \rm m}$ as a function of the burst age, while keeping the NIR colour.
In order to reproduce galaxies with more enhanced PAH emission ($F_{\rm rest\ 7.7\ \mu \rm m}/F_{\rm rest\ 3.5\ \mu \rm m}\simeq 14$), one need a younger starburst ($t_{\rm burst}<200$~Myr) or a starburst with smaller $\Theta$ ($\simeq 2$).
Although the NIR SED is almost insensitive to the burst age and optical depth, such starburst fails to reproduce the NIR colour due to non-negligible hot dust continuum contribution at $3.5\ \mu$m or too large optical depth for reddening even at NIR.
The OPT SEDs provide an independent constraint on the SBURT models.
This is because all these SBURT models predict blue OPT SEDs due to young and blue stars in the star clusters, whereas the OPT SEDs are typically as red as those of very old populations (Fig.~\ref{rest_SEDs_model_temp}).
If the observed range of $F_{\rm rest\ 7.7\ \mu \rm m}/F_{\rm rest\ 3.5\ \mu \rm m}$ is due to a range of the burst age, the OPT SEDs also change accordingly, unless the optical depth is also adjusted to compensate the effect of the burst age.

\subsubsection{Comparison with the burst-on-old composite models}\label{sed_fit_composite}

We have shown above that SBURT models for reproducing pronounced PAH bump cannot consistently reproduce the OPT--NIR--MIR SEDs.
This is particularly true for the PAH-enhanced population at $z\gtrsim 0.35$:
one needs a starburst that shows nearly maximum $F_{\rm rest\ 7.7\ \mu \rm m}/F_{\rm rest\ 3.5\ \mu \rm m}$ to enhance the PAH bump as observed, but such starburst inevitably predicts much bluer OPT SEDs than the observations.
To solve this problem, we explored two-component composite models that combine very young ($t_{\rm burst}\lesssim 100$~Myr) and old ($\simeq 10$~Gyr) populations for the PAH-enhanced population.
If the very old component with relatively red OPT SED dominates the OPT--NIR SED, and the very young starburst with strong PAH emission dominates the MIR SED, we may be able to enhance the PAH bump for a given OPT--NIR SED by increasing the relative flux contribution of the starburst component.
To also reproduce the observed narrow NIR colour range, however, the very young component must be highly reddened not to significantly contribute to the $3.5\ \mu$m flux.
A very young starburst with small compactness parameter can be very optically thick, and seems appropriate for the composite.
Figure~\ref{rest_SEDs_model_temp} shows one example of such composite models, in which an E (13~Gyr old) SWIRE template is combined with a very young ($t_{\rm burst}=70$~Myr), compact ($\Theta=1.0$), and thus optically very thick ($\tau(\mathrm{V})\simeq 16.5$) SBURT model.
The parameters of this composite model ($t_{\rm burst}$ and $\Theta$ of the young starburst, and the relative flux contribution of the young starburst with respect to the old population) were found so that the composite SED reproduces the typical rest-frame colours of the PAH-enhanced population ($\log F_{\rm rest\ 7.7\ \mu \rm m}/F_{\rm rest\ 3.5\ \mu \rm m}\ge 1.1$ and $-0.3\le \log F_{\rm rest\ 3.5\ \mu \rm m}/F_{\rm rest\ 2.0\ \mu \rm m}\le -0.2$; see Fig.~\ref{rest_mir_nir_colour_redshift_model_temp}).
We here focused on these two rest-frame NIR and MIR colours because they best characterise the NIR--MIR SED of the PAH galaxies (Sect.~\ref{sed_variation_result}), and the OPT SED is not sensitive to the parameters of this burst-on-old composite.
We caution that very similar composite SEDs can be constructed with different mixtures of populations with slightly different parameters.
For example, the $F_{\rm rest\ 7.7\ \mu \rm m}/F_{\rm rest\ 3.5\ \mu \rm m}$ ratio of the very young component changes with the burst age, but the ratio of the composite SED can be unchanged if their relative flux contribution is adjusted.

This composite SED better reproduces the PAH-enhanced population than any single-component SBURT models (Figs.~\ref{rest_mir_nir_colour_redshift_model_temp},~\ref{rest_mir_nir_colours_colours_model_temp},~\ref{rest_SEDs_model_temp}).
We note that the SWIRE SED template of NGC~6090, which fits the SEDs of the PAH-enhanced population in all OPT--NIR--MIR wavelengths (Fig.~\ref{rest_SEDs_model_temp}; Sect.~\ref{sed_fit_swire}), resembles this composite SED very much.
We caution that, although this composite model can reproduce the most representative characteristics of the observe SEDs of the PAH-enhanced population, as well as the SFR--sSFR correlation among the PAH galaxies in general (Sect.~\ref{SFR_sSFR_discussion}), we could not confirm if these galaxies are indeed made of this kind of composite.
In particular, the presence of the highly absorbed starburst could not be confirmed based on our spectral analysis (Sect.~\ref{pah_fit_test}).
A weakness of the composite model is that it still cannot reproduce the relatively depressed $F_{\rm rest\ 11.3\ \mu \rm m}/F_{\rm rest\ 7.7\ \mu \rm m}$ for the PAH-enhanced population at $z\gtrsim 0.35$, for the same reason as for the single-component SBURT models (Sect.~\ref{sed_fit_sburt})\footnote{
This composite SED actually shows slightly larger $F_{\rm rest\ 11.3\ \mu \rm m}/F_{\rm rest\ 7.7\ \mu \rm m}$ than the SBURT model of $t_{\rm burst}=200$~Myr, $\Theta\simeq 5.0$ that can well describe the NIR--MIR SED of the PAH-enhanced population (Figs.~\ref{rest_mir_nir_colour_redshift_model_temp},~\ref{rest_mir_nir_colours_colours_model_temp},~\ref{rest_SEDs_model_temp}).
This is because the very young component in the composite model adds the $11.3\ \mu$m flux due to more pronounced red dust continuum.}.

\section{Discussion}

\subsection{PAH-enhanced population}

\subsubsection{Star formation rate and specific star formation rate}\label{SFR_sSFR_discussion}

We found that the PAH-enhanced population at $z\gtrsim 0.35$ shows vigorous star formation.
Their PAH luminosity ($\log \nu L_{\nu\ \mathrm{photo}}\ (7.7\ \mu \mathrm{m})\ (L_{\sun})>10.3$) corresponds to $\log L_{\rm IR}\ (L_{\sun})>11.0$, or SFR$\gtrsim 30\ M_{\rm \sun}\ \mathrm{yr}^{-1}$, and they are in a class of LIRGs (Sect.~\ref{pah_luminosity_comp_section}).
This population seems to emerge at $z\gtrsim 0.35$.
As we noted earlier (Sect.~\ref{sed_variation_result}), the selection bias caused by the source detection with the S9W ($9.0\ \mu$m) filter cannot explain this trend.
Also, the difference in the fractions of the PAH-enhanced population between the mid-$z$ and higher-$z$ redshift bins looks statistically significant.
We further note that the survey volume is smaller by a factor of about two in $0.2<z<0.35$ than in $0.35<z<0.5$ for the same sky coverage and the same redshift interval of $dz=0.15$, causing detecting a rare population statistically more difficult in the lower redshift.
However, such survey volume difference unlikely reproduces sudden change of the distribution of $F_{\rm rest\ 7.7\ \mu \rm m}/F_{\rm rest\ 3.5\ \mu \rm m}$ of the PAH galaxies at $z\simeq 0.3$--0.35 (Fig.~\ref{rest_mir_nir_colour_redshift_model_temp}).
Therefore, we concluded that abundance of the PAH-enhanced LIRGs becomes higher at $z\gtrsim 0.35$.
This trend could be a part of the cosmic SFRD evolution whose peak comes at $z\simeq 1$--2 (e.g., \citealt{menendez07,farrah08,pope08,elbaz11,nordon12}; see also \citealt{goto10,goto11a,goto11b}).

The fact that the PAH-luminous galaxies tend to show the PAH enhancement indicates that sSFR is larger in galaxies with larger SFR (Sect.~\ref{sed_variation_result}).
More specifically, vigorous starburst galaxies at $z>0.35$ with SFR$\gtrsim 30\ M_{\rm \sun}\ \mathrm{yr}^{-1}$ show enhanced sSFR by up to a factor of 2 than an upper envelope of the sSFR distribution of modest starburst galaxies at $z<0.35$ (Fig.~\ref{nuLnu77_vs_L35}).
When compared to main sequence galaxies of $z\sim 0$ SDSS sample \citep{elbaz07}, the PAH-luminous and PAH-enhanced galaxies show larger sSFR than the local blue galaxies, whereas the PAH galaxies without PAH enhancement in the mid-$z$ ($0.1<z<0.35$) redshift bin more closely follow the sequence.
\cite{takagi10} identified their PAH-selected galaxies as $S11\ (11.0\ \mu\mathrm{m})/S7\ (7.0\ \mu\mathrm{m})>8$ at $z\sim 0.5$ in their AKARI/IRC NEP-Deep photometric sample, and argued that they are sources with larger sSFR.
The PAH-enhanced population in this paper shows $S11/S7>6$ at $z>0.35$, about a half of which actually shows $S11/S7>8$ (Fig.~\ref{redshift_colour_swire}).
\cite{hanami12} found that sSFR in their MIR-detected star-formation-dominated population in their NEP-Deep photometric sample is higher at higher redshifts up to $z\sim 2$ at all stellar masses.
Such redshift trend about sSFR in the {\it SPICY} sample as well as the NEP-Deep photometric samples seems in agreement with the fact that sSFR in star-forming galaxies is on average higher at higher redshifts (e.g., \citealt{elbaz07,elbaz11,schreiber15}).

The positive correlation between SFR and sSFR for the {\it SPICY} PAH galaxies can be naturally explained if they are burst-on-old composite.
If one adds a burst of star formation on top of the old population, both SFR and sSFR of this galaxy increase.
We introduced such composite to explain the PAH-enhanced population at $z\gtrsim 0.35$ (Sect.~\ref{sed_fit_composite}), but we can also generate a range of composites by adding different amount of the burst component, naturally generating the positive SFR--sSFR correlation.
\cite{takagi10} have already demonstrated need for such composite population for their PAH-selected galaxies.
They performed the SBURT SED fitting on their AKARI/IRC NEP-Deep photometric sample up to $z\sim 1$, and showed that the single-burst models often predict weaker PAH features than the observations.
They fitted the OPT--NIR--MIR SEDs of their PAH-selected sample, and obtained reasonably good fits for about half of their sample galaxies, but failed for the other half because of their stronger-than-modeled PAH~$7.7\ \mu$m.
Their fit generally favoured older ($t_{\rm burst}\simeq 400$~Myr) starbursts than our favoured middle-aged ones ($t_{\rm burst}\simeq 200$~Myr; Sect.~\ref{sed_fit_sburt}) for the NIR--MIR SEDs, because they fitted the OPT SED together with the NIR--MIR one.
The redder OPT SED caused the failure to reproduce the enhanced PAH features at the same time.
Their results can be interpreted that such PAH-enhanced (U)LIRGs at $0.35\lesssim z\lesssim 1$ are experiencing younger ($t_{\rm burst}<400$~Myr) starburst on top of old stellar components.
We note that LIRGs are often interacting galaxies, if not major mergers (e.g., \citealt{goals,kartaltepe10,kartaltepe12,stierwalt13}), and the burst-on-old composite seems a natural consequence in the interacting LIRGs, because the starburst there is triggered during the course of the interaction.
NGC~6090 is also a merging LIRG, and it is not surprising that its SED resembles our composite SED (Sect.~\ref{sed_fit_composite}).
We caution that, although the composite model is the preferred explanation of the SFR--sSFR correlation, we cannot confirm this composite based on our spectral analysis (see also Sect.~\ref{sed_fit_composite}).

\subsubsection{$F_{\rm rest\ 11.3\ \mu \rm m}/F_{\rm rest\ 7.7\ \mu \rm m}$ variation}\label{113_77_variation}

We have shown that both the NGC~6090 SED template and the composite SED have problems in reproducing the relatively depressed $F_{\rm rest\ 11.3\ \mu \rm m}/F_{\rm rest\ 7.7\ \mu \rm m}$ for the PAH-enhanced population at $z\gtrsim 0.35$ (Sect.~\ref{SED_fit_result}).
In such starburst templates/models, both PAH~$11.3\ \mu$m and PAH~$7.7\ \mu$m luminosities, as well as hot dust continuum beneath these PAH features, increase in a similar way with SFR, causing only a small $F_{\rm rest\ 11.3\ \mu \rm m}/F_{\rm rest\ 7.7\ \mu \rm m}$ variation.
The observations indeed show relatively small variation in this flux ratio at $z<0.35$ (Figs.~\ref{rest_mir_nir_colour_redshift_model_temp},~\ref{rest_mir_nir_colours_colours_model_temp}).

We first examined effect of possible AGN contribution to the observed $F_{\rm rest\ 11.3\ \mu \rm m}/F_{\rm rest\ 7.7\ \mu \rm m}$ variation, and found this possibility very low.
Although we already removed AGN-dominated sources based on the NIR$/$NIR--MIR$/$NIR colour-colour diagram (Sect.~\ref{photo_classification_results}), a possibility remains that AGNs partly contribute to the total observed SEDs.
We found that the unobscured AGNs, which typically show red continuum-dominated SEDs (Fig.~\ref{observed_SEDs_AGNcan}), should not dominate at NIR.
This is because the $F_{\rm rest\ 3.5\ \mu \rm m}/F_{\rm rest\ 2.0\ \mu \rm m}$ ratio shows little scatter and is consistent with the colour of the stellar SEDs (Sect.~\ref{sed_variation_result}), and the rest-frame SEDs show a clear dip around $\simeq 4\ \mu$m (Sect.~\ref{rest_frame_seds}).
Such constraints at NIR, however, do not apply to buried AGNs (e.g., \citealt{imanishi07}), because their SEDs are almost completely hidden at OPT--NIR wavelengths.
The contribution of such buried AGNs seems also insignificant, because we found that the MIR SED is dominated by the PAH features around $7.7\ \mu$m (Sect.~\ref{sed_variation_result}).
In addition, contribution of the red MIR continuum of both unobscured and buried AGNs would increase both $F_{\rm rest\ 7.7\ \mu \rm m}/F_{\rm rest\ 3.5\ \mu \rm m}$ and $F_{\rm rest\ 11.3\ \mu \rm m}/F_{\rm rest\ 7.7\ \mu \rm m}$, contradicting the observed trend for the PAH-enhanced population.

\begin{figure}[t] 
\centering
\resizebox{\hsize}{!}{\includegraphics{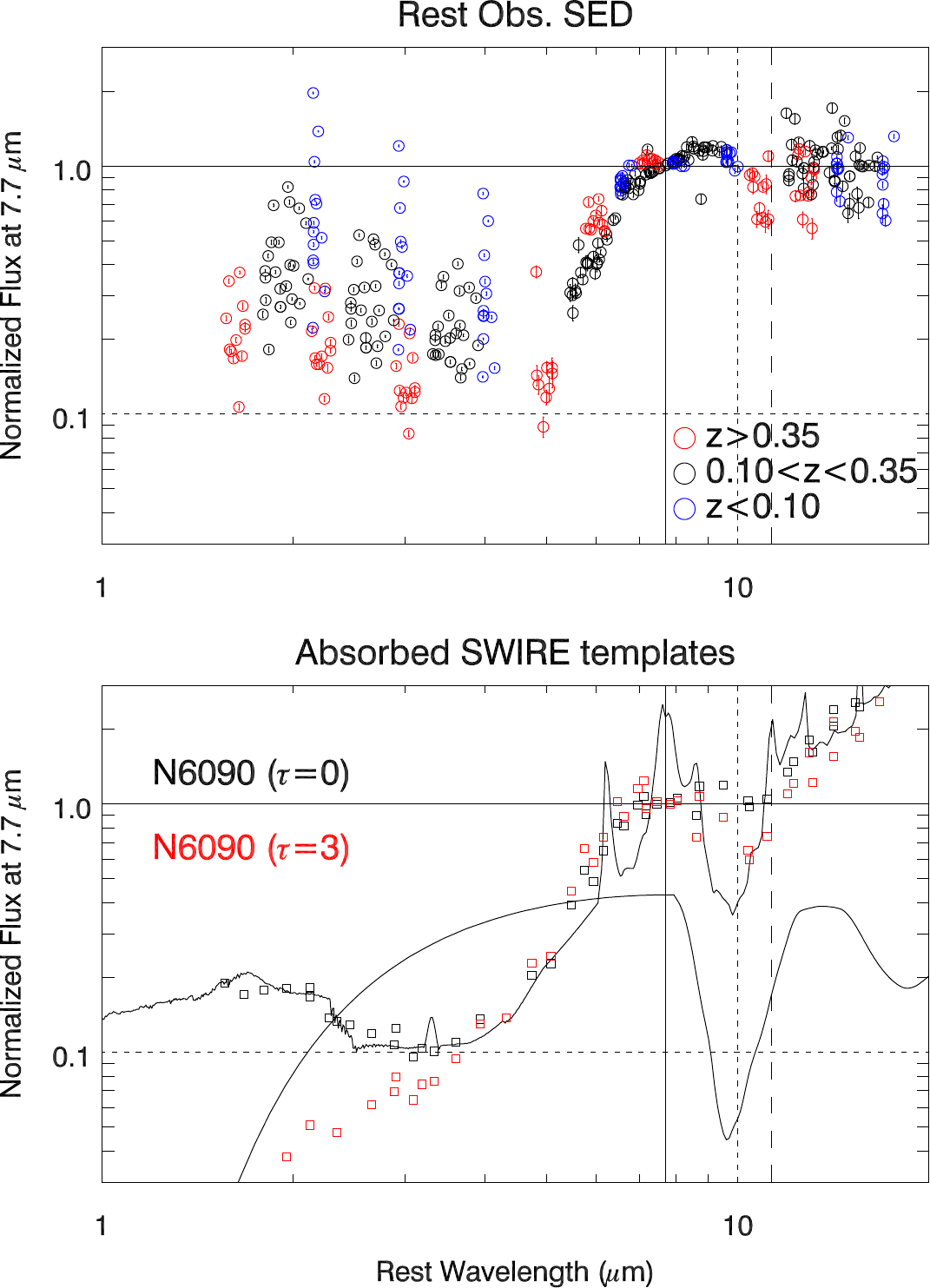}}
\caption{Comparison between the observed MIR-normalised rest-frame SEDs of the PAH galaxies (taken from Fig.~\ref{normalized_rest_SEDs}; {\it top}) and the absorbed/unabsorbed NGC~6090 SEDs ({\it bottom}).
The unabsorbed ($\tau(9.7\ \mu$m$)=0.0$) NGC~6090 SED (taken from Fig.~\ref{rest_SEDs_model_temp}) and the similar but absorbed ($\tau(9.7\ \mu$m$)=3.0$) one are shown in black and red, respectively, in the {\it bottom} panel.
We used the reconstructed SEDs from the synthetic catalogues.
In addition, a full-resolution original (unabsorbed) NGC~6090 SED and an extinction curve for $\tau(9.7\ \mu$m$)=3.0$ (with an arbitrary offset) are shown for reference.
The normalisation wavelength ($7.7\ \mu$m) and the wavelength of $11.3\ \mu$m are indicated by black solid and broken vertical lines, respectively, in each panel.
}
\label{absorbed_sed}
\end{figure}

We examined other possible mechanisms to explain the observed $F_{\rm rest\ 11.3\ \mu \rm m}/F_{\rm rest\ 7.7\ \mu \rm m}$ trend, although we could not identify the true cause.
One possibility is that the broad $9.7\ \mu$m silicate absorption absorbs more at $11.3\ \mu$m than at $7.7\ \mu$m.
We did not fit the {\it SPICY} spectra with the absorption (Sect.~\ref{pahfit_section}), but we instead compared the observed rest-frame SEDs with the absorbed NGC~6090 SED template.
We adopted an extinction curve of \cite{chiar06} and a screen geometry, and followed the same procedure to reconstruct the rest-frame SED of the template by using synthetic photometry (Sect.~\ref{SED_fit_result}).
Figure~\ref{absorbed_sed} illustrates how the $11.3\ \mu$m and the $7.7\ \mu$m fluxes are absorbed for the same amount of extinction.
We found that the observations can be reproduced if the $9.7\ \mu$m optical depth is as large as $\simeq 3$.
Galaxies with such deep silicate absorption are known to show much smaller ($\lesssim 1/10$) PAH equivalent widths than usual star-forming galaxies \citep{spoon07}.
The PAH-enhanced population shows prominent PAH~$6.2\ \mu$m whose equivalent width is typically $1\ \mu$m (Sect.~\ref{spectral_characteristics_results}), indicating that their MIR spectra are similar to PAH-rich star-forming galaxies without heavy silicate absorption.
Although it seems unlikely based on the photometric analysis that most of the PAH-enhanced population suffers from such strong silicate absorption, we cannot test this possibility based on the spectroscopy due to limited wavelength coverage to enable the PAH fit with extinction for most of our sample.
Another possibility to modify $F_{\rm rest\ 11.3\ \mu \rm m}/F_{\rm rest\ 7.7\ \mu \rm m}$ is to modify the PAH~$11.3\ \mu$m/$7.7\ \mu$m inter-band flux ratio by charging the PAH particles.
\citet{draine01} showed that this inter-band ratio in the warm ionised medium, as in PDRs, are quite different (smaller by up to about five times) from those in the cold neutral medium, such as in the general ISM (see also \citealt{li01}).
Although this mechanism in principle can reproduce the observations, we cannot test this possibility due to lack of wavelength coverage for the redshifted PAH~$11.3\ \mu$m at $z\gtrsim 0.2$.

\subsection{Advantages and limits of our slitless spectroscopic survey at MIR and its joint analysis with multi-band photometry}\label{method_advantage_limit}

We have conducted the {\it SPICY} slitless spectroscopic survey at MIR, and studied extragalactic sources together with OPT--NIR--MIR multi-band photometry.
We summarise what we could successfully perform with our unique dataset, and what we could not due to the limitations.
First, we could identify AGN-dominated sources with little ambiguity, by employing either our simple NIR colour selection and multi-band SED analysis (Sect.~\ref{photo_classification_results}).
Second, we could measure redshifts by using the simple PAH fit from the slitless survey data.
We demonstrated that the PAH redshift is reasonably accurate and robust (1\% or less in $\mathrm{d}z/(1+z)$, without outliers, at $z<0.5$), enabling good spectroscopic PAH luminosity measurement for the blindly selected sources at MIR.
Third, we have shown that the photometric SED analysis by using the IRC filter set can provide good photometric PAH luminosity estimates without assuming models or intrinsic SED shapes.
We then calibrated the photometric PAH luminosities to the spectroscopic ones, and found tight correlations without redshift dependence between them (Sect.~\ref{pah_luminosity_comp_section}).
We thereby demonstrated that uncertainty associated with the $K$-correction is small for the photometric luminosity measurements.
Because photometric observations can be deeper and reach much fainter sources below the spectroscopy sensitivity limit, the photometric approach may be essential for studying much fainter and/or more distant sources in the future.
Spectroscopy following-up in the MIR of at least the brighter subsample should remain important to calibrate the photometric analysis results.
Fourth, we could measure the stellar mass with little ambiguity by using the rest-frame NIR photometry of the stellar SED.
By combining with the PAH luminosities, we could discuss star formation activities in terms of both SFR and sSFR up to $z \simeq 0.5$ without assuming the SED types with little ambiguity due to dust extinction.
We emphasise that we could do such study for unbiassed MIR-selected spectroscopic sample.
We could also characterise the NIR--MIR SEDs of the PAH galaxies (tight NIR colour distribution, almost universal MIR bump shape, and systematic change of the MIR/NIR SED bump ratio), and broadly classify types of the galaxies (e.g., Sc, starburst, LIRG) only with photometric information (Sect.~\ref{SED_fit_result}).

Despite the successes and advantages discussed above, our extragalactic survey has also suffered from many limitations.
First, our redshift coverage was only limited to $z\simeq 0.5$ in order to detect the PAH~$7.7\ \mu$m in our wavelength coverage below $13\ \mu$m.
To go beyond the current redshift limit to approach when SFRD showed its peak ($z \sim 2$), we need both longer wavelength coverage and more sensitivity to detect prominent PAH features.
Second, we could not detect the PAH~$11.3\ \mu$m for the PAH-enhanced population at $z>0.35$ to distinguish physical reasons for the photometric $F_{\rm rest\ 11.3\ \mu \rm m}/F_{\rm rest\ 7.7\ \mu \rm m}$ variation (Sect.~\ref{113_77_variation}).
Third, we could not fit the $9.7\ \mu$m silicate absorption for most cases.
To reliably fit this broad absorption, one needs to cover up to rest-frame $\sim 14\ \mu$m or longer so that hot dust continuum beyond the PAH complex around 11.3--$13\ \mu$m is included in the fit.
Most of these limitations are caused by the limited long wavelength coverage of the instrument ($\simeq 13\ \mu$m).
The higher spectral resolution than in our survey ($R\simeq 50$) would also help to fit the silicate absorption by better constraining the PAH profiles, especially (red tail of) the PAH~$8.6\ \mu$m and (blue tail of) the PAH~$11.3\ \mu$m, with more complicated and realistic spectral models of the PAHs.
Fourth, although we could reliably measure the PAH luminosities, an ambiguity remains in converting the PAH luminosity to total IR luminosity and, thus, SFR (e.g., \citealt{nordon12}).
The SED analysis over entire MIR and FIR wavelengths is needed.
Fifth, we had no information about distributions of spectral or even photometric properties of individual galaxies due to lack of spatial resolution.
We anticipate that such spatial information would provide us additional independent clues to identify physical processes that are responsible for the observed global characteristics.
By mapping the PAH fluxes as well as the stellar continuum, we can find distributions of, e.g., SFR, sSFR, and possible PAH inter-band flux ratios.
We can then correlate these properties with distributions of, e.g., usual disk star-forming regions and galaxy interaction/merger-induced star formation, as well as the nuclear activities (AGN or starburst).
Such study would enable finding links to physical mechanisms that govern the galaxy evolution.
Most of this limitation can be removed with larger-aperture and more-sensitive next-generation space infrared telescopes such as JWST \citep{JWST18} and SPICA \citep{nakagawa17}.

\section{Conclusions}\label{conclusion_section}

In this paper, we have reported results from our MIR spectroscopic survey toward the NEP with the IRC on the AKARI space telescope.
The survey, known as ``slitless SpectroscoPIC surveY of galaxies'' ({\it SPICY}), is a slitless spectroscopic survey of a $9\ \mu$m flux-selected sample.
This is the very first blind spectroscopic survey conducted at MIR wavelengths.
The biggest advantage of the slitless spectroscopy is that the survey can be unbiased, because sources are selected simply based on their flux at the same wavelengths as for the spectroscopy.
This enables us to study the MIR evolution of galaxies in more robust way.
We analysed the {\it SPICY} MIR spectra at 5--$13\ \mu$m, together with multi (13)-band photometric data at OPT--NIR--MIR wavelengths (0.37--$18\ \mu$m) from the combined AKARI/CFHT NEP photometric catalogue, to investigate star formation properties of galaxies at $z=0.0$--0.5.

We summarise the main results and conclusions below.

\begin{enumerate}

\item We collected 5--$13\ \mu$m low-resolution ($R\simeq 50$) spectra of $S9W\ (9.0\ \mu\mathrm{m})>0.3$~mJy sources within 14 IRC FOVs ($\simeq 14\times 10'\times 10'$ regions).
We performed simple PAH fit for the 6.2, 7.7, 8.6, and $11.3\ \mu$m PAH features on the detected extragalactic sources, excluding sources with damaged spectra (due to contamination and/or truncation that can randomly happen on the slitless spectroscopy images).
We then found 48 galaxies with typical PAH features for star-forming galaxies (the PAH galaxies), with redshifts ranging $z=0.0$--0.5.
We also identified 11 AGN candidates by using red NIR colours at 2--$3\ \mu$m (and also red MIR$/$NIR colours at 3--$7\ \mu$m as a secondary condition), so that the colours are consistent with those of the red continuum-dominated AGN SED template and all PAH galaxies are classified as non-AGNs.
We confirmed that the AGN candidates show red continuum-dominated SEDs at NIR--MIR wavelengths.
The remaining non-AGN and non-PAH galaxies are mostly E--Sa with intrinsically weak PAH features, or faint Sc--starburst galaxies below the spectroscopy sensitivity, according to their broad-band colours.
Basic information about the PAH galaxies, including their source positions, OPT--NIR--MIR photometry, as well as the PAH fit results (redshifts and luminosities of the PAH $6.2\ \mu$m and $7.7\ \mu$m), were reported.
Similar basic information about the bright AGN candidates were also reported.

\item We constructed rest-frame SEDs of the PAH galaxies at OPT--NIR--MIR wavelengths, with redshifts from the PAH fit.
We found that the rest-frame SEDs look much simpler and universal within the sample, although the observed colours show extreme diversity as functions of spectral types (Sc, starburst, or LIRG) and redshift.
This made our analysis easier and probably more fundamental, in contrast to ones based on the observed-frame information.
The SED is composed of an OPT--NIR (0.4--$4\ \mu$m) bump (the stellar component) and a MIR (5--$18\ \mu$m) one with a peak around $7.7\ \mu$m.
Both the NIR slope of the OPT--NIR bump and shape of the MIR bump are almost identical within the sample.
To characterise the SED shape, we defined and measured rest-frame flux ratios of $F_{\rm rest\ 11.3\ \mu \rm m}/F_{\rm rest\ 7.7\ \mu \rm m}$, $F_{\rm rest\ 7.7\ \mu \rm m}/F_{\rm rest\ 3.5\ \mu \rm m}$, and $F_{\rm rest\ 3.5\ \mu \rm m}/F_{\rm rest\ 2.0\ \mu \rm m}$.
The $F_{\rm rest\ 7.7\ \mu \rm m}/F_{\rm rest\ 3.5\ \mu \rm m}$ ratio, which represents the relative strength of the MIR bump over the OPT--NIR one, systematically changes with redshift in a range of 2--14.
This ratio increases with spectroscopic PAH luminosities, $L_\mathrm{PAH}$ ($7.7\ \mu$m) and $L_\mathrm{PAH}$ ($6.2\ \mu$m), indicating that the MIR bump is essentially composed of the PAH features around $7.7\ \mu$m, and specific SFR (SFR per stellar mass; sSFR) is higher for sources with higher SFR.
In contrast, the $F_{\rm rest\ 3.5\ \mu \rm m}/F_{\rm rest\ 2.0\ \mu \rm m}$ ratio shows tight distribution and is consistent with the stellar colour.
The $F_{\rm rest\ 11.3\ \mu \rm m}/F_{\rm rest\ 7.7\ \mu \rm m}$ ratio decreases by a factor of $\lesssim 2$ as $F_{\rm rest\ 7.7\ \mu \rm m}/F_{\rm rest\ 3.5\ \mu \rm m}$ increases by a factor of $\sim 2$ at $z\gtrsim 0.35$.

\item We photometrically measured the monochromatic luminosities at the peak of the PAH~$7.7\ \mu$m ($\nu L_{\rm \nu\ \mathrm{photo}}$ ($7.7\ \mu$m)), and compared them with those of the spectroscopy (spectroscopic PAH~$7.7\ \mu$m luminosity, $L_\mathrm{PAH}$ ($7.7\ \mu$m), and spectroscopic monochromatic luminosities at the peak of the PAH~$7.7\ \mu$m, $\nu L_{\rm \nu\ \mathrm{spec}}$ ($7.7\ \mu$m)).
We found tight correlations among them, with little systematic offset as a function of redshift, and reported the scaling relations.

\item We identified PAH-enhanced population showing $\log \nu L_{\nu\ \mathrm{photo}}\ (7.7\ \mu\mathrm{m})>0.50+\log \nu L_{\nu\ \mathrm{photo}}\ (3.5\ \mu\mathrm{m})$ (or, equivalently, $F_{\rm rest\ 7.7\ \mu \rm m}/F_{\rm rest\ 3.5\ \mu \rm m}>7.0$) at $z\gtrsim 0.35$.
Here, $\nu L_{\nu\ \mathrm{photo}}\ (3.5\ \mu\mathrm{m})$ is a monochromatic photometric luminosities at $3.5\ \mu$m.
They show elevated SFR ($\gtrsim 30\ M_{\rm \sun}\ \mathrm{yr}^{-1}$), being comparable to LIRGs, and also elevated sSFR (by about a factor of two when compared to the upper envelope of the sSFR distribution at $z<0.35$).
We found no such PAH-enhanced population at lower redshifts, and we argued that this is neither due to source selection biases nor by chance due to small number statistic.

\item We found that the SWIRE SED templates of normal/star-forming galaxies, Sc, starburst (M82), and LIRG (NGC~6090), can mostly reproduce both the rest-frame SEDs and the flux ratios of the PAH galaxies.
We also found that the SBURT starburst models \citep{takagi03} for older ($t_{\rm burst}\gtrsim 200$~Myr) starburst with modest optical depth can mostly reproduce the observations.
For the PAH-luminous and PAH-enhanced population at $z\gtrsim 0.35$ in particular, both the NGC~6090 template and the SBURT model for middle-aged ($t_{\rm burst}\simeq 200$~Myr) starburst with modest optical depth can reproduce the observed NIR--MIR SEDs.
This SBURT model, however, has a problem in reproducing relatively red OPT SEDs at the same time as the enhanced MIR bump.
We developed a composite SED that combines a very young ($\simeq 70$~Myr), compact ($\Theta=1.0$) and, hence, optically very thick SBURT model and a very old (13~Gyr old) SWIRE template, and found that it can reproduce the overall OPT--NIR--MIR SEDs of the PAH-enhanced population.
This suggests that the PAH-enhanced population shows very young and vigorous star-formation activities on the old stellar system,
although we could not confirm the presence of the optically thick component from our spectral analysis.
We note that the positive correlation between SFR and sSFR for the PAH galaxies in general can also be explained by the composite models, by changing relative contribution of the very young component over the old one.
Both this composite SED and the NGC~6090 SED template still have difficulties in reproducing the relatively depressed $F_{\rm rest\ 11.3\ \mu \rm m}/F_{\rm rest\ 7.7\ \mu \rm m}$ for the PAH-enhanced population.

\end{enumerate}

\begin{acknowledgements}
YO acknowledges supports from the Japan Society for the Promotion of Science (JSPS) grant number 17740122, and National Science Counsel (NSC) and Ministry of Science and Technology (MOST) of Taiwan 100-2112-M-001-001-MY3, 104-2112-M-001-034-, 105-2112-M-001-024-, and 106-2112-M-001-008-.
MI acknowledges the support from the National Research Foundation of Korea (NRFK) grant, No. 2017R1A3A3001362, funded by the Korean government (MSIP).
GJW gratefully acknowledges support from The Leverhulme Trust.
TM is supported by UNAM-DGAPA PAPIIT IN104216 and CONACyT 252531.
TN acknowledges support from the JSPS KAKENHI Grant Number 26247030.
This research is based on observations with AKARI, a JAXA project with the participation of ESA, and is partly supported with the Grant-in-Aid for Scientific Research (21340042) from the JSPS.
This work is based in part on observations made with the {\it Spitzer} Space Telescope, obtained from the NASA/IPAC Infrared Science Archive, both of which are operated by the Jet Propulsion Laboratory, California Institute of Technology under a contract with the National Aeronautics and Space Administration.
This research used observations made with the Gran Telescopio Canarias (GTC), installed in the Spanish Observatorio del Roque de los Muchachos of the Instituto de Astrof\'isica de Canarias, in the island of La Palma.

\end{acknowledgements}

\onecolumn

\begin{table}
\caption{\label{table1}Numbers of {\it SPICY} sources in different classifications discussed in this paper.
In each classification, numbers of the bright samples (Sect.~\ref{photo_classification_results}) are shown in parentheses.
For the PAH galaxies, their numbers for each redshift bin (Sect.~\ref{spectral_characteristics_results}) are also shown.
Numbers of the PAH-enhanced galaxies (Sect.~\ref{sed_variation_result}) are shown in parentheses.
}
\centering
\begin{tabular}{cccc}
\hline
\hline
classification & $N$ & redshift bin & $N$ (PAH galaxies) \\
                     & total (bright) & & subtotal (PAH enhanced) \\
\hline
AGN candidates & 11 (6) & & \\
non-PAH galaxies & 112 (39) & & \\
 & & nearby & 13 (0) \\ \cline{3-4}
PAH galaxies & 48 (24) & mid-$z$ & 23 (1) \\ \cline{3-4}
 & & higher-$z$ & 12 (9) \\
\hline
total & 171 (93) & & 48 (10) \\
\hline
\end{tabular}
\end{table}

\begin{longtable}{lcccccc}
\caption{\label{table2}Source identification, cross-identification with the AKARI/IRC NEP-Wide catalogue, and the {\it SPICY} S9W ($9.0\ \mu$m) photometry of the PAH galaxies, together with cross-identification with the CFHT optical NEP photometric catalogue.
FIELD indicates one of the two survey fields defined in \cite{hwang07} (1 and 2 for the NEP-E and NEP-W fields, respectively).
}\\
\hline
\hline
{\it SPICY} ID & \multicolumn{3}{c}{AKARI/IRC NEP-Wide catalogue} & {\it SPICY} S9W flux & \multicolumn{2}{c}{CFHT NEP catalogue} \\
 & ID & RA J2000 (hms) & DEC J2000 (dms) & (mJy) & ID & FIELD \\
\hline
\endfirsthead
\caption{continued.}\\
\hline
\hline
{\it SPICY} ID & \multicolumn{3}{c}{AKARI/IRC NEP-Wide catalogue} & {\it SPICY} S9W flux & \multicolumn{2}{c}{CFHT NEP catalogue} \\
 & ID & RA J2000 (hms) & DEC J2000 (dms) & (mJy) & ID & FIELD \\
\hline
\endhead
\hline
\endfoot

F00-1 & 65030479 & 17 55 19.91 &  +66 41 45.2 & $      1.25 \pm      0.04 $ & 48924 & 2 \\
F01-7 & 65047593 & 17 58 40.87 &  +66 22 14.1 & $      1.63 \pm      0.05 $ & 23532 & 2 \\
F01-13 & 65042346 & 17 57 39.74 &  +66 20 53.2 & $      0.64 \pm      0.02 $ & 20669 & 2 \\
F01-15 & 65046006 & 17 58 21.61 &  +66 24 45.8 & $      0.59 \pm      0.02 $ & 25461 & 2 \\
F01-19 & 65042020 & 17 57 35.87 &  +66 22 04.4 & $      0.40 \pm      0.02 $ & 21980 & 2 \\
F02-0 & 65056911 & 18 00 28.93 &  +66 12 29.6 & $      1.37 \pm      0.04 $ & 9818 & 1 \\
F02-4 & 65059749 & 18 00 59.84 &  +66 10 56.5 & $      0.91 \pm      0.03 $ & 7723 & 1 \\
F02-6 & 65063554 & 18 01 42.08 &  +66 13 00.4 & $      0.77 \pm      0.03 $ & 10219 & 1 \\
F02-8 & 65060828 & 18 01 11.75 &  +66 11 39.6 & $      0.67 \pm      0.02 $ & 8535 & 1 \\
F02-9 & 65057478 & 18 00 35.22 &  +66 13 44.9 & $      0.65 \pm      0.02 $ & 10871 & 1 \\
F02-15 & 65060902 & 18 01 12.43 &  +66 16 54.4 & $      0.46 \pm      0.02 $ & 14047 & 1 \\
F02-20 & 65058156 & 18 00 42.29 &  +66 08 20.4 & $      0.39 \pm      0.02 $ & 4847 & 1 \\
F02-24 & 65057873 & 18 00 39.35 &  +66 13 52.3 & $      0.33 \pm      0.01 $ & 11076 & 1 \\
F03-4 & 65019934 & 17 53 06.59 &  +66 34 28.4 & $      0.93 \pm      0.03 $ & 38180 & 2 \\
F03-15 & 65020331 & 17 53 11.78 &  +66 36 08.5 & $      0.33 \pm      0.01 $ & 40557 & 2 \\
F04-0 & 65037228 & 17 56 39.96 &  +66 48 00.6 & $      5.48 \pm      0.17 $ & 56817 & 2 \\
F04-3 & 65038399 & 17 56 53.90 &  +66 39 22.3 & $      1.83 \pm      0.06 $ & 45137 & 2 \\
F04-5 & 65039972 & 17 57 11.55 &  +66 45 23.0 & $      1.15 \pm      0.04 $ & 54615 & 2 \\
F04-6 & 65041345 & 17 57 27.79 &  +66 45 56.6 & $      0.76 \pm      0.03 $ & 53989 & 2 \\
F04-7 & 65038112 & 17 56 50.77 &  +66 44 44.7 & $      0.74 \pm      0.03 $ & 53152 & 2 \\
F04-8 & 65041041 & 17 57 24.35 &  +66 42 36.5 & $      0.61 \pm      0.02 $ & 49849 & 2 \\
F04-13 & 65040771 & 17 57 21.13 &  +66 42 34.6 & $      0.50 \pm      0.02 $ & 50195 & 2 \\
F04-18 & 65040361 & 17 57 15.99 &  +66 43 40.0 & $      0.44 \pm      0.02 $ & 51197 & 2 \\
F04-23 & 65035546 & 17 56 19.78 &  +66 47 36.1 & $      0.41 \pm      0.02 $ & 55840 & 2 \\
F04-24 & 65040327 & 17 57 15.56 &  +66 48 21.6 & $      0.38 \pm      0.01 $ & --- & --- \footnote{This source is damaged due to a nearby bright and saturated star on the CFHT image, and is not catalogued.} \\
F05-10 & 65054492 & 18 00 01.60 &  +66 54 02.4 & $      0.75 \pm      0.03 $ & 59396 & 1 \\
F05-23 & 65054321 & 17 59 59.66 &  +66 50 09.2 & $      0.33 \pm      0.01 $ & 52788 & 1 \\
F07-11 & 65026631 & 17 54 33.33 &  +66 47 47.7 & $      0.55 \pm      0.02 $ & 56082 & 2 \\
F07-12 & 65020847 & 17 53 18.84 &  +66 44 39.7 & $      0.53 \pm      0.02 $ & 52573 & 2 \\
F08-1 & 65027087 & 17 54 38.96 &  +66 23 16.2 & $      4.01 \pm      0.12 $ & --- & ---\footnote{This source falls on one of CCD gaps on the CFHT image, and is not catalogued.} \\
F08-5 & 65030601 & 17 55 21.36 &  +66 24 26.4 & $      2.88 \pm      0.09 $ & 24990 & 2 \\
F08-11 & 65031482 & 17 55 31.62 &  +66 23 42.1 & $      0.80 \pm      0.03 $ & 23958 & 2 \\
F08-14 & 65030371 & 17 55 18.77 &  +66 25 30.9 & $      0.74 \pm      0.03 $ & 26271 & 2 \\
F08-17 & 65028594 & 17 54 57.58 &  +66 25 52.6 & $      0.54 \pm      0.02 $ & 26549 & 2 \\
F08-23 & 65027474 & 17 54 43.76 &  +66 21 04.0 & $      0.49 \pm      0.02 $ & 20996 & 2 \\
F09-2 & 65023172 & 17 53 48.51 &  +66 39 20.9 & $      4.21 \pm      0.13 $ & 45556 & 2 \\
F10-8 & 65038265 & 17 56 52.46 &  +66 27 36.0 & $      0.53 \pm      0.02 $ & 28701 & 2 \\
F10-10 & 65041559 & 17 57 30.25 &  +66 30 06.1 & $      0.48 \pm      0.02 $ & 31653 & 2 \\
F10-13 & 65037342 & 17 56 41.54 &  +66 27 04.1 & $      0.46 \pm      0.02 $ & 27999 & 2 \\
F10-19 & 65041158 & 17 57 25.61 &  +66 29 46.5 & $      0.35 \pm      0.01 $ & 31398 & 2 \\
F11-7 & 65038292 & 17 56 52.78 &  +66 06 40.8 & $      1.56 \pm      0.05 $ & 3548 & 2 \\
F11-11 & 65040931 & 17 57 22.90 &  +66 05 09.5 & $      0.94 \pm      0.03 $ & 1673 & 2 \\
F11-19 & 65040367 & 17 57 16.08 &  +66 04 54.4 & $      0.64 \pm      0.02 $ & 2082 & 2 \\
F11-23 & 65043924 & 17 57 58.08 &  +66 07 09.0 & $      0.53 \pm      0.02 $ & 4163 & 2 \\
F20-3 & 65045989 & 17 58 21.42 &  +66 28 55.3 & $      4.12 \pm      0.13 $ & 30856 & 2 \\
F20-10 & 65051479 & 17 59 26.77 &  +66 29 39.3 & $      0.51 \pm      0.02 $ & 31374 & 2 \\
F21-1 & 65032713 & 17 55 46.52 &  +66 38 39.9 & $      2.10 \pm      0.07 $ & 44027 & 2 \\
F21-20 & 65031307 & 17 55 29.48 &  +66 40 35.5 & $      0.32 \pm      0.01 $ & 46860 & 2 \\
\end{longtable}

\begin{longtable}{lccccccc}
\caption{\label{table3}Results of the PAH fit of the {\it SPICY} PAH galaxies.
The optical spectroscopic redshifts and their measurement quality flags (4=secure; identified by more than two features, 3=acceptable and almost good; identified by two features, according to \citealt{shim13}) are also shown if available.
When there are two independent measurements, we show both results.
Sources of the redshift are 1 for \cite{shim13}, 2 for \cite{oi14}, and 3 for our own GTC observations (see text).}\\
\hline
\hline
 & \multicolumn{4}{c}{{\it SPICY} PAH fit} & \multicolumn{3}{c}{optical spectroscopy} \\
ID & redshift & $\log L$ (PAH~$6.2\ \mu$m) & $\log L$ (PAH~$7.7\ \mu$m) & $\log \nu L \nu _{\rm spec}$ ($7.7\ \mu$m) & redshift & quality & source \\
 & ($z$) & ($L_{\sun}$) & ($L_{\sun}$) & ($L_{\sun}$) & ($z$) & & \\
\hline
\endfirsthead
\caption{continued.}\\
\hline
\hline
ID & redshift & $\log L$ (PAH~$6.2\ \mu$m) & $\log L$ (PAH~$7.7\ \mu$m) & $\log \nu L \nu _{\rm spec}$ ($7.7\ \mu$m) & redshift & quality & source \\
 & ($z$) & ($L_{\sun}$) & ($L_{\sun}$) & ($L_{\sun}$) & ($z$) & & \\
\hline
\endhead
\hline
\endfoot
F00-1 & $     0.225 \pm     0.001 $ & $     9.042_{-0.052}^{+0.047}$ & $     9.853_{-0.022}^{+0.021} $ & $    10.670_{-0.012}^{+0.012} $ & 0.2270,0.227 & 4,4 & 1,2 \\
F01-7 & $     0.069 \pm     0.001 $ & $     7.516_{-0.386}^{+0.201}$ & $     9.074_{-0.023}^{+0.022} $ & $     9.586^{+0.013}_{-0.014} $ & 0.0880 & 4 & 1  \\
F01-13 & $     0.117 \pm     0.002 $ & $     8.226_{-0.118}^{+0.093}$ & $     8.802_{-0.053}^{+0.048} $ & $     9.673^{+0.030}_{-0.032} $ & 0.1182 & 4 & 1  \\
F01-15 & $     0.087 \pm     0.002 $ & $     8.157_{-0.088}^{+0.073}$ & $     8.610_{-0.067}^{+0.058} $ & $     9.263^{+0.035}_{-0.038} $ & 0.0870 & 4 & 1  \\
F01-19 & $     0.492 \pm     0.002 $ & $     9.356_{-0.311}^{+0.179}$ & $     9.889_{-0.145}^{+0.109} $ & $    10.901^{+0.061}_{-0.071} $ & 0.4876 & 4 & 1  \\
F02-0 & $     0.087 \pm     0.001 $ & $     8.253_{-0.059}^{+0.052}$ & $     8.886_{-0.028}^{+0.027} $ & $     9.754^{+0.016}_{-0.017} $ & --- & --- \\
F02-4 & $     0.295 \pm     0.002 $ & $     9.272_{-0.069}^{+0.060}$ & $     9.914_{-0.037}^{+0.034} $ & $    10.747^{+0.017}_{-0.018} $ & --- & --- \\
F02-6 & $     0.145 \pm     0.002 $ & $     8.424_{-0.112}^{+0.089}$ & $     9.115_{-0.043}^{+0.039} $ & $    9.989^{+0.021}_{-0.022} $ & 0.1367 & 4 & 1  \\
F02-8 & $     0.256 \pm     0.003 $ & $     8.929_{-0.113}^{+0.090}$ & $     9.629_{-0.056}^{+0.050} $ & $    10.359^{+0.026}_{-0.027} $ & --- & --- \\
F02-9 & $     0.294 \pm     0.001 $ & $     9.217_{-0.076}^{+0.065}$ & $     9.752_{-0.049}^{+0.044} $ & $    10.592^{+0.024}_{-0.025} $ & --- & --- \\
F02-15 & $     0.232 \pm     0.001 $ & $     <8.528$ & $     9.516_{-0.072}^{+0.062} $ & $    10.176^{+0.031}_{-0.034} $ & --- & --- \\
F02-20 & $     0.477 \pm     0.003 $ & $     9.206_{-0.269}^{+0.165}$ & $     9.790_{-0.141}^{+0.106} $ & $    10.844^{+0.058}_{-0.068} $ & --- & --- \\
F02-24 & $     0.186 \pm     0.002 $ & $     8.410_{-0.146}^{+0.109}$ & $     9.116_{-0.102}^{+0.083} $ & $     9.821^{+0.042}_{-0.046} $ & --- & --- \\
F03-4 & $     0.183 \pm     0.001 $ & $     8.695_{-0.077}^{+0.065}$ & $     9.496_{-0.035}^{+0.032} $ & $    10.255^{+0.018}_{-0.019} $ & --- & --- \\
F03-15 & $     0.227 \pm     0.002 $ & $     8.370_{-0.188}^{+0.131}$ & $     8.787_{-0.195}^{+0.134} $ & $     9.848^{+0.072}_{-0.087} $ & 0.2225 & 4 & 1  \\
F04-0 & $     0.087 \pm     0.001 $ & $     8.900_{-0.022}^{+0.021}$ & $     9.506_{-0.015}^{+0.014} $ & $    10.336^{+0.010}_{-0.010} $ & 0.0880 & 4 & 1  \\
F04-3 & $     0.169 \pm     0.001 $ & $     9.042_{-0.039}^{+0.036}$ & $     9.589_{-0.026}^{+0.025} $ & $    10.411^{+0.014}_{-0.015} $ & 0.1672 & 4 & 1  \\
F04-5 & $     0.056 \pm     0.001 $ & $     7.855_{-0.062}^{+0.054}$ & $     8.463_{-0.026}^{+0.025} $ & $     9.267^{+0.014}_{-0.014} $ & 0.050 & 4 & 3 \\
F04-6 & $     0.424 \pm     0.002 $ & $     9.710_{-0.094}^{+0.077}$ & $    10.287_{-0.040}^{+0.036} $ & $    11.124^{+0.022}_{-0.023} $ & 0.4173 & 4 & 1  \\
F04-7 & $     0.086 \pm     0.001 $ & $     8.174_{-0.074}^{+0.063}$ & $     8.667_{-0.042}^{+0.038} $ & $     9.415^{+0.023}_{-0.024} $ & 0.0864 & 4 & 1  \\
F04-8 & $     0.278 \pm     0.001 $ & $     8.977_{-0.099}^{+0.080}$ & $     9.582_{-0.064}^{+0.056} $ & $    10.457^{+0.028}_{-0.030} $ & 0.2752 & 4 & 1  \\
F04-13 & $     0.193 \pm     0.002 $ & $     8.664_{-0.118}^{+0.093}$ & $     9.375_{-0.058}^{+0.052} $ & $    10.099^{+0.025}_{-0.027} $ & 0.1871 & 3 & 1  \\
F04-18 & $     0.306 \pm     0.002 $ & $     8.884_{-0.181}^{+0.127}$ & $     9.754_{-0.083}^{+0.069} $ & $    10.475^{+0.033}_{-0.036} $ & 0.3010 & 4 & 1  \\
F04-23 & $     0.452 \pm     0.005 $ & $     9.393_{-0.251}^{+0.158}$ & $    10.211_{-0.091}^{+0.075} $ & $    10.936^{+0.041}_{-0.045} $ & 0.4500 & 4 & 1  \\
F04-24 & $     0.266 \pm     0.005 $ & $     8.864_{-0.228}^{+0.149}$ & $     9.300_{-0.151}^{+0.112} $ & $    10.095^{+0.057}_{-0.066} $ & --- & --- \\
F05-10 & $     0.076 \pm     0.002 $ & $     8.001_{-0.089}^{+0.074}$ & $     8.555_{-0.042}^{+0.038} $ & $     9.293^{+0.022}_{-0.023} $ & --- & --- \\
F05-23 & $     0.393 \pm     0.004 $ & $     9.355_{-0.191}^{+0.132}$ & $     9.548_{-0.157}^{+0.115} $ & $    10.530^{+0.060}_{-0.070} $ & --- & --- \\
F07-11 & $     0.394 \pm     0.004 $ & $     8.858_{-1.110}^{+0.284}$ & $     9.907_{-0.065}^{+0.057} $ & $    10.783^{+0.033}_{-0.036} $ & 0.391 & 4 & 2 \\
F07-12 & $     0.281 \pm     0.003 $ & $     8.809_{-0.216}^{+0.144}$ & $     9.487_{-0.098}^{+0.080} $ & $    10.377^{+0.039}_{-0.043} $ & 0.2862 & 4 & 1 \\
F08-1 & $     0.088 \pm     0.001 $ & $     8.884_{-0.023}^{+0.022}$ & $     9.413_{-0.016}^{+0.016} $ & $    10.253^{+0.010}_{-0.010} $ & 0.090 & 4 & 2 \\
F08-5 & $     0.143 \pm     0.001 $ & $     9.126_{-0.026}^{+0.024}$ & $     9.752_{-0.014}^{+0.014} $ & $    10.507^{+0.009}_{-0.009} $ & 0.143 & 4 & 2 \\
F08-11 & $     0.248 \pm     0.002 $ & $     8.882_{-0.117}^{+0.092}$ & $     9.481_{-0.093}^{+0.077} $ & $    10.199^{+0.037}_{-0.040} $ & 0.262 & 4 & 2 \\
F08-14 & $     0.420 \pm     0.003 $ & $     9.423_{-0.175}^{+0.125}$ & $    10.112_{-0.076}^{+0.065} $ & $    10.923^{+0.036}_{-0.039} $ & 0.429 & 4 & 2 \\
F08-17 & $     0.200 \pm     0.002 $ & $     8.571_{-0.136}^{+0.103}$ & $     9.252_{-0.062}^{+0.055} $ & $    10.077^{+0.029}_{-0.032} $ & 0.1991,0.200 & 4,4 & 1,2 \\
F08-23 & $     0.088 \pm     0.002 $ & $     7.572_{-0.624}^{+0.246}$ & $     8.071_{-0.141}^{+0.106} $ & $     9.159^{+0.065}_{-0.076} $ & 0.0880 & 3 & 1 \\
F09-2 & $     0.108 \pm     0.001 $ & $     9.048_{-0.022}^{+0.021}$ & $     9.703_{-0.013}^{+0.013} $ & $    10.440^{+0.008}_{-0.008} $ & --- & --- \\
F10-8 & $     0.465 \pm     0.003 $ & $     9.483_{-0.205}^{+0.139}$ & $     9.992_{-0.092}^{+0.076} $ & $    11.062^{+0.041}_{-0.045} $ & 0.472 & 4 & 3\\
F10-10 & $     0.293 \pm     0.004 $ & $     8.978_{-0.189}^{+0.131}$ & $     9.536_{-0.081}^{+0.068} $ & $    10.369^{+0.037}_{-0.040} $ & 0.3025,0.303 & 4,4 & 1,2 \\
F10-13 & $     0.473 \pm     0.004 $ & $     9.482_{-0.179}^{+0.126}$ & $     10.043_{-0.112}^{+0.089} $ & $    10.932^{+0.051}_{-0.058} $ & 0.472 & 4 & 3 \\
F10-19 & $     0.409 \pm     0.004 $ & $     9.630_{-0.140}^{+0.106}$ & $     9.753_{-0.097}^{+0.080} $ & $    10.808^{+0.045}_{-0.050} $ & 0.423 & 4 & 2 \\
F11-7 & $     0.175 \pm     0.001 $ & $     9.140_{-0.037}^{+0.034}$ & $     9.645_{-0.024}^{+0.023} $ & $    10.444^{+0.012}_{-0.013} $ & --- & --- \\
F11-11 & $     0.180 \pm     0.001 $ & $     8.715_{-0.086}^{+0.072}$ & $     9.125_{-0.055}^{+0.049} $ & $    10.202^{+0.026}_{-0.028} $ & --- & --- \\
F11-19 & $     0.084 \pm     0.002 $ & $     7.963_{-0.126}^{+0.097}$ & $     8.498_{-0.069}^{+0.060} $ & $     9.425^{+0.031}_{-0.034} $ & 0.0864 & 4 & 1 \\
F11-23 & $     0.081 \pm     0.002 $ & $     7.734_{-0.195}^{+0.134}$ & $     8.357_{-0.066}^{+0.057} $ & $     9.315^{+0.033}_{-0.036} $ & --- & --- \\
F20-3 & $     0.090 \pm     0.001 $ & $     8.851_{-0.023}^{+0.022}$ & $     9.443_{-0.015}^{+0.015} $ & $    10.273^{+0.010}_{-0.010} $ & 0.0875 & 4 & 1 \\
F20-10 & $     0.448 \pm     0.003 $ & $     9.391_{-0.185}^{+0.129}$ & $    10.243_{-0.052}^{+0.047} $ & $    11.141^{+0.027}_{-0.029} $ & 0.4588 & 4 & 1 \\
F21-1 & $     0.044 \pm     0.001 $ & $     7.866_{-0.039}^{+0.036}$ & $     8.555_{-0.016}^{+0.016} $ & $     9.339^{+0.010}_{-0.010} $ & --- & --- \\
F21-20 & $     0.405 \pm     0.003 $ & $     9.262_{-0.152}^{+0.113}$ & $     9.990_{-0.150}^{+0.111} $ & $    10.630^{+0.068}_{-0.080} $ & 0.3930,0.393 & 3,4 & 1,2 \\
\end{longtable}

\begin{landscape}
\begin{longtable}{lcccccccc}
\caption{\label{table4}NIR--MIR photometry in the AKARI/IRC NEP-Wide catalogue of the {\it SPICY} PAH galaxies.
}\\
\hline
\hline
ID & $N2$ & $N3$ & $N4$ & $S7$ & $S9W$ & $S11$ & $L15$ & $L18W$ \\
 & (mJy) & (mJy) & (mJy) & (mJy) & (mJy) & (mJy) & (mJy) & (mJy) \\
\hline
\endfirsthead
\caption{continued.}\\
\hline
\hline
ID & $N2$ & $N3$ & $N4$ & $S7$ & $S9W$ & $S11$ & $L15$ & $L18W$ \\
 & (mJy) & (mJy) & (mJy) & (mJy) & (mJy) & (mJy) & (mJy) & (mJy) \\
\hline
\endhead
\hline
\endfoot
F00-1 & $     0.375 \pm     0.005 $ & $     0.257 \pm     0.004 $ & $     0.225 \pm     0.004 $ & $     0.560 \pm     0.016 $ & $     1.200 \pm     0.019 $ & $     1.630 \pm     0.028 $ & $     1.360 \pm     0.037 $ & $     1.260 \pm     0.031 $ \\
F01-7 & $     3.010 \pm     0.011 $ & $     1.890 \pm     0.009 $ & $     1.310 \pm     0.007 $ & $     1.850 \pm     0.025 $ & $     2.300 \pm     0.023 $ & $     2.480 \pm     0.034 $ & --- & --- \\
F01-13 & $     0.397 \pm     0.005 $ & $     0.270 \pm     0.004 $ & $     0.198 \pm     0.003 $ & $     0.408 \pm     0.015 $ & $     0.677 \pm     0.015 $ & $     0.766 \pm     0.022 $ & $     0.634 \pm     0.029 $ & $     0.674 \pm     0.030 $ \\
F01-15 & $     0.284 \pm     0.005 $ & $     0.198 \pm     0.003 $ & $     0.146 \pm     0.003 $ & $     0.486 \pm     0.015 $ & $     0.609 \pm     0.015 $ & $     0.684 \pm     0.022 $ & --- & --- \\
F01-19 & $     0.179 \pm     0.004 $ & $     0.159 \pm     0.003 $ & $     0.115 \pm     0.003 $ & --- & $     0.412 \pm     0.014 $ & $     0.756 \pm     0.023 $ & $     0.677 \pm     0.035 $ & $     0.557 \pm     0.028 $ \\
F02-0 & $     0.834 \pm     0.008 $ & $     0.568 \pm     0.006 $ & $     0.401 \pm     0.005 $ & $     1.220 \pm     0.021 $ & $     1.600 \pm     0.021 $ & $     1.710 \pm     0.030 $ & $     1.500 \pm     0.042 $ & $     1.290 \pm     0.035 $ \\
F02-4 & $     0.445 \pm     0.005 $ & $     0.331 \pm     0.004 $ & $     0.234 \pm     0.003 $ & $     0.355 \pm     0.014 $ & $     0.957 \pm     0.017 $ & $     1.260 \pm     0.026 $ & $     1.070 \pm     0.036 $ & $     1.030 \pm     0.030 $ \\
F02-6 & $     0.638 \pm     0.006 $ & $     0.417 \pm     0.005 $ & $     0.306 \pm     0.004 $ & $     0.502 \pm     0.016 $ & $     0.927 \pm     0.018 $ & $     1.090 \pm     0.027 $ & $     1.080 \pm     0.037 $ & $     0.674 \pm     0.027 $ \\
F02-8 & $     0.545 \pm     0.006 $ & $     0.413 \pm     0.005 $ & $     0.281 \pm     0.004 $ & $     0.292 \pm     0.013 $ & $     0.603 \pm     0.015 $ & $     0.935 \pm     0.025 $ & $     0.817 \pm     0.035 $ & $     0.650 \pm     0.030 $ \\
F02-9 & $     0.272 \pm     0.004 $ & $     0.204 \pm     0.004 $ & $     0.160 \pm     0.003 $ & $     0.258 \pm     0.014 $ & $     0.637 \pm     0.016 $ & $     0.816 \pm     0.022 $ & $     0.671 \pm     0.030 $ & $     0.664 \pm     0.029 $ \\
F02-15 & $     0.272 \pm     0.004 $ & $     0.179 \pm     0.003 $ & $     0.138 \pm     0.003 $ & $     0.226 \pm     0.014 $ & $     0.441 \pm     0.014 $ & $     0.691 \pm     0.026 $ & $     0.515 \pm     0.029 $ & $     0.356 \pm     0.028 $ \\
F02-20 & $     0.222 \pm     0.004 $ & $     0.208 \pm     0.003 $ & $     0.149 \pm     0.003 $ & $     0.242 \pm     0.013 $ & $     0.462 \pm     0.015 $ & $     0.658 \pm     0.023 $ & $     0.605 \pm     0.028 $ & $     0.719 \pm     0.026 $ \\
F02-24 & $     0.276 \pm     0.005 $ & $     0.186 \pm     0.004 $ & $     0.121 \pm     0.003 $ & $     0.155 \pm     0.010 $ & $     0.369 \pm     0.014 $ & $     0.428 \pm     0.020 $ & $     0.324 \pm     0.026 $ & $     0.262 \pm     0.025 $ \\
F03-4 & $     0.390 \pm     0.005 $ & $     0.320 \pm     0.004 $ & $     0.222 \pm     0.003 $ & $     0.397 \pm     0.016 $ & $     0.922 \pm     0.017 $ & $     1.120 \pm     0.025 $ & $     0.961 \pm     0.040 $ & $     0.986 \pm     0.036 $ \\
F03-15 & $     0.114 \pm     0.004 $ & $     0.072 \pm     0.003 $ & $     0.072 \pm     0.003 $ & $     0.145 \pm     0.010 $ & $     0.302 \pm     0.015 $ & $     0.427 \pm     0.020 $ & $     0.289 \pm     0.029 $ & $     0.251 \pm     0.024 $ \\
F04-0 & $     2.340 \pm     0.011 $ & $     1.530 \pm     0.007 $ & $     1.160 \pm     0.006 $ & $     4.450 \pm     0.033 $ & $     6.040 \pm     0.028 $ & $     6.690 \pm     0.043 $ & $     6.280 \pm     0.063 $ & $     5.610 \pm     0.052 $ \\
F04-3 & $     0.737 \pm     0.006 $ & $     0.487 \pm     0.004 $ & $     0.387 \pm     0.004 $ & $     0.836 \pm     0.018 $ & $     1.770 \pm     0.019 $ & $     2.210 \pm     0.030 $ & $     2.070 \pm     0.043 $ & $     1.850 \pm     0.037 $ \\
F04-5 & $     0.623 \pm     0.006 $ & $     0.437 \pm     0.005 $ & $     0.298 \pm     0.004 $ & $     1.220 \pm     0.020 $ & $     1.210 \pm     0.019 $ & $     1.160 \pm     0.025 $ & --- & --- \\
F04-6 & $     0.128 \pm     0.003 $ & $     0.138 \pm     0.003 $ & $     0.100 \pm     0.003 $ & $     0.140 \pm     0.009 $ & $     0.882 \pm     0.017 $ & $     1.280 \pm     0.026 $ & $     0.740 \pm     0.039 $ & $     0.908 \pm     0.031 $ \\
F04-7 & $     0.487 \pm     0.006 $ & $     0.306 \pm     0.004 $ & $     0.206 \pm     0.004 $ & $     0.718 \pm     0.018 $ & $     0.846 \pm     0.021 $ & $     0.940 \pm     0.032 $ & $     0.844 \pm     0.040 $ & $     0.862 \pm     0.043 $ \\
F04-8 & $     0.141 \pm     0.003 $ & $     0.108 \pm     0.003 $ & $     0.135 \pm     0.004 $ & $     0.238 \pm     0.012 $ & $     0.629 \pm     0.015 $ & $     0.850 \pm     0.024 $ & $     0.817 \pm     0.030 $ & $     0.906 \pm     0.031 $ \\
F04-13 & $     0.565 \pm     0.006 $ & $     0.348 \pm     0.004 $ & $     0.277 \pm     0.004 $ & $     0.253 \pm     0.013 $ & $     0.801 \pm     0.023 $ & $     0.505 \pm     0.020 $ & $     0.527 \pm     0.027 $ & $     0.530 \pm     0.034 $ \\
F04-18 & $     0.134 \pm     0.004 $ & $     0.106 \pm     0.003 $ & $     0.093 \pm     0.003 $ & $     0.164 \pm     0.013 $ & $     0.407 \pm     0.014 $ & $     0.568 \pm     0.021 $ & $     0.867 \pm     0.038 $ & $     0.909 \pm     0.034 $ \\
F04-23 & $     0.117 \pm     0.004 $ & $     0.111 \pm     0.003 $ & $     0.081 \pm     0.003 $ & --- & $     0.419 \pm     0.014 $ & $     0.773 \pm     0.027 $ & $     0.426 \pm     0.031 $ & $     0.425 \pm     0.030 $ \\
F04-24 & $     0.246 \pm     0.005 $ & $     0.206 \pm     0.004 $ & $     0.165 \pm     0.003 $ & $     0.240 \pm     0.016 $ & $     0.424 \pm     0.015 $ & $     0.547 \pm     0.021 $ & $     0.773 \pm     0.037 $ & $     0.663 \pm     0.027 $ \\
F05-10 & $     0.557 \pm     0.006 $ & $     0.373 \pm     0.004 $ & $     0.242 \pm     0.004 $ & $     0.681 \pm     0.017 $ & $     0.822 \pm     0.018 $ & $     0.892 \pm     0.025 $ & $     0.570 \pm     0.028 $ & $     0.477 \pm     0.024 $ \\
F05-23 & $     0.132 \pm     0.003 $ & $     0.116 \pm     0.003 $ & $     0.073 \pm     0.003 $ & $     0.087 \pm     0.011 $ & $     0.325 \pm     0.013 $ & $     0.619 \pm     0.023 $ & $     0.365 \pm     0.031 $ & $     0.585 \pm     0.041 $ \\
F07-11 & $     0.218 \pm     0.004 $ & $     0.171 \pm     0.003 $ & $     0.121 \pm     0.003 $ & $     0.146 \pm     0.014 $ & $     0.520 \pm     0.015 $ & $     0.944 \pm     0.025 $ & $     1.040 \pm     0.048 $ & $     0.918 \pm     0.036 $ \\
F07-12 & $     0.279 \pm     0.005 $ & $     0.204 \pm     0.003 $ & $     0.145 \pm     0.003 $ & $     0.202 \pm     0.013 $ & $     0.564 \pm     0.015 $ & $     0.680 \pm     0.022 $ & $     0.784 \pm     0.039 $ & $     0.840 \pm     0.033 $ \\
F08-1 & $     2.080 \pm     0.010 $ & $     1.330 \pm     0.007 $ & --- & $     4.390 \pm     0.032 $ & $     5.100 \pm     0.028 $ & $     5.200 \pm     0.038 $ & $     5.110 \pm     0.056 $ & $     4.630 \pm     0.047 $ \\
F08-5 & $     0.878 \pm     0.007 $ & $     0.649 \pm     0.005 $ & $     0.503 \pm     0.005 $ & $     1.590 \pm     0.022 $ & $     3.080 \pm     0.023 $ & $     3.800 \pm     0.035 $ & $     3.590 \pm     0.049 $ & $     3.170 \pm     0.043 $ \\
F08-11 & $     0.342 \pm     0.005 $ & $     0.240 \pm     0.004 $ & $     0.180 \pm     0.003 $ & $     0.319 \pm     0.014 $ & $     0.852 \pm     0.017 $ & $     0.967 \pm     0.025 $ & $     1.140 \pm     0.042 $ & $     1.390 \pm     0.034 $ \\
F08-14 & $     0.479 \pm     0.006 $ & $     0.413 \pm     0.005 $ & $     0.272 \pm     0.004 $ & $     0.198 \pm     0.012 $ & $     0.761 \pm     0.016 $ & $     1.320 \pm     0.026 $ & $     1.060 \pm     0.035 $ & $     1.010 \pm     0.028 $ \\
F08-17 & $     0.200 \pm     0.004 $ & $     0.144 \pm     0.003 $ & $     0.129 \pm     0.003 $ & $     0.263 \pm     0.013 $ & $     0.563 \pm     0.015 $ & $     0.701 \pm     0.022 $ & $     0.735 \pm     0.036 $ & $     0.667 \pm     0.039 $ \\
F08-23 & $     1.190 \pm     0.008 $ & $     0.728 \pm     0.005 $ & $     0.468 \pm     0.004 $ & $     0.540 \pm     0.018 $ & $     0.616 \pm     0.017 $ & $     0.682 \pm     0.025 $ & $     0.475 \pm     0.039 $ & $     0.389 \pm     0.032 $ \\
F09-2 & $     1.540 \pm     0.009 $ & $     1.050 \pm     0.007 $ & $     0.832 \pm     0.005 $ & $     2.710 \pm     0.025 $ & $     4.500 \pm     0.025 $ & $     5.250 \pm     0.039 $ & $     5.180 \pm     0.057 $ & $     4.760 \pm     0.048 $ \\
F10-8 & $     0.215 \pm     0.004 $ & $     0.191 \pm     0.003 $ & $     0.129 \pm     0.003 $ & $     0.158 \pm     0.013 $ & $     0.664 \pm     0.018 $ & $     1.230 \pm     0.028 $ & $     1.090 \pm     0.037 $ & $     0.915 \pm     0.030 $ \\
F10-10 & $     0.161 \pm     0.004 $ & $     0.130 \pm     0.003 $ & $     0.097 \pm     0.003 $ & $     0.143 \pm     0.010 $ & $     0.459 \pm     0.015 $ & $     0.598 \pm     0.022 $ & $     0.519 \pm     0.026 $ & $     0.486 \pm     0.023 $ \\
F10-13 & $     0.154 \pm     0.004 $ & $     0.142 \pm     0.003 $ & $     0.105 \pm     0.003 $ & $     0.121 \pm     0.011 $ & $     0.472 \pm     0.015 $ & $     0.888 \pm     0.027 $ & $     0.694 \pm     0.042 $ & $     0.990 \pm     0.043 $ \\
F10-19 & $     0.095 \pm     0.003 $ & $     0.085 \pm     0.003 $ & $     0.064 \pm     0.002 $ & $     0.070 \pm     0.008 $ & $     0.366 \pm     0.015 $ & $     0.585 \pm     0.021 $ & $     0.330 \pm     0.030 $ & $     0.498 \pm     0.032 $ \\
F11-7 & $     0.513 \pm     0.006 $ & $     0.331 \pm     0.004 $ & $     0.268 \pm     0.004 $ & $     0.740 \pm     0.016 $ & $     1.660 \pm     0.020 $ & $     2.100 \pm     0.029 $ & $     2.040 \pm     0.043 $ & $     1.760 \pm     0.035 $ \\
F11-11 & $     0.232 \pm     0.004 $ & $     0.159 \pm     0.003 $ & $     0.139 \pm     0.003 $ & $     0.484 \pm     0.017 $ & $     0.926 \pm     0.017 $ & $     1.180 \pm     0.025 $ & $     1.030 \pm     0.038 $ & $     1.170 \pm     0.035 $ \\
F11-19 & $     0.800 \pm     0.007 $ & $     0.521 \pm     0.005 $ & $     0.336 \pm     0.004 $ & $     0.635 \pm     0.017 $ & $     0.799 \pm     0.018 $ & $     0.798 \pm     0.024 $ & $     0.745 \pm     0.036 $ & $     0.541 \pm     0.025 $ \\
F11-23 & $     0.376 \pm     0.005 $ & $     0.255 \pm     0.004 $ & $     0.176 \pm     0.003 $ & $     0.469 \pm     0.015 $ & $     0.525 \pm     0.015 $ & $     0.536 \pm     0.021 $ & $     0.504 \pm     0.027 $ & $     0.509 \pm     0.027 $ \\
F20-3 & $     1.010 \pm     0.007 $ & $     0.823 \pm     0.006 $ & $     0.640 \pm     0.005 $ & $     3.730 \pm     0.031 $ & $     4.670 \pm     0.026 $ & $     4.690 \pm     0.039 $ & --- & --- \\
F20-10 & $     0.179 \pm     0.004 $ & $     0.154 \pm     0.003 $ & $     0.110 \pm     0.003 $ & $     0.080 \pm     0.008 $ & $     0.569 \pm     0.015 $ & $     0.963 \pm     0.024 $ & $     0.609 \pm     0.032 $ & $     1.030 \pm     0.030 $ \\
F21-1 & $     0.660 \pm     0.006 $ & $     0.463 \pm     0.005 $ & $     0.324 \pm     0.004 $ & $     2.120 \pm     0.023 $ & $     2.110 \pm     0.021 $ & $     2.100 \pm     0.029 $ & $     2.750 \pm     0.048 $ & $     2.780 \pm     0.043 $ \\
F21-20 & $     0.125 \pm     0.003 $ & $     0.113 \pm     0.003 $ & $     0.077 \pm     0.002 $ & --- & $     0.266 \pm     0.012 $ & $     0.466 \pm     0.020 $ & $     0.388 \pm     0.033 $ & $     0.257 \pm     0.024 $ \\
\end{longtable}
\end{landscape}

\begin{longtable}{lcccccc}
\caption{\label{table5}Optical photometry and source extraction flag in the CFHT optical NEP catalogue of the {\it SPICY} PAH galaxies.
The flags are 0 for clear detection, 1 for having neighbours, 2 for being blended, and 3 for both 1 and 2 (see \citealt{hwang07} for more).
}\\
\hline
\hline
ID & $u^{\rm *}$ & $g'$ & $r'$ & $i'$ & $z'$ & flag \\
 & (mJy) & (mJy) & (mJy) & (mJy) & (mJy) & \\
\hline
\endfirsthead
\caption{continued.}\\
\hline
\hline
ID & $u^{\rm *}$ & $g'$ & $r'$ & $i'$ & $z'$ & flag \\
 & (mJy) & (mJy) & (mJy) & (mJy) & (mJy) & \\
\hline
\endhead
\hline
\endfoot
F00-1 & --- & $    0.0541 \pm    0.0001 $ & $    0.1138 \pm    0.0001 $ & $    0.1629 \pm    0.0003 $ & $    0.1986 \pm    0.0007 $ & 0 \\
F01-7 & --- & $    0.5749 \pm    0.0000 $ & $    1.2348 \pm    0.0000 $ & $    1.6520 \pm    0.0000 $ & $    1.9898 \pm    0.0018 $ & 2 \\
F01-13 & --- & $    0.0908 \pm    0.0001 $ & $    0.1716 \pm    0.0002 $ & $    0.2359 \pm    0.0004 $ & $    0.2965 \pm    0.0008 $ & 0 \\
F01-15 & --- & $    0.1310 \pm    0.0001 $ & $    0.1932 \pm    0.0002 $ & $    0.2417 \pm    0.0002 $ & $    0.2729 \pm    0.0008 $ & 0 \\
F01-19 & --- & $    0.0119 \pm    0.0001 $ & $    0.0365 \pm    0.0001 $ & $    0.0541 \pm    0.0002 $ & $    0.0751 \pm    0.0006 $ & 0 \\
F02-0 & $    0.1306 \pm    0.0002 $ & $    0.3128 \pm    0.0003 $ & $    0.4948 \pm    0.0005 $ & $    0.6020 \pm    0.0006 $ & $    0.7171 \pm    0.0013 $ & 2 \\
F02-4 & $    0.0068 \pm    0.0001 $ & $    0.0221 \pm    0.0001 $ & $    0.0677 \pm    0.0001 $ & $    0.1022 \pm    0.0003 $ & $    0.1444 \pm    0.0005 $ & 0 \\
F02-6 & $    0.0645 \pm    0.0002 $ & $    0.1796 \pm    0.0002 $ & $    0.3148 \pm    0.0003 $ & $    0.3945 \pm    0.0004 $ & $    0.5119 \pm    0.0009 $ & 0 \\
F02-8 & $    0.0147 \pm    0.0001 $ & $    0.0412 \pm    0.0001 $ & $    0.1002 \pm    0.0002 $ & $    0.1437 \pm    0.0004 $ & $    0.2040 \pm    0.0006 $ & 0 \\
F02-9 & $    0.0116 \pm    0.0001 $ & $    0.0260 \pm    0.0001 $ & $    0.0660 \pm    0.0002 $ & $    0.0944 \pm    0.0003 $ & $    0.1378 \pm    0.0006 $ & 2 \\
F02-15 & $    0.0196 \pm    0.0001 $ & $    0.0478 \pm    0.0001 $ & $    0.1045 \pm    0.0002 $ & $    0.1474 \pm    0.0003 $ & $    0.1772 \pm    0.0007 $ & 0 \\
F02-20 & $    0.0031 \pm    0.0001 $ & $    0.0049 \pm    0.0001 $ & $    0.0164 \pm    0.0001 $ & $    0.0236 \pm    0.0002 $ & $    0.0705 \pm    0.0004 $ & 3 \\
F02-24 & $    0.0076 \pm    0.0001 $ & $    0.0270 \pm    0.0001 $ & $    0.0694 \pm    0.0001 $ & $    0.0988 \pm    0.0003 $ & $    0.1008 \pm    0.0005 $ & 0 \\
F03-4 & --- & $    0.0735 \pm    0.0001 $ & $    0.1164 \pm    0.0001 $ & $    0.1455 \pm    0.0003 $ & $    0.1560 \pm    0.0007 $ & 2 \\
F03-15 & --- & $    0.0102 \pm    0.0001 $ & $    0.0254 \pm    0.0001 $ & $    0.0395 \pm    0.0002 $ & $    0.0440 \pm    0.0004 $ & 0 \\
F04-0 & --- & $    0.5618 \pm    0.0000 $ & $    1.0224 \pm    0.0000 $ & $    1.3957 \pm    0.0000 $ & $    1.5981 \pm    0.0015 $ & 0 \\
F04-3 & --- & $    0.0697 \pm    0.0001 $ & $    0.1646 \pm    0.0002 $ & $    0.2468 \pm    0.0002 $ & $    0.3045 \pm    0.0008 $ & 2 \\
F04-5 & --- & --- & $    0.4626 \pm    0.0004 $ & $    0.5577 \pm    0.0005 $ & --- & 0 \\
F04-6 & --- & --- & $    0.0147 \pm    0.0001 $ & $    0.0234 \pm    0.0002 $ & --- & 0 \\
F04-7 & --- & $    0.2694 \pm    0.0002 $ & $    0.3908 \pm    0.0004 $ & $    0.5077 \pm    0.0005 $ & $    0.5526 \pm    0.0015 $ & 0 \\
F04-8 & --- & --- & $    0.0264 \pm    0.0001 $ & $    0.0425 \pm    0.0002 $ & --- & 0 \\
F04-13 & --- & --- & $    0.1768 \pm    0.0002 $ & $    0.2521 \pm    0.0002 $ & --- & 0 \\
F04-18 & --- & --- & $    0.0387 \pm    0.0001 $ & $    0.0521 \pm    0.0001 $ & --- & 0 \\
F04-23 & --- & $    0.0095 \pm    0.0001 $ & $    0.0196 \pm    0.0001 $ & $    0.0256 \pm    0.0002 $ & $    0.0306 \pm    0.0005 $ & 0 \\
F04-24\footnote{This source is damaged due to a nearby bright and saturated star on the CFHT image, and is not catalogued.} & --- & --- & --- & --- & --- & --- \\
F05-10 & $    0.0990 \pm    0.0003 $ & $    0.2359 \pm    0.0002 $ & $    0.3898 \pm    0.0004 $ & $    0.4699 \pm    0.0004 $ & $    0.5649 \pm    0.0010 $ & 0 \\
F05-23 & $    0.0042 \pm    0.0001 $ & $    0.0086 \pm    0.0001 $ & $    0.0232 \pm    0.0002 $ & $    0.0396 \pm    0.0003 $ & $    0.0435 \pm    0.0005 $ & 0 \\
F07-11 & --- & $    0.0183 \pm    0.0001 $ & $    0.0520 \pm    0.0001 $ & $    0.0707 \pm    0.0002 $ & $    0.0898 \pm    0.0005 $ & 0 \\
F07-12 & --- & $    0.0309 \pm    0.0001 $ & $    0.0771 \pm    0.0001 $ & $    0.1097 \pm    0.0002 $ & $    0.1338 \pm    0.0006 $ & 0 \\
F08-1\footnote{This source falls on one of CCD gaps on the CFHT image, and is not catalogued.} & --- & --- & --- & --- & --- & --- \\
F08-5 & --- & $    0.1572 \pm    0.0001 $ & $    0.3045 \pm    0.0003 $ & $    0.4321 \pm    0.0004 $ & $    0.5143 \pm    0.0009 $ & 2 \\
F08-11 & --- & $    0.0217 \pm    0.0001 $ & $    0.0656 \pm    0.0001 $ & $    0.1070 \pm    0.0003 $ & $    0.1392 \pm    0.0006 $ & 0 \\
F08-14 & --- & $    0.0283 \pm    0.0001 $ & $    0.0792 \pm    0.0001 $ & $    0.1161 \pm    0.0003 $ & $    0.1444 \pm    0.0008 $ & 2 \\
F08-17 & --- & $    0.1020 \pm    0.0001 $ & $    0.1008 \pm    0.0002 $ & $    0.1435 \pm    0.0003 $ & $    0.1274 \pm    0.0007 $ & 0 \\
F08-23 & --- & $    0.2399 \pm    0.0002 $ & $    0.5181 \pm    0.0005 $ & $    0.7053 \pm    0.0006 $ & $    0.8904 \pm    0.0008 $ & 0 \\
F09-2 & --- & $    0.3688 \pm    0.0000 $ & $    0.5932 \pm    0.0000 $ & $    1.1663 \pm    0.0000 $ & $    1.0028 \pm    0.0009 $ & 0 \\
F10-8 & --- & $    0.0083 \pm    0.0001 $ & $    0.0274 \pm    0.0001 $ & $    0.0400 \pm    0.0002 $ & $    0.0583 \pm    0.0006 $ & 0 \\
F10-10 & --- & $    0.0075 \pm    0.0001 $ & $    0.0227 \pm    0.0001 $ & $    0.0337 \pm    0.0002 $ & $    0.0468 \pm    0.0005 $ & 0 \\
F10-13 & --- & $    0.0043 \pm    0.0001 $ & $    0.0101 \pm    0.0001 $ & $    0.0142 \pm    0.0002 $ & $    0.0143 \pm    0.0005 $ & 3 \\
F10-19 & --- & $    0.0041 \pm    0.0001 $ & $    0.0127 \pm    0.0001 $ & $    0.0218 \pm    0.0002 $ & $    0.0210 \pm    0.0006 $ & 0 \\
F11-7 & --- & $    0.0707 \pm    0.0001 $ & $    0.1521 \pm    0.0001 $ & $    0.2180 \pm    0.0002 $ & $    0.2635 \pm    0.0007 $ & 0 \\
F11-11 & --- & $    0.0633 \pm    0.0001 $ & $    0.1073 \pm    0.0001 $ & $    0.1426 \pm    0.0003 $ & $    0.1586 \pm    0.0006 $ & 0 \\
F11-19 & --- & $    0.1337 \pm    0.0001 $ & $    0.3020 \pm    0.0003 $ & $    0.4274 \pm    0.0004 $ & $    0.5521 \pm    0.0010 $ & 0 \\
F11-23 & --- & $    0.0682 \pm    0.0001 $ & $    0.1460 \pm    0.0001 $ & $    0.2055 \pm    0.0002 $ & $    0.2604 \pm    0.0007 $ & 0 \\
F20-3 & --- & $    0.1723 \pm    0.0002 $ & $    0.3287 \pm    0.0003 $ & $    0.4639 \pm    0.0004 $ & $    0.5749 \pm    0.0011 $ & 0 \\
F20-10 & --- & $    0.0130 \pm    0.0001 $ & $    0.0347 \pm    0.0001 $ & $    0.0492 \pm    0.0002 $ & $    0.0760 \pm    0.0006 $ & 0 \\
F21-1 & --- & $    0.1057 \pm    0.0001 $ & $    0.2423 \pm    0.0002 $ & $    0.3330 \pm    0.0003 $ & $    0.4305 \pm    0.0004 $ & 0 \\
F21-20 & --- & $    0.0049 \pm    0.0001 $ & $    0.0168 \pm    0.0001 $ & $    0.0272 \pm    0.0002 $ & $    0.0354 \pm    0.0006 $ & 0 \\
\end{longtable}

\begin{longtable}{lccccccc}
\caption{\label{table6}Similar table as Table~\ref{table2}, but for the bright AGN candidates.
Optical spectroscopic classification results by \cite{shim13}, if available, are shown in the ``class'' column. ``TYPE1'' indicates type-1 AGN.
}\\
\hline
\hline
{\it SPICY} ID & \multicolumn{3}{c}{AKARI/IRC NEP-Wide catalogue} & {\it SPICY} S9W flux & \multicolumn{2}{c}{CFHT NEP catalogue} & class \\
 & ID & RA J2000 (hms) & DEC J2000 (dms) & (mJy) & ID & FIELD & \\
\hline
\endfirsthead
\caption{continued.}\\
\hline
\hline
{\it SPICY} ID & \multicolumn{3}{c}{AKARI/IRC NEP-Wide catalogue} & {\it SPICY} S9W flux & \multicolumn{2}{c}{CFHT NEP catalogue} & \\
 & ID & RA J2000 (hms) & DEC J2000 (dms) & (mJy) & ID & FIELD & \\
\hline
\endhead
\hline
\endfoot
F00-3 & 65028336 & 17 54 54.05 &  +66 34 18.0 & $      1.23 \pm      0.04 $ & 37688 & 2 & TYPE1 \\
F04-20 & 65041740 & 17 57 32.54 &  +66 40 28.5 & $      0.43 \pm      0.02 $ & 46530 & 2 & --- \\
F05-1 & 65055361 & 18 00 11.66 &  +66 52 15.0 & $      4.25 \pm      0.13 $ & 55298 & 1 & TYPE1 \\
F05-13 & 65055061 & 18 00 08.16 &  +66 55 01.1 & $      0.62 \pm      0.02 $ & 58237 & 1 & --- \\
F10-20 & 65040243 & 17 57 14.66 &  +66 31 13.2 & $      0.34 \pm      0.02 $ & 32913 & 2 & TYPE1 \\
F11-14 & 65037307 & 17 56 41.06 &  +66 05 25.7 & $      0.83 \pm      0.03 $ & 1859 & 2 & TYPE1 \\
\end{longtable}

\begin{landscape}
\begin{longtable}{lcccccccc}
\caption{\label{table7}Similar table as Table~\ref{table4}, but for the bright AGN candidates.
}\\
\hline
\hline
ID & $N2$ & $N3$ & $N4$ & $S7$ & $S9W$ & $S11$ & $L15$ & $L18W$ \\
 & (mJy) & (mJy) & (mJy) & (mJy) & (mJy) & (mJy) & (mJy) & (mJy) \\
\hline
\endfirsthead
\caption{continued.}\\
\hline
\hline
ID & $N2$ & $N3$ & $N4$ & $S7$ & $S9W$ & $S11$ & $L15$ & $L18W$ \\
 & (mJy) & (mJy) & (mJy) & (mJy) & (mJy) & (mJy) & (mJy) & (mJy) \\
\hline
\endhead
\hline
\endfoot
F00-3 & $     0.125 \pm     0.004 $ & $     0.185 \pm     0.003 $ & $     0.282 \pm     0.004 $ & $     0.931 \pm     0.017 $ & $     1.210 \pm     0.018 $ & $     1.630 \pm     0.027 $ & $     2.480 \pm     0.045 $ & $     2.760 \pm     0.040 $ \\
F04-20 & $     0.074 \pm     0.003 $ & $     0.138 \pm     0.003 $ & $     0.149 \pm     0.003 $ & $     0.385 \pm     0.014 $ & $     0.520 \pm     0.015 $ & $     0.590 \pm     0.021 $ & $     0.707 \pm     0.030 $ & $     0.825 \pm     0.029 $ \\
F05-1 & $     0.425 \pm     0.005 $ & $     0.561 \pm     0.005 $ & $     0.944 \pm     0.005 $ & $     3.030 \pm     0.028 $ & $     4.390 \pm     0.024 $ & $     5.960 \pm     0.038 $ & $    11.500 \pm     0.074 $ & $    14.400 \pm     0.066 $ \\
F05-13 & $     0.065 \pm     0.004 $ & $     0.116 \pm     0.003 $ & $     0.177 \pm     0.003 $ & $     0.461 \pm     0.015 $ & $     0.612 \pm     0.016 $ & $     0.766 \pm     0.026 $ & $     0.865 \pm     0.032 $ & $     0.850 \pm     0.029 $ \\
F10-20 & $     0.149 \pm     0.004 $ & $     0.226 \pm     0.004 $ & $     0.254 \pm     0.004 $ & $     0.301 \pm     0.013 $ & $     0.346 \pm     0.013 $ & $     0.390 \pm     0.021 $ & $     0.606 \pm     0.039 $ & $     0.701 \pm     0.033 $ \\
F11-14 & $     0.150 \pm     0.004 $ & $     0.210 \pm     0.003 $ & $     0.334 \pm     0.004 $ & $     0.712 \pm     0.017 $ & $     0.983 \pm     0.017 $ & $     1.170 \pm     0.025 $ & $     1.830 \pm     0.042 $ & $     2.070 \pm     0.040 $ \\
\end{longtable}
\end{landscape}

\begin{longtable}{lcccccc}
\caption{\label{table8}Similar table as Table~\ref{table5}, but for the bright AGN candidates.
}\\
\hline
\hline
ID & $u^{\rm *}$ & $g'$ & $r'$ & $i'$ & $z'$ & flag \\
 & (mJy) & (mJy) & (mJy) & (mJy) & (mJy) & \\
\hline
\endfirsthead
\caption{continued.}\\
\hline
\hline
ID & $u^{\rm *}$ & $g'$ & $r'$ & $i'$ & $z'$ & flag \\
 & (mJy) & (mJy) & (mJy) & (mJy) & (mJy) & \\
\hline
\endhead
\hline
\endfoot
F00-3 & --- & $    0.0595 \pm    0.0001 $ & $    0.1303 \pm    0.0001 $ & $    0.0650 \pm    0.0001 $ & $    0.1361 \pm    0.0004 $ & 0 \\
F04-20 & --- &  --- & $    0.0054 \pm    0.0001 $ & $    0.0128 \pm    0.0001 $ & --- & 0 \\
F05-1 & $    0.0105 \pm    0.0001 $ & $    0.0274 \pm    0.0001 $ & $    0.1075 \pm    0.0001 $ & $    0.1694 \pm    0.0002 $ & $    0.1982 \pm    0.0004 $ & 0 \\
F05-13 & $    0.0002 \pm    0.0001 $ & $    0.0005 \pm    0.0000 $ & $    0.0010 \pm    0.0001 $ & $    0.0020 \pm    0.0001 $ & $    0.0057 \pm    0.0002 $ & 0 \\
F10-20 & --- & $    0.0056 \pm    0.0001 $ & $    0.0198 \pm    0.0001 $ & $    0.0302 \pm    0.0002 $ & $    0.0391 \pm    0.0005 $ & 0 \\
F11-14 & --- & $    0.0493 \pm    0.0000 $ & $    0.0712 \pm    0.0001 $ & $    0.0935 \pm    0.0002 $ & $    0.1079 \pm    0.0004 $ & 0 \\
\end{longtable}

\begin{longtable}{lccccc}
\caption{\label{table9}Monochromatic photometric luminosities and rest-frame flux ratios of the {\it SPICY} PAH galaxies.}\\
\hline
\hline
ID & $\log \nu L \nu _{\rm phot}$ ($3.5\ \mu$m) & $\log \nu L \nu _{\rm phot}$ ($7.7\ \mu$m) & \multicolumn{3}{c}{Rest-frame flux ratios} \\
 & ($L_{\sun}$) & ($L_{\sun}$) & $F_{\rm rest\ 3.5\ \mu \rm m}/F_{\rm rest\ 2.0\ \mu \rm m}$ & $F_{\rm rest\ 7.7\ \mu \rm m}/F_{\rm rest\ 3.5\ \mu \rm m}$ & $F_{\rm rest\ 11.3\ \mu \rm m}/F_{\rm rest\ 7.7\ \mu \rm m}$ \\
\hline
\endfirsthead
\caption{continued.}\\
\hline
\hline
ID & $\log \nu L \nu _{\rm phot}$ ($3.5\ \mu$m) & $\log \nu L \nu _{\rm phot}$ ($7.7\ \mu$m) & \multicolumn{3}{c}{Rest-frame flux ratios} \\
 & ($L_{\sun}$) & ($L_{\sun}$) & $F_{\rm rest\ 3.5\ \mu \rm m}/F_{\rm rest\ 2.0\ \mu \rm m}$ & $F_{\rm rest\ 7.7\ \mu \rm m}/F_{\rm rest\ 3.5\ \mu \rm m}$ & $F_{\rm rest\ 11.3\ \mu \rm m}/F_{\rm rest\ 7.7\ \mu \rm m}$ \\
\hline
\endhead
\hline
\endfoot
F00-1 & $     9.770^{+0.010}_{-0.010} $ & $    10.234^{+0.009}_{-0.009} $ & $     0.611 \pm     0.018 $ & $     6.398 \pm     0.203 $ & $     1.035 \pm     0.041 $ \\
F01-7 & $     9.571^{+0.003}_{-0.003} $ & $     9.373^{+0.008}_{-0.008} $ & $     0.454 \pm     0.004 $ & $     1.394 \pm     0.027 $ & --- \\
F01-13 & $     9.188^{+0.010}_{-0.010} $ & $     9.334^{+0.016}_{-0.017} $ & $     0.521 \pm     0.015 $ & $     3.082 \pm     0.139 $ & $     1.038 \pm     0.063 $ \\
F01-15 & $     8.808^{+0.012}_{-0.012} $ & $     9.015^{+0.018}_{-0.019} $ & $     0.537 \pm     0.018 $ & $     3.545 \pm     0.178 $ & --- \\
F01-19 & $    10.112^{+0.019}_{-0.020} $ & $    10.662^{+0.017}_{-0.018} $ & $     0.578 \pm     0.032 $ & $     7.810 \pm     0.475 $ & $     0.838 \pm     0.072 $ \\
F02-0 & $     9.253^{+0.007}_{-0.007} $ & $     9.430^{+0.010}_{-0.010} $ & $     0.509 \pm     0.010 $ & $     3.307 \pm     0.091 $ & $     1.055 \pm     0.036 $ \\
F02-4 & $    10.031^{+0.010}_{-0.011} $ & $    10.406^{+0.011}_{-0.012} $ & $     0.559 \pm     0.016 $ & $     5.220 \pm     0.188 $ & $     0.931 \pm     0.048 $ \\
F02-6 & $     9.555^{+0.009}_{-0.009} $ & $     9.672^{+0.009}_{-0.010} $ & $     0.499 \pm     0.012 $ & $     2.882 \pm     0.085 $ & $     1.147 \pm     0.051 $ \\
F02-8 & $    10.004^{+0.009}_{-0.009} $ & $    10.103^{+0.014}_{-0.015} $ & $     0.555 \pm     0.014 $ & $     2.766 \pm     0.110 $ & $     1.074 \pm     0.067 $ \\
F02-9 & $     9.856^{+0.014}_{-0.015} $ & $    10.216^{+0.016}_{-0.016} $ & $     0.616 \pm     0.025 $ & $     5.031 \pm     0.248 $ & $     0.902 \pm     0.062 $ \\
F02-15 & $     9.590^{+0.015}_{-0.015} $ & $     9.862^{+0.019}_{-0.020} $ & $     0.529 \pm     0.022 $ & $     4.122 \pm     0.233 $ & $     1.012 \pm     0.088 $ \\
F02-20 & $    10.199^{+0.016}_{-0.016} $ & $    10.578^{+0.018}_{-0.019} $ & $     0.581 \pm     0.026 $ & $     5.263 \pm     0.295 $ & $     1.005 \pm     0.071 $ \\
F02-24 & $     9.373^{+0.017}_{-0.017} $ & $     9.505^{+0.021}_{-0.022} $ & $     0.477 \pm     0.022 $ & $     2.984 \pm     0.189 $ & $     0.933 \pm     0.095 $ \\
F03-4 & $     9.628^{+0.010}_{-0.010} $ & $     9.896^{+0.010}_{-0.010} $ & $     0.618 \pm     0.018 $ & $     4.077 \pm     0.136 $ & $     1.044 \pm     0.051 $ \\
F03-15 & $     9.265^{+0.027}_{-0.028} $ & $     9.649^{+0.026}_{-0.028} $ & $     0.630 \pm     0.048 $ & $     5.319 \pm     0.474 $ & $     0.909 \pm     0.112 $ \\
F04-0 & $     9.703^{+0.003}_{-0.003} $ & $    10.007^{+0.004}_{-0.004} $ & $     0.504 \pm     0.004 $ & $     4.426 \pm     0.051 $ & $     1.129 \pm     0.015 $ \\
F04-3 & $     9.772^{+0.006}_{-0.007} $ & $    10.101^{+0.006}_{-0.006} $ & $     0.537 \pm     0.010 $ & $     4.694 \pm     0.095 $ & $     1.142 \pm     0.031 $ \\
F04-5 & $     8.749^{+0.008}_{-0.008} $ & $     8.930^{+0.011}_{-0.011} $ & $     0.518 \pm     0.011 $ & $     3.339 \pm     0.103 $ & --- \\
F04-6 & $     9.940^{+0.020}_{-0.021} $ & $    10.741^{+0.011}_{-0.011} $ & $     0.642 \pm     0.038 $ & $    13.897 \pm     0.747 $ & $     0.643 \pm     0.041 $ \\
F04-7 & $     8.966^{+0.010}_{-0.010} $ & $     9.153^{+0.016}_{-0.017} $ & $     0.448 \pm     0.012 $ & $     3.384 \pm     0.151 $ & $     1.091 \pm     0.068 $ \\
F04-8 & $     9.714^{+0.016}_{-0.017} $ & $    10.170^{+0.015}_{-0.016} $ & $     0.942 \pm     0.048 $ & $     6.287 \pm     0.326 $ & $     1.063 \pm     0.066 $ \\
F04-13 & $     9.730^{+0.009}_{-0.009} $ & $     9.789^{+0.018}_{-0.019} $ & $     0.497 \pm     0.012 $ & $     2.522 \pm     0.119 $ & $     0.756 \pm     0.057 $ \\
F04-18 & $     9.654^{+0.019}_{-0.020} $ & $    10.092^{+0.021}_{-0.022} $ & $     0.715 \pm     0.042 $ & $     6.030 \pm     0.406 $ & $     1.540 \pm     0.114 $ \\
F04-23 & $     9.901^{+0.025}_{-0.026} $ & $    10.564^{+0.019}_{-0.020} $ & $     0.612 \pm     0.044 $ & $    10.119 \pm     0.741 $ & $     0.609 \pm     0.062 $ \\
F04-24 & $     9.800^{+0.014}_{-0.014} $ & $     9.941^{+0.022}_{-0.023} $ & $     0.698 \pm     0.029 $ & $     3.044 \pm     0.184 $ & $     1.438 \pm     0.113 $ \\
F05-10 & $     8.931^{+0.009}_{-0.009} $ & $     9.017^{+0.015}_{-0.016} $ & $     0.475 \pm     0.011 $ & $     2.681 \pm     0.109 $ & $     0.952 \pm     0.054 $ \\
F05-23 & $     9.724^{+0.027}_{-0.028} $ & $    10.371^{+0.018}_{-0.018} $ & $     0.500 \pm     0.037 $ & $     9.751 \pm     0.738 $ & $     0.629 \pm     0.062 $ \\
F07-11 & $     9.961^{+0.017}_{-0.018} $ & $    10.572^{+0.013}_{-0.013} $ & $     0.556 \pm     0.028 $ & $     8.984 \pm     0.454 $ & $     1.086 \pm     0.062 $ \\
F07-12 & $     9.782^{+0.013}_{-0.014} $ & $    10.099^{+0.018}_{-0.019} $ & $     0.553 \pm     0.021 $ & $     4.568 \pm     0.240 $ & $     1.187 \pm     0.087 $ \\
F08-1 & --- & $     9.956^{+0.004}_{-0.004} $ & --- & --- & $     1.034 \pm     0.015 $ \\
F08-5 & $     9.754^{+0.006}_{-0.006} $ & $    10.182^{+0.004}_{-0.004} $ & $     0.597 \pm     0.010 $ & $     5.884 \pm     0.092 $ & $     1.169 \pm     0.020 $ \\
F08-11 & $     9.766^{+0.012}_{-0.012} $ & $    10.140^{+0.013}_{-0.013} $ & $     0.554 \pm     0.019 $ & $     5.199 \pm     0.214 $ & $     1.195 \pm     0.064 $ \\
F08-14 & $    10.349^{+0.011}_{-0.012} $ & $    10.764^{+0.010}_{-0.010} $ & $     0.516 \pm     0.016 $ & $     5.717 \pm     0.203 $ & $     0.816 \pm     0.036 $ \\
F08-17 & $     9.430^{+0.016}_{-0.016} $ & $     9.770^{+0.015}_{-0.016} $ & $     0.651 \pm     0.030 $ & $     4.809 \pm     0.245 $ & $     1.186 \pm     0.078 $ \\
F08-23 & $     9.350^{+0.005}_{-0.005} $ & $     9.039^{+0.020}_{-0.021} $ & $     0.422 \pm     0.006 $ & $     1.074 \pm     0.052 $ & $     0.973 \pm     0.076 $ \\
F09-2 & $     9.738^{+0.004}_{-0.004} $ & $    10.081^{+0.004}_{-0.004} $ & $     0.550 \pm     0.006 $ & $     4.852 \pm     0.065 $ & $     1.187 \pm     0.017 $ \\
F10-8 & $    10.103^{+0.019}_{-0.019} $ & $    10.825^{+0.012}_{-0.012} $ & $     0.526 \pm     0.027 $ & $    11.577 \pm     0.602 $ & $     0.851 \pm     0.045 $ \\
F10-10 & $     9.648^{+0.019}_{-0.020} $ & $    10.075^{+0.020}_{-0.021} $ & $     0.633 \pm     0.036 $ & $     5.881 \pm     0.387 $ & $     0.949 \pm     0.081 $ \\
F10-13 & $    10.049^{+0.021}_{-0.022} $ & $    10.678^{+0.017}_{-0.018} $ & $     0.609 \pm     0.037 $ & $     9.554 \pm     0.599 $ & $     0.948 \pm     0.076 $ \\
F10-19 & $     9.726^{+0.025}_{-0.027} $ & $    10.372^{+0.018}_{-0.019} $ & $     0.640 \pm     0.049 $ & $     9.742 \pm     0.718 $ & $     0.628 \pm     0.066 $ \\
F11-7 & $     9.640^{+0.009}_{-0.009} $ & $    10.113^{+0.006}_{-0.006} $ & $     0.530 \pm     0.013 $ & $     6.531 \pm     0.162 $ & $     1.170 \pm     0.032 $ \\
F11-11 & $     9.375^{+0.014}_{-0.014} $ & $     9.889^{+0.010}_{-0.010} $ & $     0.601 \pm     0.024 $ & $     7.180 \pm     0.283 $ & $     1.096 \pm     0.050 $ \\
F11-19 & $     9.160^{+0.007}_{-0.007} $ & $     9.095^{+0.015}_{-0.016} $ & $     0.457 \pm     0.008 $ & $     1.893 \pm     0.074 $ & $     1.009 \pm     0.058 $ \\
F11-23 & $     8.834^{+0.010}_{-0.011} $ & $     8.887^{+0.020}_{-0.021} $ & $     0.499 \pm     0.014 $ & $     2.481 \pm     0.132 $ & $     1.017 \pm     0.074 $ \\
F20-3 & $     9.462^{+0.005}_{-0.005} $ & $     9.925^{+0.005}_{-0.005} $ & $     0.670 \pm     0.009 $ & $     6.387 \pm     0.097 $ & --- \\
F20-10 & $    10.004^{+0.019}_{-0.020} $ & $    10.647^{+0.013}_{-0.013} $ & $     0.575 \pm     0.031 $ & $     9.668 \pm     0.525 $ & $     0.774 \pm     0.049 $ \\
F21-1 & $     8.578^{+0.007}_{-0.007} $ & $     8.964^{+0.007}_{-0.007} $ & $     0.525 \pm     0.010 $ & $     5.352 \pm     0.119 $ & $     1.078 \pm     0.026 $ \\

F21-20 & $     9.782^{+0.024}_{-0.025} $ & $    10.281^{+0.021}_{-0.022} $ & $     0.561 \pm     0.038 $ & $     6.948 \pm     0.522 $ & $     0.808 \pm     0.088 $ \\
\end{longtable}

\appendix

\section{Design of the {\it SPICY} survey in detail}\label{survey_detail}

The {\it SPICY} survey is a blind slitless spectroscopic extragalactic survey at MIR by using the IRC onboard AKARI (Sect.~\ref{survey_section}).
We targeted the NEP located in the Continuous Visibility Zone (CVZ) of the AKARI satellite in sun-synchronous orbit.
Due to severe scheduling constraints posed mostly by the all sky survey of the AKARI mission, the NEP was the only field on the sky where AKARI was able to perform repeated observations of the same field for deeper observations \citep{maruma06}.
For the {\it SPICY} survey, we visited the same pointing coordinates 9 or 10 times to achieve our sensitivity goal (Sect.~\ref{survey_section}).
Another CVZ field, the South Ecliptic Pole (SEP), shows much more Galactic cirrus and is close to the LMC, so is less-suited to extragalactic deep surveys.
This survey program was conducted as one of AKARI's key projects (``mission program''), and $\simeq 110$ pointing opportunities were assigned for it during the cold phase of the AKARI mission.
This survey was executed during July 2006 and August 2007.

The {\it SPICY} survey footprint was designed to efficiently combine spectra taken with the separate cameras of the IRC, under the following technical limitations.
The IRC (see \citealt{onaka07} for full details) is composed of three independent cameras that operate simultaneously, and they are called ``NIR'', ``MIR-S'', and ``MIR-L'' for wavelength ranges of 2--$5\ \mu$m, 6--$13\ \mu$m, and 13--$27\ \mu$m, respectively.
Each camera covers an $\simeq 10'\times 10'$ FOV, and the NIR and MIR-S share the same FOV via a beam splitter.
The MIR-L has its own FOV to cover slightly offset ($\simeq 20'$) sky area.
By overlapping the FOV coverage of the MIR-L onto ones of the NIR and MIR-S, we achieved the combined wavelength coverage between $2\ \mu$m and $27\ \mu$m.
Note, however, that we utilised only the MIR-S spectra for the purpose of this paper (Sect.~\ref{survey_section}).
To enable this, we utilised the fact that the IRC FOV rotates by 1 degree per day at the NEP due to AKARI's sun-synchronous orbit.
We scheduled all observations for a pointing coordinate as close in time as possible to each other to minimise such FOV rotation.
This set of the observations defined one observing field, called a tile.
After competing the very first tile, the second tile was observed about 3 months later, when the FOV had rotated by 90 degrees and the new square tile could be directly overlaid on the previous square (but 90 degrees rotated) tile.
Here, the new pointing coordinates were set in order to overlay the new MIR-L tile on the previous NIR/MIR-S tile.
This generated a new NIR/MIR-S tile $\simeq 20'$ off the first tile.
This new NIR/MIR-S coverage of the second tile was further observed about another 3 month later with the MIR-L.
We repeated this sequence every $\simeq 3$ months, to extend a chain of the tiles where all NIR, MIR-S, and MIR-L FOVs overlap with each other.
The survey tiles thus created were distributed in a non-contiguous way around the NEP, showing a complicated shape of folded chains of 14 tiles (Fig.~\ref{SPICY_tiles}).
In reality, however, the chain of the tiles was interrupted a few times before completing the survey, due to telescope scheduling constraints.

\end{document}